# THÈSE DE DOCTORAT DE

L'ÉCOLE NATIONALE SUPERIEURE MINES-TELECOM ATLANTIQUE BRETAGNE PAYS DE LA LOIRE - IMT ATLANTIQUE
COMUE UNIVERSITÉ BRETAGNE LOIRE

ÉCOLE DOCTORALE N° 601
*Mathématiques et Sciences et Technologies*
*de l'Information et de la Communication*
Spécialité : Informatique

Par

**Ambroise LAFONT**

# Signatures and models for syntax and operational semantics in the presence of variable binding

**Thèse présentée et soutenue à Nantes , le 2 Décembre 2019**
**Unité de recherche : LS2N**

**Rapporteurs avant soutenance :**

| | |
|---|---|
| Marcelo FIORE | Professor, University of Cambridge |
| Peter LeFanu LUMSDAINE | Assistant Professor, Stockholm University |

**Composition du Jury :**

| | | |
|---|---|---|
| Président : | Thomas EHRHARD | Chercheur CNRS senior, Université Paris Diderot |
| Examinateurs : | Benedikt AHRENS | Birmingham Fellow, University of Birmingham |
| | Delia KESNER | Professeur, Université Paris Diderot |
| Dir. de thèse : | Nicolas TABAREAU | Directeur de Recherche, INRIA Rennes |
| Co-dir. de thèse : | Tom HIRSCHOWITZ | Chargé de Recherche, Université Savoie Mont Blanc |

À mon ami Nicolas Tholance, disparu trop tôt.

# Remerciements

Nicolas Tabareau et Tom Hirschowitz encadrent mon travail depuis trois ans. Leur bienveillance et leur disponibilité ont été inestimables pour la réalisation de cette thèse et au-delà.

Cette thèse se nourrit d'articles écrits avec l'aide de Benedikt Ahrens, Marco Maggesi et André Hirschowitz, que je tiens à remercier pour leur collaboration. La rédaction proprement dite de ce manuscrit a bénéficié de nombreuses suggestions d'André Hirschowitz, dont j'admire (autant que je redoute) l'inlassable exigeance. À ce propos, je signale que ces remerciements n'ont pas comparu devant ce relecteur expert, ce qui explique leur médiocrité.

Je tiens à remercier Benedikt de m'avoir accepté comme disciple lorsque j'ai commencé ma thèse, malgré mon style chaotique de programmation Coq, bien loin du standard allemand. Son départ de Nantes m'a privé d'une joyeuse compagnie.

Je voudrais remercier tous ceux qui m'ont remercié dans leur thèse, en particulier Gabriel Pallier. Afin d'encourager cette initiative, je remercie également par avance Théo et Xavier, mes deux co-bureaux thésards, ainsi que Meven et Loïc, fraîchement nominés doctorants.

Dans un souci écologique, et au risque de faire pleurer de jalousie certains membres méritoires, c'est en vrac que je remercie le reste de l'équipe Gallinette. Pour faire bref, je me suis bien fendu la poire au sein de cette troupe croissante de gais lurons. J'ai également beaucoup apprécié ces oisives discussions catégoriques, alors que le devoir m'appelaît à d'autres tâches plus urgentes.

Puis je voudrais exprimer ma gratitude envers mes parents, et tous mes ancêtres (je pense notamment au premier d'entre eux qui eut 23 paires de chromosomes : quelle terrible solitude !), sans qui je ne serai pas là aujourd'hui, mes frères et sœurs qui m'ont supporté toutes ces années, mon oncle Daniel de Rauglaudre, qui a éveillé mon intérêt pour l'informatique.

Comme de bien entendu, c'est par toi, Anne-Cécile cœur, que j'achèverai cette litanie. Mais d'abord, veux-tu être ma femme ?

# TABLE OF CONTENTS

CHAPITRE 1

# INTRODUCTION

## 1.1 Résumé long (en français)

Cette thèse s'intéresse à la mathématisation de la notion de *langage de programmation*, en portant une attention particulière à la notion de *substitution*.

La recherche dans le domaine des langages de programmation s'appuie traditionnellement sur une définition de *syntaxe* modulo renommage des variables liées, avec la *sémantique opérationnelle* associée. Nous nous intéressons à des outils mathématiques permettant de générer automatiquement la syntaxe et la sémantique à partir de données élémentaires.

En ce qui concerne la syntaxe, la spécification de structures algébriques avec variables liées est un enjeu majeur. Deux lignes principales de recherche sont en concurrence : les *ensembles nominaux* [GP99] et les *algèbres de substitution* [FPT99]. Dans cette thèse, nous explorons une variante des algèbres de substitution, proposée par [HM07 ; HM10], qui s'appuie sur la notion de *module sur une monade*. Nous abordons également dans cet esprit la mathématisation de la sémantique opérationnelle. Pour cela, nous introduisons d'abord les *monades de réduction*, puis leur généralisation, les *monades opérationnelles* : elles constituent notre approximation mathématique de la notion informatique de langage de programmation.

Dans cette thèse, nous nous intéressons à la spécification des objets de la catégorie des monades (chapitres 3 et 4), de la catégorie des *monades de réduction* (chapitre 5), et de la catégorie des *monades opérationnelles* (chapitre 6), notre objectif étant de définir un *langage formel*[1], modélisé par un objet d'une catégorie adéquate.

Caractériser un objet d'un certain type (i.e., d'une certaine catégorie $\mathcal{C}$) par une propriété d'initialité est l'objectif de ce que l'on appelle *sémantique initiale* ou *spécification algébrique* [JGW78], popularisés par [BM97]. La méthodologie générale de la

1. Ici, le mot "langage" englobe les types de donnée, les langages de programmation et les calculs logiques, ainsi que les langages pour structures algébriques considérés en algèbre universelle.

sémantique initiale peut être décrite selon les étapes suivantes :

1. Introduire une notion de signature (pour la catégorie $\mathsf{C}$).
2. Construire une notion de modèle associée, s'organisant en une catégorie munie d'un foncteur vers la catégorie $\mathsf{C}$.
3. Definir l'objet spécifié par la signature comme le modèle initial, s'il existe (la signature est alors dite *effective*).
4. Trouver une condition suffisante pour qu'une signature soit effective[2].

Les modèles d'une signature constituent le domaine atteint par le principe de récurrence, lequel est induit par l'initialité de l'objet spécifié par la signature.

Dans le chapitre 2, nous définissons une notion générale de signature permettant de caractériser un objet d'une catégorie arbitraire $\mathsf{C}$, avec la notion de modèle associée. Une telle signature est donnée par une liste de familles d'*arités* spécifiant des opérations ou des équations. Les chapitres suivants en fournissent divers cas particuliers, de la spécification de la syntaxe à la spécification de la sémantique, en identifiant à chaque fois une classe de signatures effectives.

Finalement, nous proposons un protocole fondé sur une signature à trois niveaux pour spécifier un langage de programmation :

1. spécification des constructions, par exemple une opération binaire $+$ ;
2. spécification des équations, par exemple $a + b = b + a$ (commutativité de l'opération binaire $+$) ;
3. spécification des réductions entre termes, par exemple $0 + a \rightsquigarrow a$, par des règles schématiques.

Les deux premiers points définissent ce que nous appelons la *syntaxe* du langage de programmation, tandis que le dernier point concerne la *sémantique* du langage : une réduction entre deux termes correspond à un chemin d'exécution du programme évoluant du premier terme vers le second. A titre d'exemple de langage intégrant les trois niveaux, nous proposons dans la section 5.6 une spécification du lambda calcul avec substitution explicite décrit dans [Kes09].

2. Dans la littérature, le mot signature est souvent réservé au cas où une telle condition suffisante est automatiquement satisfaite.

Afin de motiver les notions mathématiques mises en jeu, nous examinons dans ce résumé le langage de programmation fonctionnel le plus simple que l'on puisse envisager : le lambda calcul pur. Dans la section 1.1.1, nous donnons une première présentation de sa syntaxe, et la dotons d'une opération de substitution. Nous expliquons ensuite, dans la section 1.1.2, comment la notion mathématique de monade permet d'en rendre compte, puis, dans la section 1.1.3, comment la notion de morphisme de modules fournit un moyen d'exprimer une propriété essentielle des constructions de la syntaxe : la *compatibilité à la substitution*. Dans la section 1.1.4, nous caractérisons la syntaxe par son principe de récurrence, que nous formulons par une propriété d'*initialité*. Nous expliquons dans la section 1.1.5 que préciser cette propriété d'initialité requiert une notion de modèle adéquate, laquelle est déterminée par la *signature* : c'est l'occasion de présenter notre définition générale de signature pour spécifier les objets d'une catégorie arbitraire. Nous examinons ensuite le cas de syntaxes vérifiant des équations (section 1.1.6), avant d'aborder, dans la section 1.1.7, la spécification de la sémantique, modélisée dans le cadre des monades de réduction par un ensemble de réductions entre chaque paire de termes.

### 1.1.1 Présentation naïve de la syntaxe du lambda calcul

Nous donnons ici une présentation de la syntaxe du lambda calcul, ainsi qu'un aperçu de quelques difficultés habituellement associées à une telle présentation. On fixe un ensemble infini $V$ de variables, et l'on caractérise récursivement l'ensemble des *termes* ou expressions valides du lambda calcul :

- chaque variable $x \in V$ est un terme du lambda calcul,
- si $t$ et $u$ sont des termes, alors $t\,u$ est un terme, appelé *application* de $t$ à $u$ ;
- si $t$ est un terme, alors $\lambda x.t$ est un terme, appelé *lambda abstraction* de $t$, où $x$ est une variable qui peut apparaître dans $t$.

L'expression $\lambda x.t$ correspond à la notation mathématique $x \mapsto t$. Il s'agit de définir une fonction dépendant de la variable $x$, le corps de cette fonction étant donné par le terme $t$. L'expression $f\,t$ correspond à la notation mathématique $f(t)$ : c'est l'application de la fonction $f$ à l'argument $t$.

En mathématique, le nom de la variable choisie pour définir une fonction est purement conventionnel : les fonctions $x \mapsto f(x)$ et $y \mapsto f(y)$ sont identiques. Transposons cette identification dans le langage du lambda calcul : nous voulons égaliser le terme $\lambda x.t$ avec le terme $\lambda y.t'$, où $t'$ est obtenu à partir du terme $t$ en remplaçant toutes les occurrences de la variable $x$ par la variable $y$. Dans cette situation, on dit que $x$ est une *variable liée* dans $\lambda x.t$, et les occurrences de $x$ dans $t$ sont alors qualifiées de liées. Les occurrences de variables qui ne sont pas liées sont dites *libres*[3].

Ici, les termes $\lambda x.t$ et $\lambda y.t'$ sont dits *$\alpha$-équivalents*. Plus généralement, deux termes sont $\alpha$-équivalents si l'on peut renommer les variables liées de l'un pour obtenir l'autre terme. La définition précise de la relation d'$\alpha$-équivalence requiert quelques précautions. Par exemple, dans le cas précédent, il est sous-entendu que la variable $y$ n'apparait pas dans $t$ ; autrement, nous identifierions (contre notre gré) les termes $\lambda x.y$ et $\lambda y.y$.

La *substitution* est un autre aspect essentiel de la syntaxe du lambda calcul : étant donné un terme $t$, si l'on remplace toutes les occurrences (libres) d'une variable $x$ par un même terme $u$, nous obtenons une nouvelle expression valide, que nous notons $t\{x := u\}$. L'opération de substitution permet d'exprimer l'intuition mathématique suivante : le résultat d'une fonction $x \mapsto t$ appliquée à un argument $u$ est obtenu en remplaçant la variable $x$ dans $t$ par $u$. Cette affirmation se transpose, pour le lambda calcul, en la *$\beta$-équation*

$$(\lambda x.t)\, u = t\{x := u\}. \tag{1.1}$$

Cette substitution dite *unaire* est un cas particulier de l'opération de *substitution parallèle* $t\{x \mapsto u_x\}$, qui remplace simultanément toutes les variables d'un terme $t$ par un terme correspondant.

La substitution ne remplace que les occurrences libres d'une variable, afin de préserver la propriété suivante : étant donné deux termes $\lambda x.t$ et $\lambda y.t'$ supposés $\alpha$-équivalents, substituer le même terme à la même variable dans chacun d'eux fournit deux termes $\alpha$-équivalents. Si la variable concernée est identique à la variable abstraite, le terme $(\lambda x.t)\{x := u\}$ est donc tout simplement égal à $\lambda x.t$ : par exemple $(\lambda x.x)\{x := u\} = \lambda x.x$ est bien $\alpha$-équivalent à $(\lambda y.y)\{x := u\} = \lambda y.y$.

3. Une variable peut avoir une occurrence liée et une occurrence libre dans le même terme : par exemple, $x$ dans $(\lambda x.x)\, x$ est liée dans $\lambda x.x$, mais apparaît librement à droite.

### 1.1.2 La monade du lambda calcul

Le concept de monade fournit une contrepartie mathématique de la notion intuitive de syntaxe munie d'une opération de substitution. Nous motivons cette définition par l'exemple du lambda calcul. Dans le point de vue que nous adoptons ici, les termes $\alpha$-équivalents sont considérés comme identiques : ainsi, $\lambda x.x = \lambda y.y$.

Au lieu de considérer un ensemble unique de termes avec un ensemble de variables $V$ fixé à l'avance, nous définissons des classes de termes qui utilisent les mêmes variables libres. Notons $L(X)$ l'ensemble des termes dont les variables libres sont choisies dans l'ensemble $X$. Remarquons qu'un terme $t \in L(X)$ se retrouve également dans $L(Y)$ pour tout inclusion $X \subset Y$ : en effet, si les variables libres sont choisies parmi les éléments d'un ensemble $X$, elles sont en particulier choisies parmi les élements de n'importe quel ensemble $Y$ qui contient $X$.

Toute variable est en particulier un terme valide; il y a donc une inclusion $\mathsf{var}_X : X \to L(X)$ pour tout ensemble $X$. D'autre part, si l'on se donne pour toute variable $x \in X$ un terme $u_x$ dont les variables libres sont choisies dans $Y$, nous obtenons, à partir de n'importe quel terme $t \in L(X)$, un terme $t\{x \mapsto u_x\} \in L(Y)$. Cette opération de substitution parallèle vérifie les propriétés suivantes :

- chaque variable est remplacée par le terme adéquat :

$$x'\{x \mapsto u_x\} = u_{x'}$$

- la substitution identité est neutre :

$$t\{x \mapsto x\} = t$$

- toute succession de substitutions est équivalente à une substitution composée :

$$t\{x \mapsto u_x\}\{y \mapsto v_y\} = t\{x \mapsto u_x\{y \mapsto v_y\}\}$$

L'inclusion des variables dans les termes et l'opération de substitution parallèle satisfaisant les équations ci-dessus définissent une *monade sur la catégorie des ensembles*. Cet objet mathématique est au cœur des développements que nous exposons dans cette thèse.

### 1.1.3 Les constructions sont des morphismes de modules

Les concepts mathématiques de modules et de morphismes de modules offrent un cadre permettant d'exprimer la compatibilité d'une construction de la syntaxe avec l'opération de substitution. Nous illustrons ceci avec l'application $t\,u$ du lambda calcul.

La compatibilité de l'application avec la substitution se traduit par la commutation

$$(t\,u)\{x \mapsto v_x\} = t\{x \mapsto v_x\}\; u\{x \mapsto v_x\}$$

Informellement, cette équation signifie qu'il n'y a pas de différence entre effectuer la substitution avant l'application et effectuer la substitution après l'application, comme l'exprime le diagramme commutatif suivant :

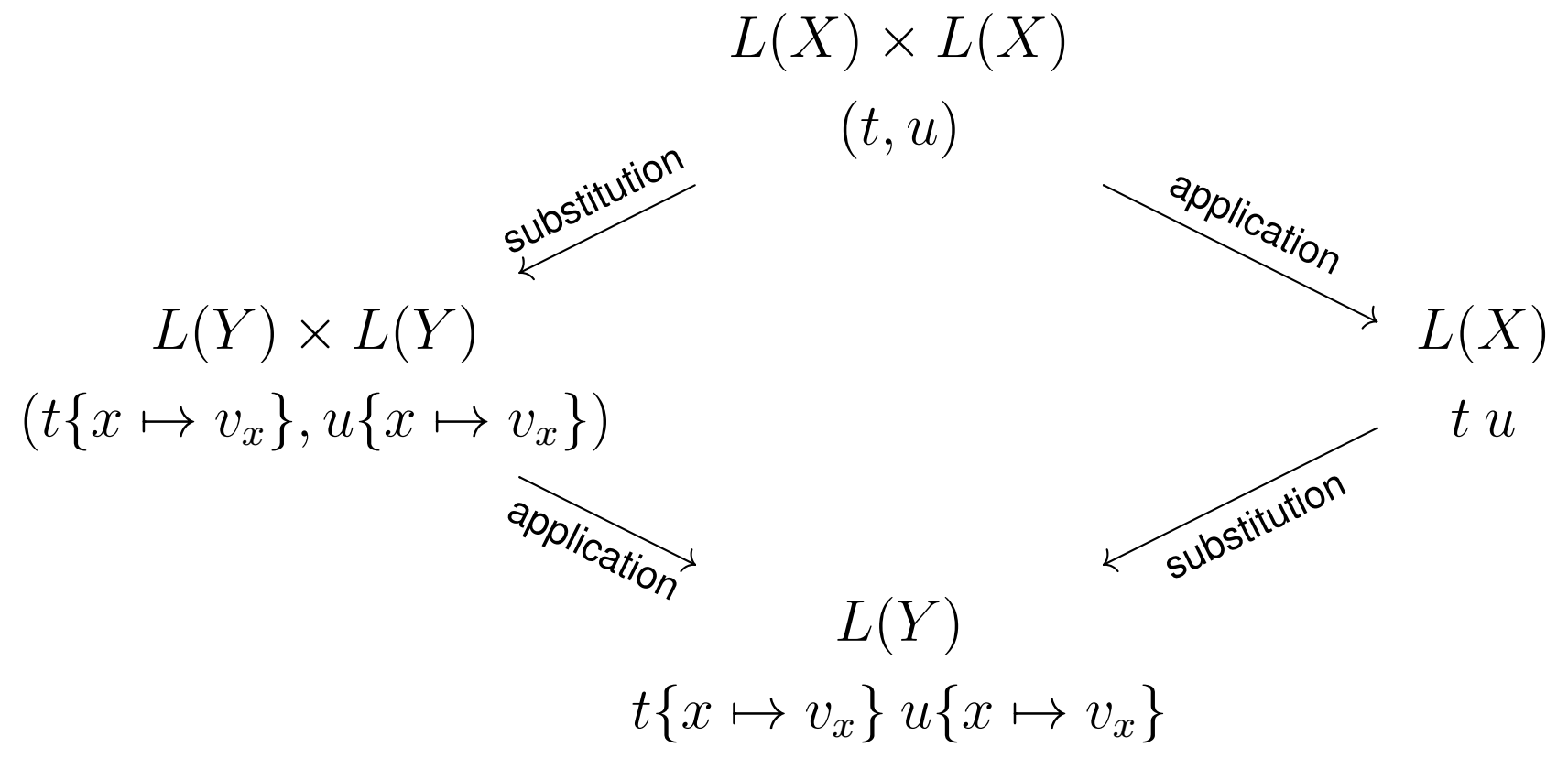

Ce constat s'appuie implicitement sur l'opération de substitution suivante dont bénéficie la collection d'ensembles $(L(X) \times L(X))_X$ :

$$(t, u)\{x \mapsto v_x\} = (t\{x \mapsto v_x\}, u\{x \mapsto v_x\})$$

Cette substitution vérifie les propriétés suivantes :

- la substitution identité est neutre :

$$(t, u)\{x \mapsto x\} = (t, u)$$

- toute succession de substitutions est équivalente à une seule substitution composée :

$$(t, u)\{x \mapsto u_x\}\{y \mapsto v_y\} = (t, u)\{x \mapsto u_x\{y \mapsto v_y\}\}$$

À ce titre, la collection des ensembles de paires de termes définit un *module sur la monade* $L$, que nous notons $L \times L$.

Les définitions de monade et de module sont similaires ; d'ailleurs toute monade définit un module sur elle-même. En fait, $L \times L$ définit aussi une monade, mais la substitution associée ne dérive pas de sa structure de module. Étant donné une paire de termes $(t, u)$ dont les variables libres sont choisies dans $X$, et pour toute variable $x \in X$, une paire de termes $(v_x, w_x)$ dont les variables libres sont choisies dans $Y$, cette substitution monadique fournit une paire de termes dont les variables libres sont dans $Y$. La substitution donnée par la structure de module ne convient pas, puisqu'elle ne s'applique que dans le cas où l'on a associé à chaque variable un terme de la monade $L$, plutôt qu'une paire de termes.

L'application du lambda calcul induit une collection de fonctions

$$L(X) \times L(X) \to L(X)$$

qui associent à toute paire $(t, u)$ le terme $t\ u$. La propriété de commutation avec la substitution sus-mentionnée en fait un *morphisme de modules* de $L \times L$ vers $L$, où $L$ est vu comme un module sur la monade homonyme.

De même, l'abstraction du lambda calcul induit une collection de fonctions

$$L(X \amalg \{\star\}) \to L(X)$$

qui, à tout terme $t$ dont les variables libres sont choisies dans l'ensemble $X$ étendu avec un nouvel élément $\star$, associe le terme $\lambda \star .t$. La famille $L(X \amalg \{\star\})_X$ est canoniquement munie d'une opération de substitution et définit donc un module. La collection des fonctions d'abstraction induit alors un morphisme de modules, en raison de la commutation avec la substitution :

$$(\lambda \star .t)\{x \mapsto u_x\} = \lambda \star . \left( t \left\{ x \mapsto \begin{cases} \star & \text{si } x = \star \\ u_x & \text{sinon.} \end{cases} \right\} \right)$$

La construction de module que nous rencontrons ici se généralise à n'importe quel module $M$ sur une monade $R$ : le *module dérivé* $M'$ se définit comme la collection d'ensembles $(M(X \amalg \{\star\}))_X$ munie d'une opération de substitution canonique.

Dans cette thèse, nous nous intéressons exclusivement à des langages de pro-grammation dont les constructions et les réductions sont compatibles avec la substitution, d'où notre intérêt pour les notions de module et de morphisme de modules.

### 1.1.4 Récursion et initialité

La présentation naïve du lambda calcul induit naturellement un principe de récurrence sur la syntaxe. Supposons, par exemple, que nous voulons calculer l'ensemble des variables libres d'un terme $t$ du lambda calcul. Pour ce faire, nous raisonnons par récurrence sur la structure du terme. Si $t$ est une variable $x$, alors le singleton $\{x\}$ constitue l'ensemble de ses variables libres. Si $t$ est une application $u\,v$, alors l'ensemble de ses variables libres est la réunion des variables libres de $u$ et $v$. Si $t$ est une lambda abstraction $\lambda x.u$, alors n'importe quelle variable libre de $u$ différente de $x$ est une variable libre de $t$.

Dans notre cadre, nous adoptons le point de vue de la sémantique initiale : le principe de récurrence est alors une conséquence d'une propriété d'*initialité*. Le lambda calcul est ainsi caractérisé comme la monade “minimale” munie d'une application et d'une lambda abstraction, dans un sens que nous allons illustrer par l'exemple du calcul des variables libres (cet exemple est étudié plus formellement dans la Section 3.5.2).

Considérons la monade $\mathcal{P}$ qui associe à $X$ l'ensemble $\mathcal{P}(X)$ de ses parties : une variable $x \in X$ induit un “terme” $\{x\} \in \mathcal{P}(X)$ ; la substitution $t\{x \mapsto u_x\}$ est calculée par la réunion $\cup_{z \in t} u_z$. L'union de deux sous-ensembles fournit une opération binaire pour $\mathcal{P}$, que nous assimilons à une “opération d'application” par analogie avec l'opération binaire d'application du lambda calcul. Cette opération associe le sous-ensemble $t \cup u$ au couple $(t, u)$. De même, une opération adéquate d'abstraction $\mathcal{P}(X \amalg \{\star\}) \to \mathcal{P}(X)$ est donnée par $t \mapsto t \cap X$, ou, de manière équivalente, par $t \mapsto t \backslash \{x\}$. Comme nous l'expliquerons dans la section suivante, ces constructions font de la monade $P$ un *modèle de la signature du lambda calcul*.

La propriété d'initialité du lambda calcul mentionnée s'instancie alors par l'existence d'une unique famille de fonctions $(\mathsf{free}_X : L(X) \to \mathcal{P}(X))_X$ vérifiant les propriétés suivantes :

- $\mathsf{free}$ préserve les variables :

$$\mathsf{free}_X(\mathsf{var}(x)) = \{x\}$$

(rappelons que pour la monade $\mathcal{P}$, la variable $x$ est vue comme le sous-ensemble $\{x\}$)

- free préserve la substitution :

$$\mathsf{free}_Y(t\{x \mapsto u_x\}) = \bigcup_{z \in \mathsf{free}_X(t)} \mathsf{free}_Y(u_z)$$

- free préserve l'application :

$$\mathsf{free}(t\ u) = \mathsf{free}(t) \cup \mathsf{free}(u)$$

- free préserve l'abstraction :

$$\mathsf{free}(\lambda x.t) = \mathsf{free}(t) \backslash \{x\}$$

Les deux premiers points caractérisent free comme un *morphisme de monades* entre $L$ et $\mathcal{P}$. La section suivante définit la notion de modèle de sorte que $L$ et $\mathcal{P}$, munies leurs opérations respectives d'application et d'abstraction, sont des modèles de la signature du lambda calcul. Le morphisme de monades free est alors un *morphisme de modèles* de cette signature, grâce aux deux derniers points.

### 1.1.5 Signatures et modèles (Chapitre 2)

Dans le chapitre 2, nous proposons une notion de signature générale, définie comme une liste de famille d'arités, pour spécifier les objets d'une catégorie arbitraire C. Une arité est la donnée d'un diagramme de foncteurs

$$\mathsf{C} \underset{u}{\overset{v}{\rightleftarrows}} \mathsf{D}\ .$$

où $u$ et $v$ sont des sections du foncteur $F : \mathsf{D} \to \mathsf{C}$, c'est-à-dire, des foncteurs qui, post-composés avec $F$, donnent le foncteur identité sur C. Un objet $c$ de C est muni d'une *action* de cette arité s'il est muni d'un morphisme $h : u(c) \to v(c)$ dont l'image par le foncteur $F$ est le morphisme identité en $c$. Notons que si les seuls morphismes que $F$ envoie sur un morphisme identité sont eux mêmes des morphismes identité, alors un objet $c$ est muni d'une action si et seulement si $u(c) = v(c)$. Nous qualifions d'*équationnelle* une arité vérifiant cette propriété, qui spécifie donc des équations.

Les objets munis d'une action d'une famille d'arités (fixée à l'avance) forment une catégorie : les morphismes sont ceux de la catégorie $\mathsf{C}$ vérifiant une condition de commutation avec les actions. C'est ainsi qu'est définie la catégorie de modèles d'une signature composée d'une seule famille d'arités. Par exemple, la catégorie d'algèbres d'un endofoncteur $G : \mathsf{C} \to \mathsf{C}$ est la catégorie de modèles de la signature composée d'une seule arité

$$\mathsf{C} \underset{(\mathsf{Id_C},\mathsf{Id_C})}{\overset{(G,\mathsf{Id_C})}{\overrightarrow{\overleftarrow{\longrightarrow}}}} \mathsf{C} \times \mathsf{C}$$

où le foncteur $\mathsf{C} \times \mathsf{C} \to \mathsf{C}$ est la deuxième projection.

Comme nous l'avons dit, une signature est une liste de familles d'arités : une première famille d'arités pour la catégorie $\mathsf{C}$ de base, puis une famille d'arités pour la catégorie de modèles induite, et ainsi de suite.

Nous illustrons ces définitions pour les *signatures monadiques* (c'est-à-dire spécifiant des monades) envisagées dans le chapitre 3, qui sont composées d'une unique arité. Considérons l'exemple de la syntaxe du lambda calcul. Dans la section précédente, nous l'avons caractérisée comme la monade initiale munie d'une application et d'une abstraction. Plus précisément, en généralisant l'exemple de la monade des parties $\mathcal{P}$ étudiée dans la section précédente, nous disons qu'une monade $R$ est munie d'une application et d'une abstraction si elle est dotée à la fois d'une opération binaire, c'est-à-dire d'un morphisme de modules $\mathsf{app}^R : R \times R \to R$, et d'un morphisme de modules $\mathsf{abs}^R : R' \to R$. De manière équivalente, c'est une monade $R$ avec un morphisme de modules de $(R \times R) \amalg R'$ vers $R$, c'est-à-dire, d'une action pour l'arité

$$\mathsf{Mon} \underset{\Theta}{\overset{\Sigma_{\mathsf{LC}}}{\overrightarrow{\overleftarrow{\longrightarrow}}}} \textstyle\int \mathsf{Mod} \tag{1.2}$$

que nous allons préciser. La catégorie $\mathsf{Mon}$ est la catégorie des monades ; la catégorie $\int \mathsf{Mod}$ rassemble toutes les catégories de modules sur des monades différentes[4] : un objet est une paire d'une monade et d'un module sur cette monade, et un morphisme entre $(R, M)$ et $(S, N)$ est une paire d'un morphisme de monades $f : R \to S$ et d'un morphisme de modules $g : M \to f^*N$ sur la monade $R$, où $f^*N$ est le module sur $R$

4. En termes catégoriques, la catégorie $\int \mathsf{Mod}$ est obtenue par la construction de Grothendieck pour le foncteur qui à toute monade associe sa catégorie de modules.

obtenu canoniquement à partir de $N$, en précomposant ses opérations de substitution par $f$. Le foncteur $\int \mathsf{Mod} \to \mathsf{Mon}$ renvoie la première projection. Une section de ce foncteur, ou *module paramétrique*, associe fonctoriellement à toute monade un module sur elle-même. Par exemple, le module paramétrique $\Theta$ associe à toute monade $R$ le module $R$ (rappelons en effet que toute monade est un module sur elle-même). Par ailleurs, le module paramétrique $\Sigma_{\mathsf{LC}}$ du lambda calcul associe à toute monade $R$ le module $(R \times R) \amalg R'$ sur $R$.

Un module paramétrique quelconque $\Sigma$ induit une arité de manière similaire à $\Sigma_{\mathsf{LC}}$ dans l'équation 1.2. Considérons la catégorie de modèles associée à la signature composée de cette unique arité. Un *modèle* $(R, \rho)$ (parfois noté simplement $R$) de cette signature est une monade $R$ munie d'un morphisme de modules $\rho : \Sigma(R) \to R$. Ainsi, la monade des parties $\mathcal{P}$ et la monade du lambda calcul $L$ induisent des modèles de $\Sigma_{\mathsf{LC}}$. En fait, le lambda calcul est le modèle initial : si $R$ est un modèle, alors il existe un unique morphisme de monades $f : L \to R$ qui préserve l'opération binaire et l'abstraction.

Plus généralement, la monade $S$ spécifiée par la signature induite par un module paramétrique $\Sigma$ est munie d'une action $\sigma : \Sigma(R) \to R$ qui en fait un modèle initial : étant donné un modèle $(R, \rho)$, il existe un unique morphisme de monades $f : S \to R$ qui préserve la structure de modèle, i.e., vérifiant pour tout $t \in \Sigma(R)(X)$

$$f(\sigma(t)) = \rho(f(t)) \tag{1.3}$$

Un tel morphisme de monades constitue un *morphisme de modèles* entre $(S, \sigma)$ et $(R, \rho)$. Notons que le membre de droite de l'équation 1.3 nécessite de donner un sens à l'expression $f(t)$ lorsque $t$ est un élément de $\Sigma(S)_X$ : par fonctorialité de $\Sigma$, c'est tout simplement $\Sigma(f)(t)$. En effet, tout morphisme de monades $f : R \to T$ induit un morphisme de modules $\Sigma(R) \to f^*\Sigma(T)$ sur la monade $R$, morphisme que nous notons abusivement $f$, ou $\Sigma(f)$. Dans le cas de l'opération binaire, on a $\Sigma(S)_X = S(X) \times S(X)$. Donc $t$ est une paire $(u, v)$ et l'on définit $\Sigma(f)(u, v)$ par $(f(u), f(v))$.

L'effectivité d'une signature induite par un module paramétrique, c'est-à-dire l'existence du modèle initial associé, n'est pas systématique[5]. C'est néanmoins le cas de tout module paramétrique que nous appelons *algébrique*, qui spécifie une syntaxe disposant d'un ensemble d'opérations n-aires, dont certaines lient des variables dans

5. La signature du contre-exemple 49 associe à toute monade $R$ le module $(\mathcal{P}(R(X)))_X$.

leurs arguments. La signature du lambda calcul provient d'un module paramétrique $\Sigma_{\mathsf{LC}}$ algébrique : l'application est une opération binaire classique, tandis que l'abstraction est une opération unaire liant une seule variable dans son unique argument.

### 1.1.6 Syntaxes avec équations (chapitres 3 et 4)

La syntaxe d'un langage est parfois définie modulo certaines équations : c'est le cas du lambda calcul différentiel [ER03a], du pi calcul avec ses règles de congruence structurelle, et du lambda calcul avec substitution explicite décrit par [Kes09]. Dans cette thèse, nous abordons la spécification de monades correspondant à des syntaxes vérifiant ce type d'équations.

Dans le chapitre 3, nous étudions les modules paramétriques que nous appellons *présentables* : ce sont, en quelque sorte, des quotients de modules paramétriques algébriques. Une signature induite par un module paramétrique présentable est également qualifiée de présentable. Nous montrons qu'une telle signature est effective (Théorème 57). Il devient possible de spécifier une opération binaire commutative (Section 3.8.1). Pour cela, il suffit de remarquer que la donnée d'une telle opération est équivalente à la donnée d'une opération prenant en argument une paire non ordonnée de termes. Décrivons maintenant le module paramétrique $\Sigma_{\mathsf{comm-bin}}$ associé : il s'agit d'un quotient du module paramétrique algébrique $\Sigma_{\mathsf{bin}}$ d'une opération binaire qui associe à toute monade $R$ le module $R \times R$. Plus précisément, à toute monade $R$, le module paramétrique $\Sigma_{\mathsf{comm-bin}}$ associe le module $(\mathcal{S}^2 R(X)))_X$, où $\mathcal{S}^2 R(X)$ est l'ensemble des paires d'éléments de $R(X) \times R(X)$ quotienté par la relation $(t, u) \sim (u, t)$, c'est-à-dire, l'ensemble des paires non ordonnées. La syntaxe bénéficie alors d'une opération qui prend en argument un couple non ordonné de termes, comme désiré.

Néanmoins, la classe des signatures présentables paraît limitée. Considérons en effet l'exemple d'une opération binaire associative : nous ne savons pas en donner une signature, présentable ou non. Remarquons que nous pouvons malgré cela donner une définition intuitive de modèle dans ce cas particulier : il s'agit d'une monade $R$ munie d'une opération binaire $b : R \times R \to R$ telle que pour tous $x, y, z$ dans $R(X)$ les expressions $b(b(x, y), z)$ et $b(x, b(y, z))$ sont égales. En d'autres termes, il s'agit d'un modèle $(R, b)$ de la signature d'une opération binaire, tel que les deux morphismes de modules de $R \times R \times R$ vers $R$, associant à tout triplet $(x, y, z)$ les deux expressions envisagées, sont égaux.

Dans le chapitre 4, nous donnons une définition d'équation généralisant cet exemple pour les modèles d'une signature monadique quelconque $\Sigma$ : il s'agit de la donnée,

- pour chaque modèle $R$ de la signature $\Sigma$, de deux morphismes de modules $e_R, e'_R$ de même domaine $A_R$ et codomaine $B_R$ (dans l'exemple ci-dessus, $e_R, e'_R : R \times R \times R \to R$ associent respectivement $b(b(t,u),v)$ et $b(t,b(u,v)))$ à un même triplet $(t,u,v) \in R^3(X)$) ;

- pour tout morphisme de modèles $f : R \to S$, de deux morphismes de modules $A_f : A_R \to A_S$ et $B_f : B_R \to B_S$ tels que les diagrammes suivants commutent :

$$\begin{array}{ccc} A_R & \xrightarrow{e_R} & B_R \\ {\scriptstyle A_f}\downarrow & & \downarrow{\scriptstyle B_f} \\ A_S & \xrightarrow[e'_S]{} & B_S \end{array} \qquad \begin{array}{ccc} A_R & \xrightarrow{e'_R} & B_R \\ {\scriptstyle A_f}\downarrow & & \downarrow{\scriptstyle B_f} \\ A_S & \xrightarrow[e'_S]{} & B_S \end{array}$$

Ces données sont soumises comme d'habitude à une condition additionnelle de fonctorialité. On dit qu'un modèle $R$ de la signature $\Sigma$ vérifie l'équation lorsque $e_R = e'_R$. En fait, nous donnons une définition équivalente d'équation dans le chapitre 4, en tant qu'arité équationnelle d'une forme particulière. La vérification de l'équation se reformule en tant qu'une action pour l'arité.

Nous considérons alors les *2-signatures* : il s'agit de signatures composées d'une arité induite par un module paramétrique et d'une famille de ces arités équationnelles. Un modèle (ou 2-modèle) d'une 2-signature est un modèle $R$ de la signature induite par le module paramétrique sous-jacent qui vérifie toutes les équations. Nous parlons de 2-signatures *algébriques* lors que le module paramétrique est algébrique, et que les équations sont *élémentaires* (Définition 104). Techniquement, il s'agit d'équations dont l'action fonctorielle du domaine envoie des morphismes surjectifs de monades sur des morphismes surjectifs de foncteurs, et dont le codomaine est de la forme $R \mapsto (((R')')^{\cdots})'$. Tout exemple de signature présentable que nous considérons dans le chapitre 3 peut être reformulé en tant que 2-signature algébrique, induisant la même catégorie de modèles (à isomorphisme près). Nous montrons que toute 2-signature algébrique est effective (Théorème 107).

### 1.1.7 Sémantique (chapitres 5 and 6)

Il est possible de spécifier par une 2-signature la syntaxe du lambda calcul quotientée par la $\beta$-équivalence (1.1) : les termes $(\lambda x.t)\,u$ et $t\{x := u\}$ sont ainsi égalisés. Cependant, cette équation est habituellement orientée, et considérée comme une étape d'exécution lorsque l'on considère le lambda calcul comme un langage de programmation fonctionnel. A ce titre, il est plus adéquat d'intégrer la $\beta$-réduction à la sémantique du langage, plutôt que d'imposer la $\beta$-équation dans la syntaxe. Ceci motive la notion de *monade de réduction*, que nous introduisons dans le chapitre 5, étendant celle de monade, pour rendre compte de la structure additionnelle de réduction. Intuitivement, une monade de réduction est une monade $R$ munie, pour chaque paire de termes $(t, u) \in R(X)$, d'un ensemble de réductions entre $t$ et $u$ que l'on note $t \blacktriangleright u$, et d'une opération de substitution associée : pour toute famille de termes $(v_x)_{x \in X}$ avec $v_x \in R(Y)$, pour toute réduction $m$ entre $t$ et $u$, cette substitution associe une réduction $m\{x \mapsto v_x\}$ entre $t\{x \mapsto v_x\}$ et $u\{x \mapsto v_x\}$. Des équations analogues à celles intervenant dans la définition de module sont requises.

Dans ce contexte, une *signature de réduction* est une signature d'une forme spécifique pour la catégorie des monades de réductions, déterminée par une 2-signature pour monades et une famille de *règles de réductions*. Par exemple, la règle de réduction pour la congruence de l'application du lambda calcul s'exprime informellement ainsi, en désignant explicitement l'application du lambda calcul par la construction $\mathsf{app}$ :

$$\frac{T \rightsquigarrow T' \qquad U \rightsquigarrow U'}{\mathsf{app}(T, U) \rightsquigarrow \mathsf{app}(T', U')}$$

Cette règle se décompose en trois paires de termes : les hypothèses $(T, T')$ et $(U, U')$, et la conclusion $(\mathsf{app}(T, U), \mathsf{app}(T', U'))$, construites à partir des "métavariables" $T$, $T'$, $U$ et $U'$. Ces paires paramétrées induisent des paires de morphismes $(h_{1,1}(R), h_{1,2}(R))$, $(h_{2,1}(R), h_{2,2}(R))$ et $(c_1(R), c_2(R))$ de $R$-modules entre $R^4$ et $R$ pour n'importe quel modèle $R$ de la signature monadique du lambda calcul. De plus, cette construction est fonctorielle. Plus précisément, si $f : R \to S$ est un morphisme de modèles, alors les diagrammes suivants sont commutatifs :

$$\begin{array}{ccc} R^4 & \xrightarrow{h_{i,j}(R)} & R \\ {\scriptstyle f^4}\downarrow & & \downarrow{\scriptstyle f} \\ S^4 & \xrightarrow[h_{i,j}(S)]{} & S \end{array} \qquad \begin{array}{ccc} R^4 & \xrightarrow{c_i(R)} & R \\ {\scriptstyle f^4}\downarrow & & \downarrow{\scriptstyle f} \\ S^4 & \xrightarrow[c_i(S)]{} & S \end{array}$$

Ainsi formulée, cette règle de congruence entre dans notre définition de *règle de réduction* détaillée en Section 5.3. Une *action* de cette règle dans une monade de réduction $R$ munie d'une opération binaire app est la donnée d'une réduction app-cong$(m_T, m_U)$ entre app$(T, T')$ et app$(U, U')$ pour tout $(T, T', U, U') \in R^4(X)$, toute réduction $m_T$ entre $T$ et $T'$, et toute réduction $m_U$ entre $U$ et $U'$. Il faut de plus que app-cong commute avec la substitution, c'est-à-dire que l'équation suivante soit vérifiée :

$$\text{app-cong}(m_T, m_U)\{x \mapsto v_x\} = \text{app-cong}(m_T\{x \mapsto v_x\}, m_U\{x \mapsto v_x\})$$

Le théorème principal (Théorème 150) du chapitre 5 affirme l'effectivité d'une signature de réduction composée d'une 2-signature effective et de n'importe quelle famille de règles de réduction. Détaillons la notion de modèle mise en jeu ici : il s'agit d'une monade de réduction munie

- d'une structure de modèle de la 2-signature pour la monade sous-jacente,
- d'une action de chaque règle de réduction de la signature.

Le modèle initial est construit à partir du modèle initial de la 2-signature, et la structure additionnelle de réductions est construite inductivement à partir des règles de réduction.

En guise d'exemples, nous proposons quelques signatures pour des variantes du lambda calcul avec $\beta$-réduction, puis, dans la section 5.6, une signature pour le lambda calcul avec substitution explicite, tel que décrit dans [Kes09] par un ensemble de constructions soumises à une équation syntaxique, auquel s'ajoutent des règles de réduction entre termes. Cette spécification se fait en trois étapes : une 1-signature pour les opérations du langage, une 2-signature prenant en compte l'équation syntaxique, et une signature de réduction spécifiant les réductions adéquates.

Dans le chapitre 6, nous généralisons ces développements, et traitons de nouveaux exemples, comme le lambda calcul en appel par valeurs avec réduction à grands pas :

la syntaxe est celle du lambda calcul classique, mais dans cette variante, un terme se réduit en une valeur, c'est-à-dire en une variable ou bien en une lambda abstraction. La notion de valeur n'est pas stable par substitution arbitraire : par exemple, $x\{x := y\,y\} = y\,y$ n'est pas une valeur. En revanche, elle est stable par substitution de valeurs : si $v$ et $(w_x)_{x\in X}$ sont des valeurs, alors $v\{x \mapsto w_x\}$ est une valeur. Nous pouvons ainsi définir la monade $\mathsf{LC}_v$ des valeurs du lambda calcul : elle est munie d'une inclusion dans la monade $L$ du lambda calcul qui nous permet de considérer cette dernière comme un module sur $\mathsf{LC}_v$.

Dans cette variante du lambda calcul, la $\beta$-réduction est formulée ainsi :

$$\frac{t \rightsquigarrow \lambda x.t' \qquad u \rightsquigarrow u' \qquad t'\{x := u'\} \rightsquigarrow v}{t\,u \rightsquigarrow v}$$

Une autre règle de réduction assure que toute valeur, en tant que terme, se réduit vers elle-même.

Les réductions sont stables par substitution : si $t \in L(X)$ se réduit en une valeur $v \in \mathsf{LC}_v(X)$ et $(u_x)_{x\in X}$ est une famille de valeurs indexée par l'ensemble des variables libres $X$, alors $t\{x \mapsto u_x\}$ se réduit en $v\{x \mapsto u_x\}$.

La collection des ensembles de réductions entre un terme et une valeur est définie par induction, en appliquant successivement les règles de réduction envisagées. Elle est munie d'une opération de substitution adéquate, comme expliqué précédemment.

Remarquons qu'un lambda terme peut être représenté par un arbre binaire dont les feuilles sont des valeurs et les nœuds correspondent aux applications. Cette représentation induit une bijection : un tel arbre binaire détermine un lambda terme de manière unique.

$(x\ (\lambda y.t))\ z \qquad \Leftrightarrow$

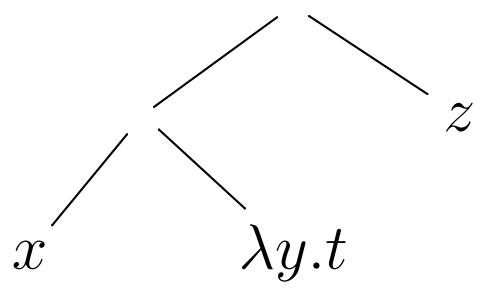

Ainsi, nous pouvons identifier le module $L$ des lambda termes avec la composition $B \cdot \mathsf{LC}_v$, où $B$ est la monade des arbres binaires : $B(X)$ est l'ensemble des arbres binaires dont les feuilles sont choisies dans $X$.

S'inspirant de cet exemple, nous définissons la notion de *monade opérationnelle*. Il s'agit :

- d'une monade $R$ (dans notre exemple, $R = \mathsf{LC}_v$) ;

- d'une paire d'*endofoncteurs* $(T_1, T_2)$ *sur* Set (dans notre exemple, $T_1 = B$ et $T_2 =$ Id) ;

- d'un ensemble de réductions entre $t$ et $u$, pour chaque paire $(t, u) \in T_1(R(X)) \times T_2(R(X))$ ;

- d'une opération de substitution appropriée sur ces ensembles de réductions.

Nous définissons ensuite les *signatures opérationnelles* permettant de spécifier ce type d'objet mathématique : ce sont des cas particuliers de signatures pour la catégorie des monades opérationnelles. Nous démontrons un résultat d'effectivité pour ces signatures (Théorème 207).

### 1.1.8 Formalisation

Les preuves des résultats principaux des chapitres 3 et 4 ont été vérifiées à l'aide de l'assistant de preuve Coq. Il s'agit d'un logiciel dans lequel il est possible de reproduire (ou *formaliser*) des définitions mathématiques ainsi que des démonstrations qui, si elles sont validées par le logiciel, sont en principe incontestables. En pratique, cette garantie est à nuancer dans la mesure où des défauts d'implémentation permettant de prouver des énoncés contradictoires sont régulièrement exhibés (et corrigés). D'autre part, la correspondance entre les énoncés formalisés dans la théorie des types de Coq et les énoncés mathématiques est une question théorique qui n'est pas évidente. Mais plus pragmatiquement, il est possible de se tromper dans la formalisation d'une définition, et le logiciel Coq ne nous est d'aucune aide dans cette étape. Il arrive, par exemple, qu'on formalise une mauvaise caractérisation d'un certain ensemble, cette caractérisation se révélant contradictoire par la suite. Cette erreur peut passer inaperçue car elle n'empêchera pas de montrer que ses éléments vérifient les propriétés qui intéressent le mathématicien (en effet, les éléments de l'ensemble vide satisfont n'importe quelle propriété). Le lecteur d'une formalisation peut donc certes se dispenser d'examiner les démonstrations acceptées par le logiciel, mais il doit vérifier consciencieusement les définitions.

Pour nos développements, nous avons choisi de nous appuyer sur la bibliothèque `UniMath` de Coq. Celle-ci présente quelques avantages, malgré un problème de "taille" (que nous mentionnerons brièvement) :

i elle comporte un certain nombre de définitions et de résultats que nous avons pu directement exploiter ;

ii elle utilise un nombre limité de fonctionnalités du langage ;

iii elle intègre l'*axiome d'univalence*.

Concernant le point i, `UniMath` propose, entre autres, une implémentation élaborée de la théorie des catégories et des ensembles quotients.

Le point ii permet de limiter la complexité de la théorie des types en laquelle il nous faut faire confiance (cependant, le problème de taille dont il est question plus loin affaiblit grandement cet argument). Il réduit également le risque d'être confronté à des erreurs d'implémentation de Coq, en se restreignant à des fonctionnalités éprouvées.

L'axiome d'univalence mentionné par le point iii en tant que tel n'est pas utilisé de manière cruciale. Il a cependant un certain nombre de conséquences utiles, par exemple l'extensionnalité fonctionnelle (deux fonctions sont égales si elles envoient chaque élément sur la même image : contrairement à ce qui se passe dans la théorie des ensembles, cet énoncé est indépendant de la théorie des types de Coq), ou encore l'extensionnalité propositionnelle (deux propositions logiquement équivalentes sont égales) qui permet de manipuler convenablement la notion de sous-ensemble.

Pour finir, la bibliothèque `UniMath` présente un important défaut dû à un problème de "taille". En effet, elle nécessite une option de compilation qui rend le système contradictoire, en rendant possible l'implémentation d'une variante du paradoxe de Russel. Cette option est en pratique utilisée dans la définition (imprédicative) des quotients. Nous avons ignoré l'existence de cette faille logique dans les raisonnements mathématiques que nous avons par la suite formalisés.

## 1.2 Long summary

This section is an Enlgish translation of the previous one.

This thesis deals with the mathematization of the notion of *language of programming*, paying particular attention to the notion of *substitution*.

Research in the field of programming languages traditionally relies on a definition of *syntax* modulo renaming of bound variables, with its associated *operational semantics*. We are interested in mathematical tools allowing to automatically generate syntax and semantics from basic data.

As regards the mathematization of syntax, the specification of algebraic structures with related variables is a major issue. Two main lines of research are in competition: *nominal sets* [GP99] and *substitution algebras* [FPT99]. In this thesis, we explore a variant of substitution algebras, proposed by [HM07; HM10], which is based on the notion of *module over a monad*. Following the same spirit, we tackle the mathematization of operational semantics. To this end, we introduce first the *reduction monads*, then their generalization, the *operational monads*: they are our mathematical approximation of the notion of programming language.

In this thesis, we look at the specification of objects in the category of monads (Chapters 3 and 4), the category of *reduction monads* (Chapter 5), and the category of *operational monads* (Chapter 6), our objective being to define a *formal language*[6], modeled by an object of a relevant category.

The concept of characterising an object of a given type (i.e., of a certain category C) through an initiality property is standard in computer science, where it is known under the terms *Initial Semantics* and *Algebraic Specification* [JGW78], and has been popularised by the movement of *Algebra of Programming* [BM97]. The general methodology of the initial semantics can be described according to the following steps:

1. Introduce a notion of signature (for the category C).
2. Construct a notion of associated model, organizing into a category with a functor to the category C.
3. Define the object specified by the signature as the initial model, if it exists (the signature is then called *effective*).

6. Here, the word “language” encompasses data types, programming languages and logic calculi, as well as languages for algebraic structures as considered in Universal Algebra.

4. Find a sufficient condition for a signature to be effective[7].

The models of a signature form the domain reached by the principle of recursion, which is induced by the initiality of the object specified by the signature.

In Chapter 2, we define a general notion of signature to characterize an object of an arbitrary category $\mathsf{C}$, with its associated category of models. Such a signature is given by a list of families of *arities* specifying operations or equations. The remaining chapters provide various special cases, from the specification of the syntax to the specification of the semantics. Each time, we identify a class of effective signatures.

Finally, we propose a protocol based on a three-level signature to specify a programming language:

1. specification of constructions, for example a binary operation $+$;

2. specification of equations, for example $a + b = b + a$ (commutativity the binary operation);

3. specification of reductions between terms, for example $0 + a \rightsquigarrow a$, by schematic rules.

The first two points define what we call the *syntax* of the programming language, while the last point concerns the *semantics* of language: a reduction between two terms is a program execution path evolving from the first term to the second. As an example of language integrating the three levels, we propose in the section 5.6 a specification of lambda calculus with explicit substitution described in [Kes09].

In order to motivate the mathematical notions involved, we examine in this summary the simplest functional programming language that can be considered: the pure lambda calculus. In section 1.2.1, we give a first presentation of its syntax, and endow it with an operation of substitution. We then explain, in section 1.2.2, how the mathematical notion of monad can account for it, then, in the section 1.2.3, how the notion of morphism of modules provides a way to express an essential property of the constructions of the syntax: *compatibility with substitution*. In the section 1.2.4, we characterize the syntax by its principle of recursion, which we formulate by a property of *initiality*. We explain in section 1.2.5 that specify this property of initiality requires an adequate notion of model, which is determined by the *signature*: this is the opportunity to present our

7. In the literature, the word signature is often reserved for the case where such a sufficient condition is automatically satisfied.

general definition of signature to specify objects of an arbitrary category. We examine then the case of syntaxes satisfying equations (section 1.2.6), before approaching, in section 1.2.7, the specification of semantics, modeled as part of the reduction monads by a set of reductions between each pair of terms.

### 1.2.1 Naive presentation of the lambda calculus syntax

We give here a presentation of the syntax of lambda calculus, as well as a overview of some of the difficulties usually associated with such a presentation. We fix an infinite set $V$ of variables, and we characterize recursively all of the valid *terms* or expressions of the lambda calculus:

- each variable $x \in V$ is a term of the lambda calculus,
- if $t$ and $u$ are terms, then $t\,u$ is a term, called *application* of $t$ to $u$ ;
- if $t$ is a term, then $\lambda x.t$ is a term, called lambda abstraction of $t$, where $x$ is a variable that can appear in $t$.

The expression $\lambda x.t$ corresponds to the mathematical notation $x \mapsto t$. It is about defining a function depending on the variable $x$, the body of this function being given by the term $t$. The expression $f\,t$ corresponds to the mathematical notation $f(t)$: it is the application of the function $f$ to the argument $t$.

In mathematics, the name of the variable chosen to define a function is purely conventional: the functions $x \mapsto f(x)$ and $y \mapsto f(y)$ are identical. Let us transpose this identification into the language of lambda calculus: we want to equalize the term $\lambda x.t$ with the term $\lambda y.t'$, where $t'$ is obtained from the term $t$ by replacing all occurrences of the variable $x$ with the variable $y$. In this situation, we say that $x$ is a *variable bound* in $\lambda x.t$, and the occurrences of $x$ in $t$ are then qualified as bound. Occurrences of variables that are not bound are said *free*.[8]

Here, the terms $\lambda x.t$ and $\lambda y.t'$ are said $\alpha$*-equivalent*. More generally, two terms are $\alpha$-equivalents if we can rename the bound variables of one to get the other term. The precise definition of the $\alpha$-equivalence requires some precautions. For example, in the previous case, it is implied that the variable $y$ does not appear in $t$; otherwise, we would identify (against our will) the terms $\lambda x.y$ and $\lambda y.y$.

8. A variable can have a bound instance and a free occurrence in the same term: for example, $x$ in $(\lambda x.x)\,x$ is bound in $\lambda x.x$, but appears freely on the right.

*Substitution* is another essential aspect of the syntax of lambda calculus: given a term $t$, if we replace all (free) occurrences of a variable $x$ by the same term $u$, we get a new valid expression, which we note $t\{x := u\}$. The substitution operation allows to express the following mathematical intuition: the result of a function $x \mapsto t$ applied to an argument $u$ is obtained by replacing the variable $x$ in $t$ by $u$. This statement is transposed, for the lambda calculus, in the *$\beta$-equation*

$$(\lambda x.t)\, u = t\{x := u\}. \tag{1.4}$$

This so-called *unary* substitution is a special case of the operation of *simultaneous substitution* $t\{x \mapsto u_x\}$, which replaces simultaneously each variable of a term $t$ by a corresponding term.

Substitution only replaces the free occurrences of a variable, so as to preserve the following property: given two terms $\lambda x.t$ and $\lambda y.t'$ assumed $\alpha$-equivalent, replacing the same variable with the same term yields two $\alpha$-equivalent terms. If the involved variable is identical to the abstracted variable, the term $(\lambda x.t)\{x := u\}$ is simply equal to $\lambda x.t$: for example $(\lambda x.x)\{x := u\} = \lambda x.x$ is indeed $\alpha$-equivalent to $(\lambda y.y)\{x := u\} = \lambda y.y$.

### 1.2.2 The lambda calculus monad

The concept of monad provides a mathematical counterpart to the intuitive notion of syntax with a substitution operation. We motivate this definition by the example of the lambda calculus. In the point of view that we adopt here, the $\alpha$-equivalent terms are considered identical: thus, $\lambda x.x = \lambda y.y$.

Instead of considering a single set of terms with a set $V$ of variables fixed in advance, we define classes of terms which use the same set of free variables. Let $L(X)$ be the set of terms whose free variables are selected from the set $X$. Note that a term $t \in L(X)$ is also found in $L(Y)$ for any inclusion $X \subset Y$: indeed, if the free variables are chosen from the elements of a set $X$, they are also chosen from the elements of any superset $Y$ of $X$.

Any variable is in particular a valid term: there is an inclusion $\mathsf{var}_X : X \to L(X)$ for any set $X$. On the other hand, if one gives for all variable $x \in X$ a term $u_x$ whose free variables are chosen in $Y$, we get, from any term $t \in L(X)$, a term $t\{x \mapsto u_x\} \in L(Y)$. This simultaneous substitution operation satisfies the following properties:

- each variable is replaced by the appropriate term:

$$x'\{x \mapsto u_x\} = u_{x'}$$

- identity substitution is neutral:

$$t\{x \mapsto x\} = t$$

- any succession of substitutions is equivalent to a composed substitution:

$$t\{x \mapsto u_x\}\{y \mapsto v_y\} = t\{x \mapsto u_x\{y \mapsto v_y\}\}$$

The inclusion of variables in terms and the simultaneous substitution operation satisfying the equations above define a *monad on the category of sets*. This mathematical object is at the heart of developments that we expose in this thesis.

### 1.2.3 Constructions are morphisms of modules

The mathematical concepts of modules and morphisms of modules provide a framework for expressing the compatibility of a construction in the syntax with the substitution operation. We illustrate this with the $t\,u$ application of the lambda calculus.

The compatibility of the application with the substitution results in the commutation

$$(t\,u)\{x \mapsto v_x\} = t\{x \mapsto v_x\}\,u\{x \mapsto v_x\}$$

Informally, this equation means that there is no difference between performing the substitution before the application and performing the substitution after the application, as shown in the following commutative diagram:

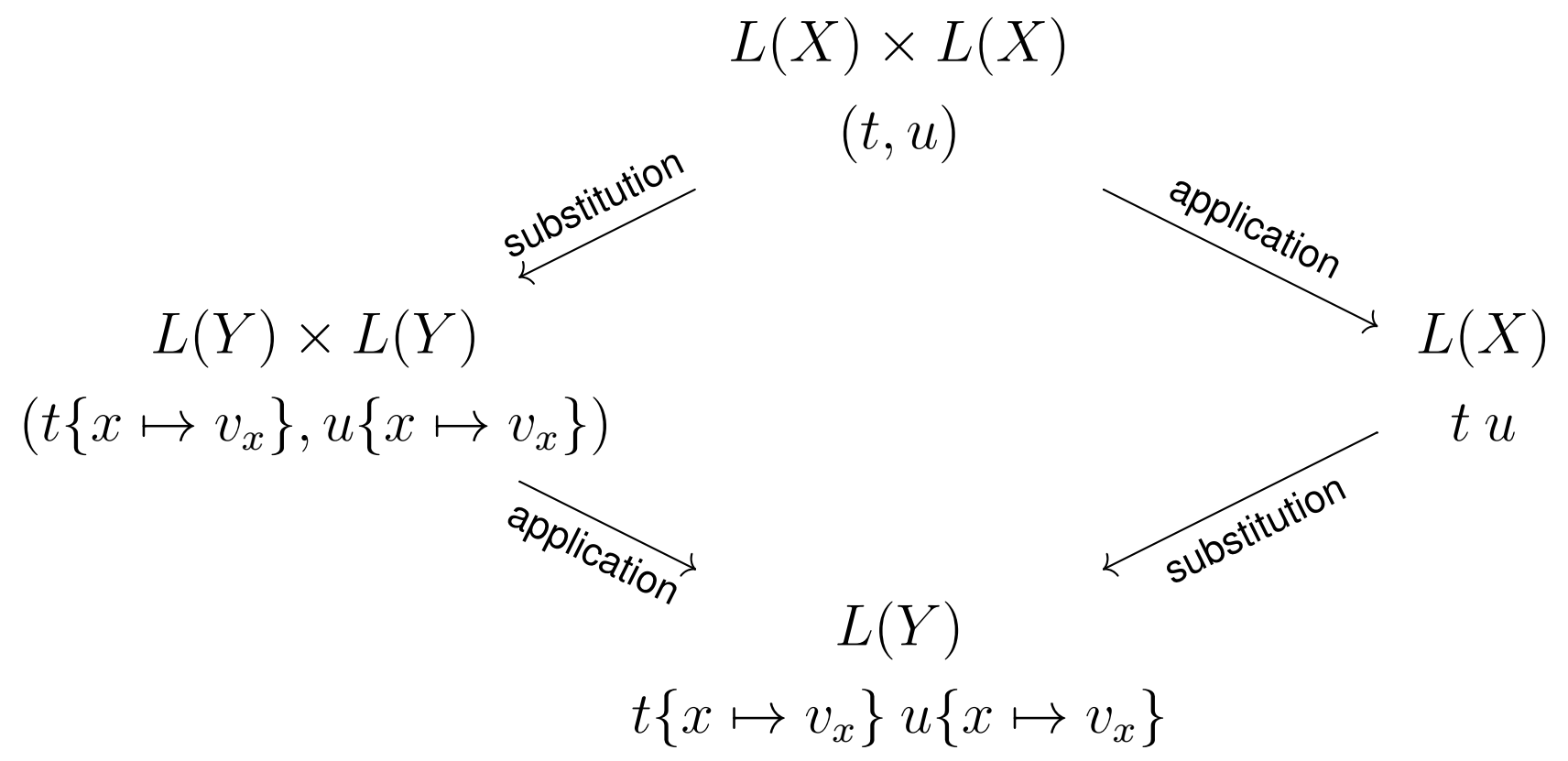

This observation implicitly relies on the following substitution operation of which the collection of sets $(L(X) \times L(X))_X$:

$$(t, u)\{x \mapsto v_x\} = (t\{x \mapsto v_x\}, u\{x \mapsto v_x\})$$

This substitution verifies the following properties:

- identity substitution is neutral:

$$(t, u)\{x \mapsto x\} = (t, u)$$

- any succession of substitutions is equivalent to a single compound substitution:

$$(t, u)\{x \mapsto u_x\}\{y \mapsto v_y\} = (t, u)\{x \mapsto u_x\{y \mapsto v_y\}\}$$

As such, the collection of sets of terms pairs defines a *module over the monad* $L$ , which we note $L \times L$.

The definitions of monad and module are similar. Every monad defines a module on itself. In fact, $L \times L$ also defines a monad, but the associated substitution does not derive from its module structure. Given a pair of terms $(t, u)$ whose free variables are chosen in $X$, and for any variable $x \in X$, a pair of terms $(v_x, w_x)$ whose free variables are chosen in $Y$, this monadic substitution provides a pair of terms whose free variables are in $Y$. The given substitution by the module structure is not appropriate, since it does not apply to the case where each variable is associated with a term in the monad $L$, rather than a pair of terms.

The application of lambda calculus induces a collection of functions

$$L(X) \times L(X) \to L(X)$$

that associate with any pair $(t, u)$ the term $t\ u$. The commutativity property with the aforementioned substitution makes it a *morphism of modules* from $L \times L$ to $L$, where $L$ is seen as a module on the homonymous monad.

Similarly, the abstraction of lambda calculus induces a collection of functions

$$L(X \amalg \{\star\}) \to L(X)$$

which to any term $t$ whose free variables are chosen in the set $X$ extended with a new $\star$ element, associates the term $\lambda \star .t$. The family $L(X \amalg \{\star\})_X$ is canonically equipped with an operation of substitution and therefore defines a module. The collection of functions of abstraction then induces a morphism of modules, due to the commutation with substitution:

$$(\lambda \star .t)\{x \mapsto u_x\} = \lambda \star . \left( t \left\{ x \mapsto \begin{cases} \star & \text{if } x = \star \\ u_x & \text{otherwise.} \end{cases} \right\} \right)$$

The module construction that we encounter here generalizes to any module $M$ on a monad $R$: the *derived module* $M'$ is defined as the collection of sets $(M(X \amalg \{\star\}))_X$ equipped with a canonical substitution operation.

In this thesis, we are interested exclusively in programming languages whose constructions and reductions are compatible with substitution, hence our interest in the notions of module and morphism of modules.

### 1.2.4 Recursion and initiality

The naive presentation of lambda calculus naturally induces a principle of recursion on the syntax. Suppose, for example, that we want calculate the set of free variables of a term $t$ of the lambda calculus. To do this, we reason by recursion on the structure of the term. If $t$ is a $x$ variable, then the set of its free variables is the singleton $\{x\}$. If $t$ is an application $u\,v$, then the set of its free variables is the union of the free variables of $u$ and $v$. If $t$ is a lambda abstraction $\lambda x.u$, then any free variable of $u$ different from $x$ is a free variable of $t$.

In our framework, we adopt the point of view of Initial Semantics: the principle of recursion is then a consequence of a property of *initiality*. The lambda calculus is thus characterized as the "minimal" monad provided with an application and a lambda abstraction, in a sense that we will illustrate with the example of computation of free variables (this example is studied more formally in Section 3.5.2).

Consider the monad $\mathcal{P}$ that associates to $X$ its powerset $\mathcal{P}(X)$: a variable $x \in X$ induces a "term" $\{x\} \in \mathcal{P}(X)$ ; the substitution $t\{x \mapsto u_x\}$ is calculated by the meeting $\cup_{z \in t} u_z$. The union of two subsets provides a binary operation for $\mathcal{P}$, which we call an "application operation" by analogy with the binary application operation of the lambda calculus. This operation associates the subset $t \cup u$ to the pair $(t, u)$. Similarly, an adequate abstraction operation $\mathcal{P}(X \amalg \{\star\}) \to \mathcal{P}(X)$ is given by $t \mapsto t \cap X$, or equivalently, by $t \mapsto t \backslash \{x\}$. As we will explain in the following section, these constructions make the monad $P$ a *model of the signature of lambda calculus*.

The initiality property of the mentioned lambda calculus then instantiates by the existence of a single family of functions $(\mathsf{free}_X : L(X) \to \mathcal{P}(X))_X$ verifying the following properties:

- $\mathsf{free}$ preserves the variables:

$$\mathsf{free}_X(\mathsf{var}(x)) = \{x\}$$

  (remember that for the monad $\mathcal{P}$, the variable $x$ is seen as the subset $\{x\}$)

- $\mathsf{free}$ preserves the substitution:

$$\mathsf{free}_Y(t\{x \mapsto u_x\}) = \bigcup_{z \in \mathsf{free}_X(t)} \mathsf{free}_Y(u_z)$$

- $\mathsf{free}$ preserves the application:

$$\mathsf{free}(t\ u) = \mathsf{free}(t) \cup \mathsf{free}(u)$$

- $\mathsf{free}$ preserves abstraction:

$$\mathsf{free}(\lambda x.t) = \mathsf{free}(t) \backslash \{x\}$$

The first two points characterize $\mathsf{free}$ as a *morphism of monads* between $L$ and $\mathcal{P}$. The following section defines the notion of model so that $L$ and $\mathcal{P}$, with their respective operations of application and abstraction, are models of the signature of the lambda calculus. The morphism of monads $\mathsf{free}$ is then a *morphism of models* of this signature, thanks to the last two points.

### 1.2.5 Signatures and models (Chapter 2)

In Chapter 2, we define a general notion of signature, as a list of families of arities, to characterize an object of an arbitrary category $\mathsf{C}$. An arity is given by a functor diagram

$$\mathsf{C} \underset{u}{\overset{v}{\rightleftarrows}} \mathsf{D}\,.$$

where $u$ and $v$ are sections of the functor $F : \mathsf{D} \to \mathsf{C}$, that is, functors which, post-composed with $F$, yield the identity functor on $\mathsf{C}$. An object $c$ of $\mathsf{C}$ is euqipped with an *action* of this arity if it is equipped with a morphism $h : u(c) \to v(c)$ whose image by $F$ is the identity morphism in $c$. Note that if the only morphisms mapped by $F$ to identity morphisms are themselves identity morphisms, then an object $c$ is equipped with an action if and only if $u(c) = v(c)$. We call *equational* such an arity.

Objects equipped with an action of a (fixed) family of arities form a category: morphisms are those of the category $\mathsf{C}$ satisfying a commutation condition with the actions. This is how the category of models of a signature consisting of a single family of arities is defined. For example, the category of algebras of an endofunctor $G : \mathsf{C} \to \mathsf{C}$ is the category of models of the signature consisting of a single arity

$$\mathsf{C} \underset{(\mathsf{Id}_\mathsf{C},\mathsf{Id}_\mathsf{C})}{\overset{(G,\mathsf{Id}_\mathsf{C})}{\rightleftarrows}} \mathsf{C} \times \mathsf{C}$$

where the functor $\mathsf{C} \times \mathsf{C} \to \mathsf{C}$ is the second projection.

As we said, a signature is a list of families of arities: a first family of arities for the base category $\mathsf{C}$, then a family of arities for the induced category of models, and so on.

We illustrate these definitions with the *monadic signatures* (that is, signatures specifying monads) that we consider in Chapter 3, consisting of a single arity. Consider the example of the syntax of lambda calculus. In the previous section, we characterized it as the initial monad equipped with an application and an abstraction. More specifically, generalizing the example of the powerset monad $\mathcal{P}$ studied in the previous section we say that a monad $R$ is equipped with application and abstraction if it has both a binary operation, that is to say a morphism of modules $\mathsf{app}^R : R \times R \to R$, and a module morphism $\mathsf{abs}^R : R' \to R$. Equivalently, it is a monad $R$ with a morphism of modules from $(R \times R) \amalg R'$ to $R$, that is, an action of the arity

$$\mathsf{Mon} \underset{\Theta}{\overset{\Sigma_{\mathsf{LC}}}{\rightrightarrows}} \int \mathsf{Mod} \tag{1.5}$$

that we now detail: $\mathsf{Mon}$ is the category of monads; $\int \mathsf{Mod}$ gather all the category of modules over different monads:[9] an object is a pair of a monad et a module over this monad, and a morphism between $(R, M)$ and $(S, N)$ is a pair of a monad morphism $f : R \to S$ and a module of morphisms $g : M \to f^*N$ over the monad $R$, where $f^*N$ is the module over $R$ built canonically out of $N$, by precomposing its substitution operations with $f$. The functor $\int \mathsf{Mod} \to \mathsf{Mon}$ is the first projection. A section of this functor, or *parametric module*, maps functorially any monad $R$ to a module over itself. For example, the parametric module $\Theta$ maps a monad $R$ to the module $R$ (recall indeed that any monad is canonically a module over itself). The parametric module $\Sigma_{\mathsf{LC}}$ of lambda calculus maps any monad $R$ to the module $(R \times R) \amalg R'$ over $R$.

A parametric module $\Sigma$ induces an arity, as does $\Sigma_{\mathsf{LC}}$ in Equation 1.5. Let us consider the category of models associated to the signature consisting of this unique arity. A *model* $(R, \rho)$ (sometimes simply noted $R$) of this signature is then a monad $R$ with a module morphism $\rho : \Sigma(R) \to R$. Thus, the powerset monad $\mathcal{P}$ and the lambda calculus monad $L$ induce models of $\Sigma_{\mathsf{LC}}$. In fact, the lambda calculus is the initial model: if $R$ is a model, then there is a unique morphism of monads $f : L \to R$ which preserves binary operation and abstraction.

More generally, the monad $S$ specified by a signature induced by a parametric module $\Sigma$ is equipped with an action $\sigma : \Sigma(R) \to R$ which makes it the initial model: given a model $(R, \rho)$, there exists a unique morphism of monads $f : S \to R$ which preserves the model structure, i.e., verifying for any $t \in \Sigma(R)(X)$

$$f(\sigma(t)) = \rho(f(t)) \tag{1.6}$$

9. Categorically speaking, the category $\int \mathsf{Mod}$ is obtained by performing the Grothendieck construction for the functor mapping a monad to its category of modules.

Such a morphism of monads constitutes what a *morphism of models* between $(S, \sigma)$ and $(R, \rho)$. Note that the right-hand side of Equation 1.6 requires making sense of the expression $f(t)$ when $t$ is an element of $\Sigma(S)_X$: by functoriality of $\Sigma$, this is just $\Sigma(f)$. Indeed, any morphism of monads $f : R \to T$ induces a morphism of modules $\Sigma(R) \to f^*\Sigma(T)$ on the monad $R$, morphism that we denote $\Sigma(f)$, or just $f$.

The effectivity of a signature induced by a parametric module, that is, the existence of the associated initial model, is not systematic.[10]This is nevertheless the case with any parametric module that we call *algebraic*, which specifies a syntax that has a set of n-ary operations, some of which bind variables in their arguments. The signature of lambda calculus comes from an algebraic parametric module: the application is a classic binary operation, while abstraction is a unary operation binding a single variable in its unique argument.

### 1.2.6 Syntaxes with equations (Chapters 3 and 4)

The syntax of a language is sometimes defined modulo some equations: this is the case of the differential lambda calculus [ER03a], of the pi calculus with its structural congruence rules, and of lambda calculus with explicit substitution as described in [Kes09]. An example of possible equation for the lambda calculus is given by the the $\beta$-equivalence $(\lambda x.t)\ u = t\{x := u\}$ for the lambda calculus, is common practice. In this thesis, we deal with the specification of monads corresponding to languages satisfying this kind of syntactic equations.

In Chapter 3, we study the parametric modules that we call *presentable*: these are, in a way, quotients of algebraic parametric modules. A signature induced by a presentable parametric module is also called presentable. We show that such a signature is effective (Theorem 57). It becomes possible to specify a commutative binary operation (Section 3.8.1). For that, it suffices to notice that the data of such an operation is equivalent to the data of an operation taking as argument an unordered pair of terms. Now let us describe the associated presentable parametric module $\Sigma_{\mathsf{comm-bin}}$: this is a quotient of the algebraic parametric module $\Sigma_{\mathsf{bin}}$ of a binary operation that associates to any monad $R$ the module $R \times R$. More specifically, to any monad $R$, the parametric module $\Sigma_{\mathsf{comm-bin}}$ associates the module $(\mathcal{S}^2 R(X)))_X$, where $\mathcal{S}^2 R(X)$ is the set of pairs of elements of $R(X) \times R(X)$ quotiented by the relation $(t, u) \sim (u, t)$, that is, the set of

10. The counterexample 49 is the signature associating to any monad $R$ the module $(\mathcal{P}(R(X)))_X$.

unordered pairs. The syntax then comes with an operation that takes as argument a non ordered of pair of terms, as desired.

Nevertheless, the class of presentable signatures seems limited. Consider the example of an associative binary operation: we do not know how to give a signature, presentable or not. Note that we can nevertheless give an intuitive definition of model in this particular case: it is a monad $R$ equipped with a binary operation $b : R \times R \to R$ such that for all $x, y, z$ in $R(X)$ the expressions $b(b(x, y), z)$ and $b(x, b(y, z))$ are equal. In other words, it is a model $(R, b)$ of the signature of a binary operation, such that the two morphisms of $R$-modules from $R \times R \times R$ to $R$, associating to any triple $(x, y, z)$ the two considered expressions are equal.

In Chapter 4, we give a definition of equations generalizing this example for models of any monadic signature $\Sigma$: this is the data,

- for each $R$ model of the signature $\Sigma$, of two morphisms of modules $e_R, e'_R$ with the same domain $A_R$ and codomain $B_R$ (in the example above, $e_R, e'_R : R \times R \times R \to R$ associate respectively $b(b(t, u), v)$ and $b(t, b(u, v))$ to the same triple $(t, u, v) \in R^3(X)$);
- for any morphism of $f : R \to S$ models, of two morphisms of modules $A_f : A_R \to A_S$ and $B_f : B_R \to B_S$ such that following diagrams commute:

$$
\begin{array}{ccc}
A_R & \xrightarrow{e_R} & B_R \\
{\scriptstyle A_f}\downarrow & & \downarrow{\scriptstyle B_f} \\
A_S & \xrightarrow[e'_S]{} & B_S
\end{array}
\qquad
\begin{array}{ccc}
A_R & \xrightarrow{e'_R} & B_R \\
{\scriptstyle A_f}\downarrow & & \downarrow{\scriptstyle B_f} \\
A_S & \xrightarrow[e'_S]{} & B_S
\end{array}
$$

These data are submitted as usual to an additional condition of functoriality. It is said that a model $R$ of the signature $\Sigma$ satisfies the equation when $e_R = e'_R$. In fact, we give an equivalent definition of equation in Chapter 4, as particular equational arities. A monad satisfies the equation if and only it is equipped with an action of the arity.

We then consider *2-signatures*: they are signatures consisting of an arity induced by a parametric module and a family of these equational arities. A model (or 2-model) of a 2-signature is a model $R$ of the signature induced by the underlying parametric module that satisfies all the equations. A 2-signature is said *algebraic* if the underlying parametric module is, and if the equations are *elementary* (Definition 104). Technically, they are equations whose functorial action of the domain sends surjective morphisms of monads on surjective morphisms of functors, and whose codomain is of the form $R \mapsto (((R')')^{\cdots})'$. All the examples of presentable signature that we consider in Chapter 3 can be reformulated as an algebraic 2-signature, inducing the same category of models (up to isomorphism). We show that any algebraic 2-signature is effective. (Theorem 107).

### 1.2.7 Semantics (Chapters 5 and 6)

It is possible to specify by a 2-signature the syntax of the lambda calculus quotiented by the $\beta$-equivalence (1.1): the terms $(\lambda x.t)\, u$ and $t\{x := u\}$ are thus equalized. However, this equation is usually oriented, and considered as an execution step when we consider lambda calculus as a functional programming language. As such, it is more appropriate to integrate the $\beta$-reduction into the semantics of language, rather than impose the $\beta$-equation in the syntax. This motivates the concept of *reduction monad*, which we introduce in the chapter 5, extending that of monad, to account for the additional structure of reduction. Intuitively, a reduction monad is a monad $R$ equipped, for each pair of terms $(t, u) \in R(X)$, with a set of reductions between $t$ and $u$ that we note $t \blacktriangleright u$, and an associated substitution operation: for any family of terms $(v_x)_{x \in X}$ with $v_x \in R(Y)$, for any reduction $m$ between $t$ and $u$, this substitution associates a reduction $m\{x \mapsto v_x\}$ between $t\{x \mapsto v_x\}$ and $u\{x \mapsto v_x\}$. Equations analogous to those involved in the definition of module are required.

In this context, a *reduction signature* consists of a signature for monads, that is to say a 2-signature, and a family of *reduction rules*. For example, the reduction rule for the congruence of the application of the lambda calculus is informally expressed as follows, by explicitly designating the application of the lambda calculated by the app construction:

$$\frac{T \rightsquigarrow T' \qquad U \rightsquigarrow U'}{\mathsf{app}(T, U) \rightsquigarrow \mathsf{app}(T', U')}$$

This rule breaks down into three pairs of terms: the hypotheses $(T, T')$ and $(U, U')$, and the conclusion $(\mathsf{app}(T, U), \mathsf{app}(T', U'))$, constructed from the "metavariables" $T$, $T'$, $U$, and $U'$. These parameterized pairs induce pairs of morphisms $(h_{1,1}(R), h_{1,2}(R))$, $(h_{2,1}(R), h_{2,2}(R))$ and $(c_1(R), c_2(R))$ of $R$-modules between $R^4$ and $R$ for any model $R$ of the monadic signature of lambda calculus. In addition, this construction is functorial. More precisely, if $f : R \to S$ is a morphism of models, then the following diagrams are

commutative:

$$\begin{array}{ccc} R^4 & \xrightarrow{h_{i,j}(R)} & R \\ {\scriptstyle f^4}\downarrow & & \downarrow{\scriptstyle f} \\ S^4 & \xrightarrow[h_{i,j}(S)]{} & S \end{array} \qquad \begin{array}{ccc} R^4 & \xrightarrow{c_i(R)} & R \\ {\scriptstyle f^4}\downarrow & & \downarrow{\scriptstyle f} \\ S^4 & \xrightarrow[c_i(S)]{} & S \end{array}$$

Thus formulated, this rule of congruence fits into our definition of *reduction rule* detailed in Section 5.3. An *action* of this rule in a reduction monad $R$ equipped with a binary operation app is given by a reduction app-cong$(m_T, m_U)$ between app$(T, T')$ and app$(U, U')$, for all $(T, T', U, U') \in R^4(X)$, any reduction $m_T$ between $T$ and $T'$, and any reduction $m_U$ between $U$ and $U'$. It is moreover necessary that app-cong commutes with substitution, that is, the following equation is verified:

$$\text{app-cong}(m_T, m_U)\{x \mapsto v_x\} = \text{app-cong}(m_T\{x \mapsto v_x\}, m_U\{x \mapsto v_x\})$$

The main theorem (Theorem 150) of Chapter 5 asserts the effectivity of a reduction signature composed of an effective 2-signature and any family of reduction rules.

Let us detail the involved notion of model here: it is a reduction monad equipped with

- a model structure of the 2-signature for the underlying monad,
- an action of each reduction rule of the signature.

The initial model is built from the initial model of the 2-signature, and the additional structure of reductions is built inductively from the reduction rules. Besides some signatures for variants of lambda calculus with $\beta$-reduction, we propose in the section 5.6 a signature for the lambda calculus with explicit substitution, as described in [Kes09], as a set of opertaions subject to a syntactic equation, with the addition of reduction rules between terms. This specification is done in three steps: a 1-signature for language operations, a 2-signature taking into account the syntactic equation, and a reduction signature specifying the relevant reductions.

In Chapter 6, we generalize these developments, and deal with new examples, such as lambda calculus in call by value with big-step reductions: the syntax is that of lambda calculus, but in this variant, a term reduces to a value, that is to say in a variable or a lambda abstraction. The notion of value is not stable by arbitray substitutions: for example, $x\{x := y\ y\} = y\ y$ is not a value. On the other hand, it is stable by substitution of values: $v$ and $(w_x)_{x \in X}$ are values, then $v\{x \mapsto w_x\}$ is a value. We can thus define the monad $\mathsf{LC}_v$ of the values of lambda calculus: it is equipped with an inclusion in the monad $L$ of lambda calculus that allows us to consider the latter as a module over $\mathsf{LC}_v$.

In this variant of lambda calculus, the $\beta$-reduction is formulated as:

$$\frac{t \rightsquigarrow \lambda x.t' \qquad u \rightsquigarrow u' \qquad t'\{x := u'\} \rightsquigarrow v}{t\ u \rightsquigarrow v}$$

Another reduction rule ensures that any value, as a term, reduces to itself.

The reductions are stable by substitution: if $t \in L(X)$ reduces to a value $v \in \mathsf{LC}_v(X)$ and $(u_x)_{x \in X}$ is a family of values indexed by the set of free variables $X$, then $t\{x \mapsto u_x\}$ reduces to $v\{x \mapsto u_x\}$.

The collection of sets of reductions between a term and a value is defined by induction, by successively applying the redution rules. It is equipped with an adequate substitution operation, as explained previously.

Note that a lambda term can be represented by a binary tree whose leaves are values and the nodes correspond to applications. This induced representation yields a bijection: such a binary tree determines a lambda term in a unique way.

$(x\ (\lambda y.t))\ z \qquad \Leftrightarrow$

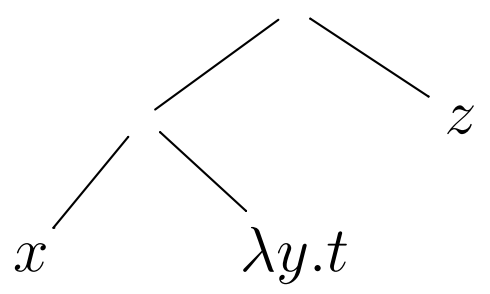

Thus we can identify the module $L$ of lambda terms with the composition $B \cdot \mathsf{LC}_v$, where $B$ is the monad of binary trees: $B(X)$ is the set of binary trees whose leaves are chosen in $X$.

Inspired by this example, we define the concept of *operational monad*. It consists of

- a monad $R$ (in our example, $R = \mathsf{LC}_v$);
- a pair $(T_1, T_2)$ of endofunctors on $\mathsf{Set}$ (in our for example, $T_1 = B$ and $T_2 = \mathsf{Id}$);
- a set of reductions between $t$ and $u$ for each pair $(t, u) \in T_1(R(X)) \times T_2(R(X))$;
- an appropriate substitution operation on these sets of reductions.

We then define the *operational signatures* that allow to specify this type of mathematical object, and prove an initiality result for these signatures (Theorem 207).

### 1.2.8 Formalization

Proofs of the main results of chapters 3 and 4 have been verified using the Coq proof assistant. This is a software in which it is possible to reproduce (or *formalize*) mathematical definitions as well as proofs which, if validated by the software, are indisputable in principle. This guarantee must in practice be qualified: software bugs allowing to prove contradictory statements are regularly found (and corrected). Furthermore, the correspondance between statements proved in Coq's type theory and mathematical statements is a non obvious theoretical question. More pragmatically, it is possible to make a mistake in the formalization of a definition, and the software Coq does not help in this step. It happens, for example, that we formalize a wrong characterization of a given set, this characterization being contradictory. This error may not be noticed because it does not prevent to show that the elements of this set satisfy some mathematical property, as the elements of the empty set satisfy any property. The reader of a formalization can thus certainly dispense to check the demonstrations, but he must pay close attention to the definitions.

For our developments, we chose to rely on the Coq library `UniMath`. This has some advantages, despite a "size" issue (which we will briefly mention):

i it contains useful definitions and results for our purpose;

ii it uses a limited amount of language features;

iii it incorporates the *axiom of univalence* and many of its consequences.

Regarding point i, `UniMath` proposes, among other things, an implementation of the theory of categories, and quotients of sets.

Point ii limits the complexity of the type theory in which we must trust, although this argument is highly questionable in view of the size issue. It also reduces the risk to be confronted with implementation bugs, by restricting itself to tried and tested features of Coq.

The axiom of univalence mentioned in point iii as such is not used crucially. It has the advantage, however, of gathering in this single axiom a number of useful consequences, for example functional extensionality (two functions are equal if they send each element on the same image), or propositional extensionality (logically equivalent propositions are equal) which makes it possible to handle the notion of subset.

Finally, the `UniMath` library has a major flaw due to a "size" issue. Indeed, it requires a compilation option that makes the system contradictory, allowing the implementation of a variant of Russel's paradox. This option is in practice used in the (impredicative) definition of quotients. We have ignored the existence of this particular inconsistency in the system in the proofs that we have formalized. We thus believe that our formalization does not exploit this logical flaw.

## 1.3 Synopsis

We give a synopsis of this thesis before presenting related work.

Chapter 2 concerns an arbitrary category $\mathsf{C}$ and presents a general notion of signature to specify objects of this category. Each of other chapters concerns a single category $\mathsf{C}$ and provides a class of effective signatures specifying objects of this category.

In Chapters 3 and 4, we have $\mathsf{C} = \mathsf{Mon}$ the category of monads on $\mathsf{Set}$. They are respectively adapted from [Ahr+19a] *High-level signatures and initial semantics* and [Ahr+19b] *Modular specification of monads through higher-order presentations*.

In Chapter 5, $\mathsf{C}$ is the category of reduction monads, which model both the syntax (as a monad) and the semantics of a language. We generalize them in Chapter 6 where we focus on the category of *operational monads*.

We suppose a certain familiarity with category theory and basic notions such as categories, functors, natural transformations, monads.

## 1.4 Computer-checked formalization

The intricate nature of our main results made it desirable to provide a mechanically checked proof of these results. We achieved this work for Chapters 3 and 4.

Our computer-checked proof is based on the `UniMath` library [VAG+], which itself is based on the proof assistant Coq [CoqDev19]. The main reasons for our choice of proof assistant are twofold: firstly, the logical basis of the Coq proof assistant, dependent type theory, is well suited for abstract algebra, in particular, for category theory. Secondly, a suitable library of category theory, ready for use by us, had already been developed [AL17].

The formalization can be consulted on `https://github.com/UniMath/largecatmodules`. A guide is given in the README.

For the purpose of this thesis, we refer to a fixed version of our library, with the short hash 50fd617. This version compiles with version 10839ee of `UniMath`.

Throughout the thesis, statements are annotated with their corresponding identifiers in the formalization. These identifiers are also hyperlinks to the online documentation stored at `https://initialsemantics.github.io/doc/50fd617/index.html`.

## 1.5 Related work

### 1.5.1 General signatures

A general notion of signature as an endofunctor with *strength*[11] (on a monoidal category) is suggested in the work on syntax of Fiore and his collaborators, starting with [FPT99]. This notion is also studied in Matthes and Uustalu [MU04] and Ghani, Uustalu, and Hamana [GUH06]. Their treatment rests on the technical device of strength that we avoid here. Any such signature gives rise to a signature in our sense (cf. Proposition 44 for the particular case of specifying monads).

Fiore and Hur's *equational systems* [FH09] yield a notion of signature which generalize signatures with strength (see [FH09, Section 6.1]). In this article, the authors provide a notion of equation for an algebra of an endofunctor, yielding a subcategory of algebras satisfying this equation as a definition of category of models, equipped with a forgetful functor to the base category. Equational systems yield particular signatures in our sense (Example 14), but contrary to the work of Fiore and Hur, we don't provide a general sufficient condition ensuring an initiality result.

### 1.5.2 Syntax and monads

In a classical paper, Barr [Bar70] explained the construction of the "free monad" generated by an endofunctor[12]. In another classical paper, Kelly and Power [KP93] explained how any finitary monad can be presented as a coequalizer of free monads[13]. There, free monads correspond to our initial models of a signature by an algebraic parametric module without any binding construction.

Fiore, Plotkin, and Turi [FPT99] develop a notion of substitution monoid. Following [ACU15], this setting can be rephrased in terms of relative monads and modules over them [Ahr16]. Accordingly, our present contributions could probably be customised for this "relative" approach.

Hamana [Ham03] proposes initial algebra semantics for "binding term rewriting systems", based on Fiore, Plotkin, and Turi's presheaf semantics of variable binding and Lüth and Ghani's monadic semantics of term rewriting systems [LG97b].

11. A (tensorial) strength for a functor $F : V \to V$ is given by a natural transformation $\beta_{v,w} : v \otimes Fw \to F(v \otimes w)$ commuting suitably with the associator and the unitor of the monoidal structure on $V$.
12. Fiore and Saville [FS17] give an enlightening generalization of the construction by Barr.
13. Their work has been applied to various more general contexts (e.g. [Sta13]).

The work by Fiore with collaborators [FPT99; FH10; FM10] and the work by Uustalu with collaborators [MU04; GUH06] share two traits: firstly, the modelling of variable binding by *nested abstract syntax*, and, secondly, the reliance on tensorial strengths in the specification of substitution.

The signatures for monads that we present here is actually closely related to that of Fiore and collaborators:

- Our notion of equations and that of model for them in Chapter 4 is very close to the notion of equational systems and that of algebra for them in [FH09]: in particular, the preservation of epimorphisms, which occurs in their construction of inductive free algebras for equational systems, appears here in our definition of elementary equation.

- In [FH10], Fiore and Hur introduce a notion of equation based on syntax with *meta-variables*: essentially, a specific syntax, say, $T := T(M, X)$ considered there depends on two contexts: a meta-context $M$, and an object-context $X$. The terms of the actual syntax are then those terms $t \in T(\emptyset, X)$ in an empty meta-context. An equation for $T$ is, simply speaking, a pair of terms in the same pair of contexts. Transferring an equation to any model of the underlying algebraic 1-signature is done by induction on the syntax with meta-variables. The authors show a monadicity theorem which straightforwardly implies an initiality result very similar to ours. That monadicity result is furthermore an instance of a more general theorem by Fiore and Mahmoud [FM10, Theorem 6.2].

- Translations between languages similar to the translation we present in Section 5.7 are also studied in [FM10]. It would be interesting to understand formal connections.

- At this stage, our work only concerns untyped syntax, but we anticipate it will generalize to the sorted setting as in [FH10] (see also the more general [FH13]).

We should mention several other mathematical approaches to syntax.

Gabbay and Pitts [GP99] employ a different technique for modelling variable binding, based on nominal sets. This *nominal* approach to binding syntax has been actively studied[14]. We highlight some contributions:

14. The approaches by Fiore and collaborators and Gabbay and Pitts [GP99] are nicely compared by Power [Pow07], who also comments on some generalization of the former approach.

- Clouston [Clo10] discusses signatures, structures (a.k.a. models), and equations over signatures in nominal style.
- Fernández and Gabbay [FG10] study signatures and equational theories as well as rewrite theories over signatures.
- Kurz and Petrisan [KP10] study closure properties of subcategories of algebras under quotients, subalgebras, and products. They characterize full subcategories closed under these operations as those that are definable by equations. They also show that the signature of the lambda calculus is effective, and study the subcategory of algebras of that signature specified by the $\beta$- and $\eta$-equations.

Yet another approach to syntax is based on Lawvere Theories. This is clearly illustrated in the paper [HP07], where Hyland and Power also outline the link with the language of monads and put in an historical perspective.

Finally, let us mention the classical approach based on Cartesian closed categories recently revisited and extended by T. Hirschowitz [Hir13].

### 1.5.3 Semantics

*Structural operational semantics* (SOS) are tackled by the seminal paper of Turi and Plotkin [TP97]. It introduced a largely accepted setting, rich of results and examples. However, this setting does not cover languages with binding constructions. The extension of this setting to the binding cases has been explored, notably by Staton [Sta08], but it does not cover lambda calculus. Several first steps towards it have been proposed [Ahr16; LG97a], where reductions are modeled by preorders. We propose a new one, based on graphs of reductions, starting with the reduction signatures in Chapter 5. They allow for the specification of

1. term constructions, including constructions that bind variables, e.g., abstraction;
2. syntactic equalities between terms; and
3. reduction rules, including reduction rules with hypotheses, e.g., congruence rules for the term constructions.

Ahrens [Ahr16] gives a notion of signature that allows for the specification of syntax with binding operations, as well as reduction rules on that syntax. The format for reduction rules considered there does not allow expressing rules with hypotheses, e.g., the aforementioned congruence rules. Instead, the congruence rules are hard-coded in [Ahr16], so that the head-$\beta$-reduction (and other limited variants) cannot be specified by that formalism. The constructors are modelled by morphisms of modules between modules into preordered sets, i.e., by families of preorder-preserving maps.

Signatures for rewriting systems, and initial semantics for them, are given by Hamana [Ham03] under the name "binding term rewriting system (BTRS)". Hamana considers preorder-valued functors. There, signatures for rewrite rules allow for rules without hypotheses only, though some rules with hypotheses, in particular, congruence rules, seem to be hard-coded in Hamana's framework (see [Ham03, Figure 3]).

CHAPTER 2

# GENERAL SIGNATURES

In this chapter, we give an abstract notion of signature for specifying objects of a category $\mathsf{C}$. All the notions of signatures presented in this manuscript are instances of these. In Chapters 3 and 4, we study the case when $\mathsf{C} = \mathsf{Mon}$ the category of monads on sets; in Chapter 5, we focus on the case when $\mathsf{C}$ is the category of reduction monads, and in Chapter 6, we consider the more general category of operational monads.

For each signature $S$, there is an associated category of models $\mathsf{C}^S$ equipped with a forgetful functor $U^S : \mathsf{C}^S \to \mathsf{C}$. For example, the category of models of the *empty signature* is just $\mathsf{C}$. By definition, the object of $\mathsf{C}$ specified by a signature $S$ is the image by $U^S$ of the initial model, if any.

For each signature $S$, we define a notion of $S$-arity, specifying operations or equations. A family $E$ of such arities induces an extended signature $S, E$, equipped with a functor $U^E$ from $\mathsf{C}^{S,E}$ to $\mathsf{C}^S$, factorizing $U^{S,E} : \mathsf{C}^{S,E} \to \mathsf{C}$.

Any signature is constructed inductively in this manner from the empty signature.

## 2.1 Arities over a category

We first define arities over a category $\mathsf{C}$, fixed in this section. Then, $S$-arities will be defined as arities over the category of models of the signature $S$.

We define what is an *action* of an arity in an object $c \in \mathsf{C}$, and the associated notion of compatible morphism of $\mathsf{C}$ between objects equipped with such an action.

We begin by introducing the general notion of arity over a category in Section 2.1.1, before considering the particular case of equational arities in Section 2.1.2, allowing to specify equations.

### 2.1.1 General arities

**Definition 1.** An **arity over** $\mathsf{C}$ is a quadruple $(\mathsf{D}, a, u, v)$ consisting of:

- a category $\mathsf{D}$;
- a functor $a : \mathsf{D} \to \mathsf{C}$;
- two sections $u, v : \mathsf{C} \to \mathsf{D}$ of $a : \mathsf{D} \to \mathsf{C}$.

**Remark 2.** We will mainly consider the case where $a$ is a Grothendieck fibration. In this case, thanks to the Grothendieck construction, an arity over $\mathsf{C}$ can be defined as a pseudo functor $a : \mathsf{C}^o \to \mathsf{Cat}$, and two natural transformations $u, v : 1 \to a$, where $1$ is the terminal functor from $\mathsf{C}$ to $\mathsf{Cat}$.

**Example 3.** Any endofunctor $F : \mathsf{C} \to \mathsf{C}$ induces an arity $(\mathsf{D}, a, u, v)$ specifying algebra structures for this endofunctor, as we will explain in Example 6:

- $a : \mathsf{C} \times \mathsf{C} \to \mathsf{C}$ is the first projection;
- $u = \langle \mathsf{Id}_{\mathsf{C}}, F \rangle$ maps an object $c$ to the pair $(c, F(c))$;
- $v$ is the diagonal functor, mapping an object $c$ to $(c, c)$.

Next, we define the notion of action of an arity in an object of $\mathsf{C}$, and the associated notion of compatible morphism of $\mathsf{C}$.

**Definition 4.** An **action of an arity** $A = (\mathsf{D}, a, u, v)$ **in an object** $c \in \mathtt{ob}\,\mathsf{C}$ is a morphism $h : u(c) \to v(c)$ such that $a(h) = \mathtt{id}_c$.

**Definition 5.** Let $A = (\mathsf{D}, a, u, v)$ be an arity over $\mathsf{C}$. Let $c_1$ and $c_2$ be two objects of $\mathsf{C}$ equipped with actions $h_1 : u(c_1) \to v(c_1)$ and $h_2 : u(c_2) \to v(c_2)$. A morphism $f : c_1 \to c_2$ is **compatible with the actions** $h_1$ **and** $h_2$ if the following diagram commutes:

$$\begin{array}{ccc} u(c_1) & \xrightarrow{h_1} & v(c_1) \\ {\scriptstyle u(f)}\downarrow & & \downarrow{\scriptstyle v(f)} \\ u(c_2) & \xrightarrow[h_2]{} & v(c_2) \end{array}$$

**Example 6.** Let $F$ be an endofunctor on $\mathsf{C}$. In Example 3, we constructed from it an arity . An action of it in an object $c \in \mathtt{ob}\,\mathsf{C}$ is a morphism $h : F(c) \to c$. A compatible morphism between $c$ and $c'$ equipped with actions $h$ and $h'$ is a morphism between the induced algebras of $F$.

### 2.1.2 Equational arities

**Definition 7.** An arity $(\mathsf{D}, a, u, v)$ over $\mathsf{C}$ is said **equational** if for every object $c \in \mathsf{C}$, the category $a^{-1}(c)$ is discrete (that is, any morphism is an identity morphism), where $a^{-1}(c)$ is the subcategory of objects of $\mathsf{D}$ mapped to $c$ and morphisms mapped to $\mathtt{id}_c$ by $a$.

**Remark 8.** Let $(\mathsf{D}, a, u, v)$ be an arity. If $a$ is a Grothendieck fibration, then this arity is equational if and only if $a$ is a discrete fibration.

**Remark 9.** An action of an equational arity $(\mathsf{D}, a, u, v)$ in an object $c \in \mathsf{C}$ is always an identity morphism. Thus, an object $c$ is equipped with an action if and only if $u_c = v_c$. Any morphism is then compatible with actions of an equational arity.

**Example 10.** An *equational system* [FH09] induces an equational arity over a category of algebras. More precisely, an equational system $\mathbb{S} = (\mathsf{C} : \Sigma \triangleright \Gamma \vdash L = R)$ consists of endofunctors $\Sigma, \Gamma : \mathsf{C} \to \mathsf{C}$, and functors $L, R : \Sigma\text{-alg} \to \Gamma\text{-alg}$ between categories of algebras preserving the underlying object, that is, such that the following diagram commutes:

$$\begin{array}{ccc} \Sigma\text{-alg} & \overset{L}{\underset{R}{\rightrightarrows}} & \Gamma\text{-alg}\,. \\ & \searrow \quad \swarrow & \\ & \mathsf{C} & \end{array}$$

Such a structure defines an equational arity $(\mathsf{D}, a, u, v)$ over the category $\Sigma\text{-alg}$ as follows.

- $\mathsf{D}$ is the category of objects $c \in \mathsf{C}$ equipped with an algebra structure for both $\Sigma$ and $\Gamma$. More formally, $\mathsf{D}$ is defined as the pullback

$$\begin{array}{ccc} \mathsf{D} & \longrightarrow & \Gamma\text{-alg}\,. \\ {\scriptstyle a}\downarrow & \lrcorner & \downarrow \\ \Sigma\text{-alg} & \longrightarrow & \mathsf{C} \end{array}$$

- $a : \mathsf{D} \to \Sigma\text{-alg}$ is induced by the definition of $\mathsf{D}$ as a pullback.

- $u$ maps a $\Sigma$-algebra $c$ to the underlying object of $c$ equipped with the same $\Sigma$-algebra structure and the $\Gamma$-algebra given by $L(c)$.

- $v$ maps a $\Sigma$-algebra $c$ to the underlying object of $c$ equipped with the same $\Sigma$-algebra structure and the $\Gamma$-algebra given by $R(c)$.

Then, the category of $\mathbb{S}$-algebras [FH09, Definition 3.4] is retrieved as the category of objects of $\Sigma\text{-alg}$ equipped with an action of this arity and compatible morphisms. More concretely, it is the full subcategory of algebras of $\Sigma$ that have the same image by $L$ and $R$.

### 2.1.3 Family of arities and their models

**Definition 11.** Let $E$ be a family of arities over $\mathsf{C}$. The **category** $\mathsf{C}^E$ **of models of** $E$ is defined as follows:

- objects are objects $c$ of $\mathsf{C}$ equipped with an action of each arity in $E$;
- morphisms are those which are compatible with the action of any arity in $E$.
- composition and identities are the obvious ones.

It is equipped with a forgetful functor $U^E : \mathsf{C}^E \to \mathsf{C}$.

## 2.2 Signatures over a category

Here we define signatures and their models.

**Definition 12.** A **signature over a category** $\mathsf{C}$ is a finite list $E_1, \ldots, E_n$ consisting of families of arities over categories such that:

- $E_1$ is a family of arities over $\mathsf{C}$.
- each $E_i$ is a family of arities over the category of models of $E_{i-1}$, for $i > 1$.

The empty list is called the **empty signature**. The **category of models** $\mathsf{C}^S$ **of a signature** $S = E_1, \ldots, E_n$ is defined as the category of models of $E_n$, if $n > 0$, or as $\mathsf{C}$ otherwise. Given a signature $S$, a $S$**-arity** is an arity over the category of models $\mathsf{C}^S$. Each signature $S = E_1, \ldots, E_n$ is equipped with a forgetful functor $U^S$ from its category of models to $\mathsf{C}$ defined as the composition:

$$\mathsf{C}^{E_n} \xrightarrow{U^{E_n}} \mathsf{C}^{E_{n-1}} \xrightarrow{U^{E_{n-1}}} \ldots \longrightarrow \mathsf{C}^{E_1} \xrightarrow{U^{E_1}} \mathsf{C}$$

An **action of a signature** $S = E_1, \ldots, E_n$ **in an object** $c \in \mathtt{ob}\,\mathsf{C}$ is an action of $E_1$ in $c$, inducing a model $m_1$ of $E_1$, then an action of $E_2$ in $c$, inducing a model $m_2$ of $E_2$, and so on.

**Example 13.** An endofunctor $\Sigma : \mathsf{C} \to \mathsf{C}$ induces a signature consisting of a single arity constructed in Example 3: it is easily verified that the category of models of this signature is the category of algebras of $\Sigma$.

**Example 14.** An equational system $\mathsf{C} : \Sigma \triangleright \Gamma \vdash L = R$ yields a signature extending the one induced by the endofunctor $\Sigma$ (as explained in Example 13) with a singleton family consisting of the equational arity induced by the equational system (Example 10). The associated category of models coincides with the category of algebras of this equational system, as defined in [FH09, Definition 3.4]. More concretely, as explained in Example 10, models are objects of $\mathsf{C}$ equipped with a $\Sigma$-algebra structure that is mapped to the same image by $L$ and $R$, and morphisms are algebra morphisms.

**Definition 15.** A signature $S$ is said **effective** if its category of models has an initial object $\hat{S}$.

In the following chapters, we consider different instances of signatures:

- In Chapter 3, signatures consist of a single arity over the category of monads.
- In Chapter 4, these signatures are extended with a family of equational arities.
- In Chapters 5 and 6, we introduce the category of reduction monads, and then of their generalization, the operational monads. They are monads with additionnal structure. Thus, signatures over the category of monads are relevant for specifying the monadic part: we explain how to make it formal in the next section. In Chapter 5, we construct arities specifying the additional structure of reduction monads. Signatures for operational monads of Chapter 6 extend them with arities for specifying the *state functors*.

## 2.3 Pulling back arities and signatures

In this section, we explain how a functor $F : \mathsf{B} \to \mathsf{C}$ can turn any signature $S$ over $\mathsf{C}$ into a *pullback signature* $F^*S$ over $\mathsf{B}$. This will be used in Chapters 5 and 6 to upgrade signatures for monads into signatures for reduction or operational monads. We explain how it works for arities.

### 2.3.1 Pullback of arities

In this section, we fix a functor $F : \mathsf{B} \to \mathsf{C}$.

**Definition 16.** Let $A = (\mathsf{D}, a, u, v)$ be an arity over $\mathsf{C}$. The **pullback of $A$ along $F$** is the arity $F^*A = (\mathsf{D}', a', u', v')$ over $\mathsf{B}$ as follows:

- $a' : \mathsf{D}' \to \mathsf{B}$ is the pullback of $a$ along $F$, considering the pullback diagram

$$\begin{array}{ccc} \mathsf{D}' & \xrightarrow{d} & \mathsf{D} \\ {\scriptstyle a'}\downarrow & \lrcorner & \downarrow{\scriptstyle a} \\ \mathsf{B} & \xrightarrow[F]{} & \mathsf{C} \end{array}$$

- $u'$ and $v'$ are the universal morphisms from $\mathsf{B}$ to $\mathsf{D}'$ factorizing the cones

$$\begin{array}{ccc} \mathsf{B} & \xrightarrow{u \cdot F} & \mathsf{D} \\ {\scriptstyle \mathtt{id}_B}\downarrow & & \downarrow{\scriptstyle a} \\ \mathsf{B} & \xrightarrow[F]{} & \mathsf{C} \end{array} \qquad \begin{array}{ccc} \mathsf{B} & \xrightarrow{v \cdot F} & \mathsf{D} \\ {\scriptstyle \mathtt{id}_B}\downarrow & & \downarrow{\scriptstyle a} \\ \mathsf{B} & \xrightarrow[F]{} & \mathsf{C} \end{array}$$

**Definition 17.** Let $E$ be a family of arities over $\mathsf{C}$. The **pullback of $E$ along $F$** is the family of arities $F^*E$ over $\mathsf{B}$ consists of the pullbacks of arities of $E$ along $F$.

The process of taking the category of models commutes with pullback:

**Proposition 18.** *Let $E$ be a family of arities over $\mathsf{C}$. Then there is a functor $p_{F,E}$ : $\mathsf{B}^{F^*E} \to \mathsf{C}^E$ inducing a pullback diagram*

$$\begin{array}{ccc} \mathsf{B}^{F^*E} & \xrightarrow{p_{F,E}} & \mathsf{C}^E \\ {\scriptstyle U^{F^*E}}\downarrow & \lrcorner & \downarrow{\scriptstyle U^E} \\ \mathsf{B} & \xrightarrow[F]{} & \mathsf{C} \end{array}$$

### 2.3.2 Pullback of signatures

In this section, we fix a functor $F : \mathsf{B} \to \mathsf{C}$.

**Definition 19.** The **pullback $F^*S$ of a signature $S = E_1, \ldots, E_n$ over $\mathsf{C}$ along $F$** is the signature $p_1^*E_1, p_2^*E_2, \ldots, p_n^*E_n$ over $\mathsf{B}$, where the functors $p_i$ are defined inductively as follows:

- $p_1 : \mathsf{B} \to \mathsf{C}$ is $F$;
- $p_i : \mathsf{B}^{p^*_{i-1}E_{i-1}} \to \mathsf{C}^{E_{i-1}}$ is $p_{p_{i-1},E_{i-1}}$ for $i > 1$.

We denote by $p_{F,S} : \mathsf{B}^{F^*S} \to \mathsf{C}^S$ the functor $F$ if $n = 0$, or $p_{p_n,E_n}$ otherwise.

Again, the process of taking the category of models commutes with pullback:

**Proposition 20.** *Let $S$ be a signature over $\mathsf{C}$. Then we have the following pullback diagram:*

$$\begin{array}{ccc} \mathsf{B}^{F^*S} & \xrightarrow{p_{F,S}} & \mathsf{C}^S \\ {\scriptstyle U^{F^*S}}\downarrow & \lrcorner & \downarrow{\scriptstyle U^S} \\ \mathsf{B} & \xrightarrow[F]{} & \mathsf{C} \end{array}$$

Pullbacks allow to combine multiple signatures over a category $\mathsf{C}$ into a single one:

**Definition 21.** We define the **product $S_1 \times \cdots \times S_n$ of signatures $S_1$, ..., $S_n$ over** $\mathsf{C}$ as the signature over $\mathsf{C}$ defined inductively on $n$ as follows:

- if $n = 0$, then the resulting signature is the empty signature;
- if $n > 0$, then the resulting signature is the pullback of $S_1 \times \cdots \times S_{n-1}$ along $U^{S_n} : \mathsf{C}^{S^n} \to \mathsf{C}$.

PART I

# Syntax

CHAPTER 3

# PRESENTABLE SIGNATURES FOR MONADS

---

The present chapter is adapted from [Ahr+19a]. We identify the class of *presentable signatures* over the category of monads and we show that they are effective. Monads account for the syntax of some untyped programming languages with variable binding.

These signatures are fairly more general than those introduced in some of the seminal papers on this topic [FPT99; HHP93; GP99], which are essentially given by a family of lists of natural numbers indicating the number of variables bound in each subterm of a syntactic construction (we call them “algebraic signatures” below).

Examples are given in Section 3.8.

This work improves on a previous attempt [HM12] in two main ways: firstly, it gives a much simpler condition for effectivity; secondly, it provides computer-checked proofs for all the main statements.

## Organisation of the chapter

Section 3.1 gives a succinct account of the notion of module over a monad, which is the crucial tool underlying our definition of presentables signatures. Signatures and models are described in Sections 3.2 and 3.3 respectively. In Section 3.4, we give our definition of a syntax, and we present our first main result, a modularity result about merging extensions of syntax. In Section 3.5, we show through examples how recursion can be recovered from initiality. Our notion of *presentation of a signature* appears in Section 3.6. There, we also state our second main result: presentable signatures generate a syntax. The proof of that result is given in Section 3.7. Finally, in Section 3.8, we give examples of presentable signatures.

## 3.1 Categories of modules over monads

The main mathematical notion underlying our presentable signatures is that of module over a monad. In this section, we recall the definition and some basic facts about modules over a monad in the specific case of the category $\mathsf{Set}$ of sets, although most definitions are generalizable. See [HM10] for a more extensive introduction on this topic.

### 3.1.1 Modules over monads

A *monad* (over $\mathsf{Set}$) is a monoid in the category $\mathsf{Set} \longrightarrow \mathsf{Set}$ of endofunctors of $\mathsf{Set}$, i.e., a triple $R = (R, \mu, \eta)$ given by a functor $R \colon \mathsf{Set} \longrightarrow \mathsf{Set}$, and two natural transformations $\mu \colon R \cdot R \longrightarrow R$ and $\eta \colon I \longrightarrow R$ such that the following equations hold:

$$\mu \circ \mu R = \mu \circ R\mu, \qquad \mu \circ \eta R = 1_R, \qquad \mu \circ R\eta = 1_R \ .$$

Given two monads $R = (R, \eta, \mu)$ and $R' = (R', \eta', \mu')$, a *morphism* $f \colon R \to R'$ *of monads* is given by a natural transformation $f : R \longrightarrow S$ between the underlying functors such that

$$f \circ \eta = \eta', \qquad f \circ \mu = \mu' \circ (f \cdot f) \ .$$

Let $R$ be a monad.

**Definition 22** (Modules)**.** A left $R$-module is given by a functor $M \colon \mathsf{Set} \longrightarrow \mathsf{Set}$ equipped with a natural transformation $\rho^M \colon M \cdot R \longrightarrow M$, called *module substitution*, which is compatible with the monad composition and identity:

$$\rho^M \circ \rho^M R = \rho^M \circ M\mu, \qquad \rho^M \circ M\eta = 1_M.$$

There is an obvious corresponding definition of right $R$-modules that we do not need to consider in this thesis. From now on, we will write “$R$-module” instead of “left $R$-module” for brevity.

**Example 23.**

- Every monad $R$ is a module over itself, which we call the *tautological* module.
- For any functor $F \colon \mathsf{Set} \longrightarrow \mathsf{Set}$ and any $R$-module $M \colon \mathsf{Set} \longrightarrow \mathsf{Set}$, the composition $F \cdot M$ is an $R$-module (in the evident way).

- For every set $W$ we denote by $\underline{\mathsf{W}}\colon \mathsf{Set} \longrightarrow \mathsf{Set}$ the constant functor $\underline{\mathsf{W}} := X \mapsto W$. Then $\underline{\mathsf{W}}$ is trivially an $R$-module since $\underline{\mathsf{W}} = \underline{\mathsf{W}} \cdot R$.
- Let $M_1$, $M_2$ be two $R$-modules. Then the product functor $M_1 \times M_2$ is an $R$-module (see Proposition 25 for a general statement).

**Definition 24** (Linearity)**.** We say that a natural transformation of $R$-modules $\tau\colon M \longrightarrow N$ is *linear*[1] if it is compatible with module substitution on either side:

$$\tau \circ \rho^M = \rho^N \circ \tau R.$$

We take linear natural transformations as morphisms among modules. It can be easily verified that we obtain in this way a category $\mathsf{Mod}(R)$.

Limits and colimits in the category of modules can be constructed pointwise:

**Proposition 25** (`LModule_Colims_of_shape`, `LModule_Lims_of_shape`)**.** $\mathsf{Mod}(R)$ *is complete and cocomplete.*

### 3.1.2 The total category of modules

We already introduced the category $\mathsf{Mod}(R)$ of modules with fixed base $R$. It is often useful to consider a larger category which collects modules with different bases. To this end, we need first to introduce the notion of pullback.

**Definition 26** (Pullback)**.** Let $f\colon R \longrightarrow S$ be a morphism of monads and $M$ an $S$-module. The module substitution $M \cdot R \xrightarrow{Mf} M \cdot S \xrightarrow{\rho^M} M$ defines an $R$-module which is called *pullback* of $M$ along $f$ and noted $f^*M$.[2]

**Definition 27** (The total module category)**.** We define the *total module category* $\int^R \mathsf{Mod}(R)$, or $\int \mathsf{Mod}$ for short, as follows[3]:

1. Given a monoidal category $\mathcal{C}$, there is a notion of (left or right) module over a monoid object in $\mathcal{C}$ (see, e.g., [Bra14, Section 4.1] for details). The term "module" comes from the case of rings: indeed, a ring is just a monoid in the monoidal category of Abelian groups. Similarly, our monads are just the monoids in the monoidal category of endofunctors on $\mathsf{Set}$, and our modules are just modules over these monoids. Accordingly, the term "linear(ity)" for morphisms among modules comes from the paradigmatic case of rings.
2. The term "pullback" is standard in the terminology of Grothendieck fibrations (see Proposition 28).
3. Our notation for the total category is modelled after the category of elements of a presheaf, and, more generally, after the Grothendieck construction of a pseudofunctor.

- its objects are pairs $(R, M)$ of a monad $R$ and an $R$-module $M$;
- a morphism from $(R, M)$ to $(S, N)$ is a pair $(f, m)$ where $f\colon R \longrightarrow S$ is a morphism of monads, and $m\colon M \longrightarrow f^*N$ is a morphism of $R$-modules.

Composition and identity morphisms are the expected ones. The category $\int \mathsf{Mod}$ comes equipped with a forgetful functor to the category of monads, given by the projection $(R, M) \mapsto R$.

**Proposition 28** (`cleaving_bmod`)**.** *The forgetful functor* $\int \mathsf{Mod} \to \mathsf{Mon}$ *is a Grothendieck fibration with fibre* $\mathsf{Mod}(R)$ *over a monad* $R$*. In particular, any monad morphism* $f : R \longrightarrow S$ *gives rise to a functor*

$$f^*\colon \mathsf{Mod}(S) \longrightarrow \mathsf{Mod}(R)$$

*given on objects by Definition 26.*

**Proposition 29** (`pb_LModule_colim_iso`, `pb_LModule_lim_iso`)**.** *For any monad morphism* $f : R \longrightarrow S$*, the functor* $f^* : \mathsf{Mod}(S) \longrightarrow \mathsf{Mod}(R)$ *preserves limits and colimits.*

### 3.1.3 Derivation

For our purposes, important examples of modules are given by the following general construction. Let us denote the final object of $\mathsf{Set}$ as $*$.

**Definition 30** (Derivation)**.** For any $R$-module $M$, the *derivative* of $M$ is the functor $M' := X \mapsto M(X + *)$. It is an $R$-module with the substitution $\rho^{M'}\colon M' \cdot R \longrightarrow M'$ defined as in the diagram

$$\begin{array}{ccc} M(R(X) + *) & \xrightarrow{\rho^{M'}_X} & M(X + *) \\ {\scriptstyle M(\ [R(i_X),\ \eta_{X+*} \circ \underline{*}]\ )} \big\downarrow & \nearrow {\scriptstyle \rho^M_{X+*}} & \\ M(R(X + *)) & & \end{array} \tag{3.1}$$

where $i_X\colon X \longrightarrow X + *$ and $\underline{*}\colon * \longrightarrow X + *$ are the canonical injections.

Derivation is a continuous and cocontinuous endofunctor on the category $\mathsf{Mod}(R)$ of modules over a fixed monad $R$. In particular, derivation can be iterated: we denote by $M^{(k)}$ the $k$-th derivative of $M$.

**Definition 31.** Given a list of nonnegative integers $(a) = (a_1, \ldots, a_n)$ and a left module $M$ over a monad $R$, we denote by $M^{(a)} = M^{(a_1,\ldots,a_n)}$ the module $M^{(a_1)} \times \cdots \times M^{(a_n)}$. Observe that, when $(a) = ()$ is the empty list, $M^{()}$ is the final module $*$.

**Definition 32.** For every monad $R$ and $R$-module $M$ we have a natural *substitution morphism* $\sigma \colon M' \times R \longrightarrow M$ defined by $\sigma_X = \rho^M_X \circ w_X$, where $w_X : M(X + *) \times R(X) \to M(R(X))$ is the map

$$w_X \colon (a, b) \mapsto M([\eta_X, \underline{b}])(a), \qquad \underline{b} \colon * \mapsto b.$$

**Lemma 33** (`substitution_laws`)**.** *The transformation $\sigma$ is linear.*

The substitution $\sigma$ allows us to interpret the derivative $M'$ as the “module $M$ with one formal parameter added”.

Abstracting over the module turns the substitution morphism into a natural transformation that is the counit of the following adjunction:

**Proposition 34** (`deriv_adj`)**.** *The endofunctor of $\mathsf{Mod}(R)$ mapping $M$ to the $R$-module $M \times R$ is left adjoint to the derivation endofunctor, the counit being the substitution morphism $\sigma$.*

## 3.2 Signatures for monads

In this section, we consider particular signatures over the category of monads consisting of a single arity in the sense of Chapter 2.

The purpose of a signature is to act on monads. An action of a signature $\Sigma$ in a monad $R$ is a morphism from the module $\Sigma(R)$ to the tautological one $R$. For instance, in the case of the signature $\Sigma$ of a binary operation, we have $\Sigma(R) := R^2 = R \times R$. Hence a signature assigns, to each monad $R$, a module over $R$ in a functorial way: this motivates the definition of *parametric modules*.

**Definition 35.** A *parametric module* is a section $\Sigma$ of the forgetful functor from the category $\int \mathsf{Mod}$ to the category $\mathsf{Mon}$, that is, a functor $\Sigma : \int \mathsf{Mod} \to \mathsf{Mon}$ making the following diagram commute:

$$\begin{array}{ccc} \mathsf{Mon} & \xrightarrow{\Sigma} & \int \mathsf{Mod} \\ & \searrow\!\!\!\!= \quad \swarrow & \\ & \mathsf{Mon} & \end{array}$$

We give first a basic example:

**Example 36.** The assignment $R \mapsto R$ yields a parametric module, which we denote by $\Theta$.

**Remark 37.** Any parametric module $\Sigma$ defines uniquely a signature over the category of monads consisting of a single arity $(\int \mathsf{Mod}, U, \Sigma, \Theta)$, where $U : \int \mathsf{Mod} \to \mathsf{Mon}$ is the canonical forgetful functor. In this Chapter, we only consider such signatures, which we take as a definition of signature in the formalization. In the following we may use the same expression to designate both the parametric module and the induced signature.

Now we give some basic examples of parametric modules, inducing signatures.

**Example 38.**

1. For any functor $F\colon \mathsf{Set} \longrightarrow \mathsf{Set}$ and any parametric module $\Sigma$, the assignment $R \mapsto F \cdot \Sigma(R)$ yields a parametric module which we denote $F \cdot \Sigma$.
2. The assignment $R \mapsto *_R$, where $*_R$ denotes the final module over $R$, yields a parametric module which we denote by $*$.
3. Given two parametric modules $\Sigma$ and $\Upsilon$, the assignment $R \mapsto \Sigma(R) \times \Upsilon(R)$ yields a parametric module which we denote by $\Sigma \times \Upsilon$. For instance, $\Theta^2 = \Theta \times \Theta$ is the parametric module of any (first-order) binary operation, and, more generally, $\Theta^n$ is the parametric module of $n$-ary operations.
4. Given two parametric modules $\Sigma$ and $\Upsilon$, the assignment $R \mapsto \Sigma(R) + \Upsilon(R)$ yields a parametric module which we denote by $\Sigma + \Upsilon$. For instance, $\Theta^2 + \Theta^2$ yields the signature of a pair of binary operations.

The last example above explains how we can combine multiple signatures induced by parametric modules into a single one. Hence we do not need to distinguish between such arities—used to specify a single syntactic construction—and families of such arities—used to specify a family of syntactic constructions. Our notion of signatures that we present here allow us to do both (via Proposition 42 for families that are not necessarily finitely indexed).

*Elementary* signatures are of a particularly simple shape:

**Definition 39.** For each sequence of nonnegative integers $s = (s_1, \ldots, s_n)$, the assignment $R \mapsto R^{(s_1)} \times \cdots \times R^{(s_n)}$ (see Definition 31) is a parametric module, which we denote by $\Theta^{(s)}$, or by $\Theta'$ in the specific case of $s = (1)$. Parametric modules of this form and their induced signatures are said *elementary*.

**Remark 40.** The product of two elementary parametric modules is elementary.

**Definition 41.** A *morphism between two parametric modules* $\Sigma_1, \Sigma_2 \colon \mathsf{Mon} \longrightarrow \int \mathsf{Mod}$ is a natural transformation $m \colon \Sigma_1 \longrightarrow \Sigma_2$ which, post-composed with the projection $\int \mathsf{Mod} \longrightarrow \mathsf{Mon}$, becomes the identity. Parametric modules form a subcategory $\mathsf{PMod}$ of the category of functors from $\mathsf{Mon}$ to $\int \mathsf{Mod}$.

Limits and colimits of parametric modules can be easily constructed pointwise:

**Proposition 42** (`Sig_Lims_of_shape`, `Sig_Colims_of_shape`, `Sig_isDistributive`)**.** *The category of parametric modules is complete and cocomplete. Furthermore, it is distributive: for any parametric module $\Sigma$ and family of parametric modules $(S_o)_{o \in O}$, the canonical morphism $\coprod_{o \in O}(S_o \times \Sigma) \to (\coprod_{o \in O} S_o) \times \Sigma$ is an isomorphism.*

**Definition 43.** An *algebraic parametric module* is a (possibly infinite) coproduct of elementary parametric modules. The induced signature is said *algebraic*.

These signatures are those which appear in [FPT99]. For instance, the algebraic signature of the lambda-calculus is induced by the parametric module $\Sigma_{\mathsf{LC}} = \Theta^2 + \Theta'$.

To conclude this section, we explain the connection between *signatures with strength* (on the category $\mathsf{Set}$) and our signatures induced by parametric modules.

Signatures with strength were introduced in [MU04] (even though they were not given an explicit name there). The relevant definitions regarding signatures with strength are summarized in [AMM18], to which we refer the interested reader.

We recall that a signature with strength [AMM18, Definition 4] is a pair of an endofunctor $H : [\mathcal{C}, \mathcal{C}] \to [\mathcal{C}, \mathcal{C}]$ together with a strength-like datum. Here, we only consider signatures with strength over the base category $\mathcal{C} := \mathsf{Set}$. Given a signature with strength $H$, we also refer to the underlying endofunctor on the functor category $[\mathsf{Set}, \mathsf{Set}]$ as $H : [\mathsf{Set}, \mathsf{Set}] \to [\mathsf{Set}, \mathsf{Set}]$.

A morphism of signatures with strength [AMM18, Definition 5] is a natural transformation between the underlying functors that is compatible with the strengths in a suitable sense. Together with the obvious composition and identity, these objects and morphisms form a category SigStrength [AMM18].

Any signature with strength $H$ gives rise to a parametric module $\tilde{H}$ [HM12, Section 7]. This parametric module associates, to a monad $R$, an $R$-module whose underlying functor is $H(UR)$, where $UR$ is the functor underlying the monad $R$. Similarly, given two signatures with strength $H_1$ and $H_2$, and a morphism $\alpha : H_1 \to H_2$ of signatures with strength, we associate to it a morphism of parametric modules $\tilde{\alpha} : \tilde{H}_1 \to \tilde{H}_2$. This morphism sends a monad $R$ to a module morphism $\tilde{\alpha}(R) : \tilde{H}_1(R) \longrightarrow \tilde{H}_2(R)$ whose underlying natural transformation is given by $\alpha(UR)$, where, as before, $UR$ is the functor underlying the monad $R$. These maps assemble into a functor:

**Proposition 44** (`sigWithStrength_to_sig_functor`)**.** *The maps sketched above yield a functor* $\tilde{(-)}$ : SigStrength $\longrightarrow$ PMod*.*

## 3.3 Categories of models

We unfold the definitions of *model of a signature* and *action of a signature* (as introduced in the general setting in Chapter 2) to the case of signatures induced by parametric modules.

**Definition 45** (Models and actions)**.** Given a parametric module $\Sigma$, the *category* $\mathsf{Mon}^\Sigma$ *of models of* $\Sigma$ is defined as the category of models of the induced signature. Objects are pairs $(R, r)$ of a monad $R$ equipped with an action of $\Sigma$ in $R$, that is, with a module morphism $r : \Sigma(R) \to R$. A *morphism from* $(R, r)$ *to* $(S, s)$ is a morphism of monads $m : R \to S$ compatible with the actions, in the sense that the following diagram of $R$-modules commutes:

$$\begin{array}{ccc} \Sigma(R) & \xrightarrow{\quad r \quad} & R \\ {\scriptstyle \Sigma(m)}\downarrow & & \downarrow{\scriptstyle m} \\ m^*(\Sigma(S)) & \xrightarrow[\quad m^*s \quad]{} & m^*S \end{array}$$

Here, the horizontal arrows come from the actions, the left vertical arrow comes from the functoriality of parametric modules, and $m\colon R \longrightarrow m^*S$ is the morphism of monads seen as morphism of $R$-modules. This is equivalent to asking that the square of underlying natural transformations commutes, i.e., $m \circ r = s \circ \Sigma(m)$.

**Example 46.** The usual $\mathsf{app}\colon \mathsf{LC}^2 \longrightarrow \mathsf{LC}$ is an action of the elementary signature $\Theta^2$ in the monad $\mathsf{LC}$ of syntactic lambda calculus. The usual $\mathsf{abs}\colon \mathsf{LC}' \longrightarrow \mathsf{LC}$ is an action of the elementary signature $\Theta'$ in the monad $\mathsf{LC}$. Then $[\mathsf{app}, \mathsf{abs}] : \mathsf{LC}^2 + \mathsf{LC}' \longrightarrow \mathsf{LC}$ is an action of the algebraic signature of the lambda calculus $\Theta^2 + \Theta'$ in the monad $\mathsf{LC}$.

In the formalisation, the category of models of a signature $\Sigma$ is recovered as the fiber category over $\Sigma$ of the displayed category [AL17] of models, see `rep_disp`. We have also formalized a direct definition (`rep_fiber_category`) and shown that the two definitions yield isomorphic categories: `catiso_modelcat`.

**Definition 47** (Pullback)**.** Let $f\colon \Upsilon \longrightarrow \Sigma$ be a morphism of parametric modules and $(R, r)$ a model of $\Sigma$. The linear morphism $\Upsilon(R) \xrightarrow{f(R)} \Sigma(R) \xrightarrow{r} R$ defines an action of $\Upsilon$ in $R$. The induced model of $\Upsilon$ is called *pullback*[4] of $(R, r)$ along $f$ and denoted by $f^*(R, r)$.

## 3.4 Syntax

We are primarily interested in parametric modules $\Sigma$ inducing effective signatures.

### 3.4.1 Effectivity and parametric modules

**Definition 48.** We call a parametric module *effective* if its induced signature is. If $\Sigma$ is effective, that is, if $\mathsf{Mon}^{\Sigma}$ has an initial object, this object is essentially unique; we call it the *syntax generated by* $\Sigma$, denoted by $\hat{\Sigma}$. By abuse of notation, we also denote by $\hat{\Sigma}$ the monad underlying the model $\hat{\Sigma}$.

In this work, we aim to identify parametric modules that are effective. This is not automatic: below, we give a parametric module that is not effective. Afterwards, we give suitable sufficient criteria for parametric modules to be effective.

**Non-example 49.** Let $\mathcal{P}$ denote the powerset functor and consider the parametric module $\mathcal{P} \cdot \Theta$ (see Example 38, Item 1): it associates, to any monad $R$, the module $\mathcal{P} \cdot R$ that sends a set $X$ to the powerset $\mathcal{P}(RX)$ of $RX$. The induced signature is not effective.

4. Following the terminology introduced in Definition 26, the term “pullback” is justified by Lemma 55.

Instead of giving a direct proof of the fact that $\mathcal{P} \cdot \Theta$ is not effective, we deduce it as a simple consequence of a stronger result that we consider interesting in itself: an analogue of Lambek's Lemma, given in Lemma 52.

The following preparatory lemma explains how to construct new models of a parametric module $\Sigma$ from old ones:

**Lemma 50.** *Let $(R, r)$ be a model of a parametric module $\Sigma$. Let $\eta : \mathsf{Id} \to R$ be the unit of the monad $R$, and let $\rho^{\Sigma(R)} : \Sigma(R) \cdot R \to \Sigma(R)$ be the module substitution of the $R$-module $\Sigma(R)$.*

- *The injection $\mathsf{Id} \to \Sigma(R) + \mathsf{Id}$ together with the natural transformation*

$$(\Sigma(R) + \mathsf{Id}) \cdot (\Sigma(R) + \mathsf{Id}) \simeq \Sigma(R) \cdot (\Sigma(R) + \mathsf{Id}) + \Sigma(R) + \mathsf{Id}$$
$$\downarrow{\scriptstyle \Sigma(R)[r,\eta]+_+_}$$
$$\Sigma(R) \cdot R + \Sigma(R) + \mathsf{Id}$$
$$\downarrow{\scriptstyle [\rho^{\Sigma(R)}, id]+_}$$
$$\Sigma(R) + \mathsf{Id}$$

  *give the endofunctor $\Sigma(R) + \mathsf{Id}$ the structure of a monad.*

- *Moreover, this monad can be given the following $\Sigma$-action:*

$$\Sigma\Big(\Sigma(R) + \mathsf{Id}\Big) \xrightarrow{\Sigma([r,\eta])} \Sigma(R) \cdot R \xrightarrow{\rho^{\Sigma(R)}} \Sigma(R) \longrightarrow \Sigma(R) + \mathsf{Id} \tag{3.2}$$

- *The natural transformation $[r, \eta] : \Sigma(R) + \mathsf{Id} \to R$ is a model morphism, that is, it commutes suitably with the $\Sigma$-actions of Diagram* (3.2) *in the source and $r : \Sigma(R) \longrightarrow R$ in the target.*

**Definition 51.** Given a model $M$ of $\Sigma$, we denote by $M^\sharp$ the $\Sigma$-model constructed in Lemma 50, and by $\epsilon_M : M^\sharp \longrightarrow M$ the morphism of models defined there.

**Lemma 52** (`iso_mod_id_model`)**.** *If $\Sigma$ is effective, then the morphism of $\Sigma$-models*

$$\epsilon_{\hat{\Sigma}} : \hat{\Sigma}^\sharp \longrightarrow \hat{\Sigma}$$

*is an isomorphism.*

We go back to considering the signature $\Sigma := \mathcal{P} \cdot \Theta$. Suppose that $\Sigma$ is effective. From Lemma 52 it follows that $\mathcal{P}\hat{\Sigma}X + X \cong \hat{\Sigma}X$. In particular, we have an injective map from $\mathcal{P}\hat{\Sigma}X$ to $\hat{\Sigma}X$—contradiction.

On the other hand, as a starting point, we can identify the following class of effective parametric modules:

**Theorem 53** (`algebraic_sig_effective`)**.** *Algebraic signatures are effective.*

This result is proved in a previous work [HM07, Theorems 1 and 2]. The construction of the syntax proceeds as follows: an algebraic parametric module induces an endofunctor on the category of endofunctors on Set. Its initial algebra (constructed as the colimit of the initial chain) is given the structure of a monad with an action of the algebraic signature, and then a routine verification shows that it is actually initial in the category of models. The computer-checked proof uses the construction of a monad from an algebraic signature formalized in [AMM18].

In Section 3.6, we show a more general effectiveness result: Theorem 57 states that *presentable* signatures, which form a superclass of algebraic signatures, are effective.

### 3.4.2 Modularity

In this section, we study the problem of how to merge two syntax extensions. Our answer, a “modularity” result (Theorem 54), was stated already in the preliminary version [HM12, Section 6], there without proof.

Suppose that we have a pushout square of effective parametric modules,

$$\begin{array}{ccc} \Sigma_0 & \longrightarrow & \Sigma_1 \\ \downarrow & & \downarrow \\ \Sigma_2 & \longrightarrow & \Sigma \end{array}$$

Intuitively, the signatures $\Sigma_1$ and $\Sigma_2$ specify two extensions of the signature $\Sigma_0$, and $\Sigma$ is the smallest extension containing both these extensions. Modularity means that the corresponding diagram of models,

$$\begin{array}{ccc} \hat{\Sigma}_0 & \longrightarrow & \hat{\Sigma}_1 \\ \downarrow & & \downarrow \\ \hat{\Sigma}_2 & \longrightarrow & \hat{\Sigma} \end{array}$$

is a pushout as well—but we have to take care to state this in the "right" category. The right category for this purpose is the following total category $\int^{\Sigma} \mathsf{Mon}^{\Sigma}$ of models:

- An object of $\int^{\Sigma} \mathsf{Mon}^{\Sigma}$ is a triple $(\Sigma, R, r)$ where $\Sigma$ is a parametric module, $R$ is a monad, and $r$ is an action of $\Sigma$ in $R$.
- A morphism in $\int^{\Sigma} \mathsf{Mon}^{\Sigma}$ from $(\Sigma_1, R_1, r_1)$ to $(\Sigma_2, R_2, r_2)$ consists of a pair $(i, m)$ of a parametric module morphism $i : \Sigma_1 \longrightarrow \Sigma_2$ and a morphism $m$ of $\Sigma_1$-models from $(R_1, r_1)$ to $(R_2, i^*(r_2))$.
- It is easily checked that the obvious composition turns $\int^{\Sigma} \mathsf{Mon}^{\Sigma}$ into a category.

Now for each signature $\Sigma$, we have an obvious inclusion from the fiber $\mathsf{Mon}^{\Sigma}$ into $\int^{\Sigma} \mathsf{Mon}^{\Sigma}$, through which we may see the syntax $\hat{\Sigma}$ of any effective signature as an object in $\int^{\Sigma} \mathsf{Mon}^{\Sigma}$. Furthermore, a morphism $i \colon \Sigma_1 \longrightarrow \Sigma_2$ of effective parametric modules yields a morphism $i_* := \hat{\Sigma}_1 \longrightarrow \hat{\Sigma}_2$ in $\int^{\Sigma} \mathsf{Mon}^{\Sigma}$. Hence our pushout square of effective parametric modules as described above yields a square in $\int^{\Sigma} \mathsf{Mon}^{\Sigma}$.

**Theorem 54** (`pushout_in_big_rep`). *Modularity holds in $\int^{\Sigma} \mathsf{Mon}^{\Sigma}$, in the sense that given a pushout square of effective parametric modules as above, the associated square in $\int^{\Sigma} \mathsf{Mon}^{\Sigma}$ is a pushout again.*

The proof uses, in particular, the following fact:

**Lemma 55** (`rep_cleaving`). *The projection $\pi : \int^{\Sigma} \mathsf{Mon}^{\Sigma} \to \mathsf{PMod}$ is a Grothendieck fibration. In particular, given a morphism $f \colon \Upsilon \longrightarrow \Sigma$ of parametric modules, the pullback map defined in Definition 47 extends to a functor*

$$f^* : \mathsf{Mon}^{\Sigma} \longrightarrow \mathsf{Mon}^{\Upsilon} \quad .$$

Note that Theorem 54 does *not* say that a pushout of effective parametric modules is effective again; it only tells us that if all of the signatures in a pushout square are effective, then the syntax generated by the pushout is the pushout of the syntaxes. In general, we do not know whether a colimit (or even a binary coproduct) of effective parametric modules is effective again.

In Section 3.6 we study *presentable* parametric modules, which we show to be effective and closed under colimits.

## 3.5 Recursion

We now show through examples how certain forms of recursion can be derived from initiality.

### 3.5.1 Example: Translation of intuitionistic logic into linear logic

We start with an elementary example of translation of syntaxes using initiality, namely the translation of second-order intuitionistic logic into second-order linear logic [Gir87, page 6]. The syntax of second-order intuitionistic logic can be defined with one unary operator $\neg$, three binary operators $\vee$, $\wedge$ and $\Rightarrow$, and two binding operators $\forall$ and $\exists$. The associated (algebraic) signature is $\Sigma_{LJ} = \Theta + 3 \times \Theta^2 + 2 \times \Theta'$. As for linear logic, there are four constants $\top, \bot, 0, 1$, two unary operators $!$ and $?$, five binary operators $\&$, $⅋$, $\otimes$, $\oplus$, $\multimap$ and two binding operators $\forall$ and $\exists$. The associated (algebraic) signature is $\Sigma_{LL} = 4 \times * + 2 \times \Theta + 5 \times \Theta^2 + 2 \times \Theta'$.

By universal property of coproduct, a model of $\Sigma_{LJ}$ is given by a monad $R$ with module morphisms:

- $r_\neg : R \longrightarrow R$
- $r_\wedge, r_\vee, r_\Rightarrow : R \times R \longrightarrow R$
- $r_\forall, r_\exists : R' \longrightarrow R$

and similarly, we can decompose an action of $\Sigma_{LL}$ into as many components as there are operators.

The translation will be a morphism of monads between the initial models (i.e. the syntaxes) $o : \hat{\Sigma}_{LJ} \longrightarrow \hat{\Sigma}_{LL}$ coming from the initiality of $\hat{\Sigma}_{LJ}$, satisfying the expected equations. Indeed, equipping $\hat{\Sigma}_{LL}$ with an action $r'_\alpha : \alpha(\hat{\Sigma}_{LL}) \longrightarrow \hat{\Sigma}_{LL}$ for each operator $\alpha$ of intuitionistic logic ($\neg$,$\vee$,$\wedge$,$\Rightarrow$,$\forall$ and $\exists$) yields a morphism of monads $o : \hat{\Sigma}_{LJ} \longrightarrow \hat{\Sigma}_{LL}$ such that $o(r_\alpha(t)) = r'_\alpha(\alpha(o)(t))$ for each $\alpha$.

The definition of $r'_\alpha$ is then straightforward to devise, following the recursive clauses

given on the right:

$$
\begin{aligned}
r'_{\neg} &= r_{\multimap} \circ (r_{!} \times r_{0}) & (\neg A)^o &:= (!A) \multimap 0 \\
r'_{\wedge} &= r_{\&} & (A \wedge B)^o &:= A^o \& B^o \\
r'_{\vee} &= r_{\oplus} \circ (r_{!} \times r_{!}) & (A \vee B)^o &:= !A^o \oplus !B^o \\
r'_{\Rightarrow} &= r_{\multimap} \circ (r_{!} \times id) & (A \Rightarrow B)^o &:= !A^o \multimap B^o \\
r'_{\exists} &= r_{\exists} \circ r_{!} & (\exists x A)^o &:= \exists x !A^o \\
r'_{\forall} &= r_{\forall} & (\forall x A)^o &:= \forall x A^o
\end{aligned}
$$

The induced action of $\Sigma_{LJ}$ in the monad $\hat{\Sigma}_{LL}$ yields the desired translation morphism $o : \hat{\Sigma}_{LJ} \to \hat{\Sigma}_{LL}$. Note that variables are automatically preserved by the translation because $o$ is a monad morphism.

### 3.5.2 Example: Computing the set of free variables

As above, we denote by $\mathcal{P}X$ the powerset of $X$. Union gives a composition operator $\mathcal{P}(\mathcal{P}X) \to \mathcal{P}X$ defined by $u \mapsto \bigcup_{s \in u} s$, which yields a monad structure on $\mathcal{P}$.

We now define an action of the signature of lambda calculus $\Sigma_{\mathsf{LC}}$ in the monad $\mathcal{P}$. We take the binary union operator $\cup \colon \mathcal{P} \times \mathcal{P} \to \mathcal{P}$ as action of the application signature $\Theta \times \Theta$ in $\mathcal{P}$; this is a module morphism since binary union distributes over union of sets. Next, given $S \in \mathcal{P}(X + *)$ we define $\mathsf{Maybe}_X^{-1}(S) = S \cap X$. This defines a morphism of modules $\mathsf{Maybe}^{-1} \colon \mathcal{P}' \to \mathcal{P}$; a small calculation using a distributivity law of binary intersection over union of sets shows that this natural transformation is indeed linear. It can hence be used to model the abstraction signature $\Theta'$ in $\mathcal{P}$.

Associated to this model of $\Sigma_{\mathsf{LC}}$ in $\mathcal{P}$ we have an initial morphism $\mathsf{free} \colon \mathsf{LC} \to \mathcal{P}$. Then, for any $t \in \mathsf{LC}(X)$, the set $\mathsf{free}(t)$ is the set of free variables occurring in $t$.

### 3.5.3 Example: Computing the size of a term

We now consider the problem of computing the "size" of a $\lambda$-term, that is, for any set $X$, a function $s_X \colon \mathsf{LC}(X) \longrightarrow \mathbb{N}$ such that

$$\begin{aligned} s_X(x) &= 0 \qquad (x \in X \text{ variable}) \\ s_X(\mathsf{abs}(t)) &= 1 + s_{X+*}(t) \\ s_X(\mathsf{app}(t,u)) &= 1 + s_X(t) + s_X(u) \end{aligned}$$

To express this map as a morphism of models, we first need to find a suitable monad underlying the target model. The first candidate, the constant functor $X \mapsto \mathbb{N}$, does not admit a monad structure; the problem lies in finding a suitable unit for the monad. (More generally, given a monad $R$ and a set $A$, the functor $X \mapsto R(X) \times A$ does not admit a monad structure whenever $A$ is not a singleton.)

This problem hints at a different approach to the original question: instead of computing the size of a term (which is $0$ for a variable), we compute a generalized size $gs$ which depends on arbitrary (formal) sizes attributed to variables. We have

$$gs : \prod_{X:\mathsf{Set}} \Big( \mathsf{LC}(X) \to (X \to \mathbb{N}) \to \mathbb{N} \Big)$$

Here, unsurprisingly, we recognize the continuation monad (see also [JG07] for the use of continuation for implementing complicated recursion schemes using initiality)

$$\mathsf{Cont}_{\mathbb{N}} := X \mapsto (X \to \mathbb{N}) \to \mathbb{N}$$

with multiplication $\lambda f.\lambda g.f(\lambda h.h(g))$.

Now we can define $gs$ through initiality by endowing the monad $\mathsf{Cont}_{\mathbb{N}}$ with a structure of $\Sigma_{\mathsf{LC}}$-model as follows.

The function $\alpha(m,n) = 1 + m + n$ induces a natural transformation

$$c_{\mathsf{app}} \colon \mathsf{Cont}_{\mathbb{N}} \times \mathsf{Cont}_{\mathbb{N}} \longrightarrow \mathsf{Cont}_{\mathbb{N}}$$

thus an action for the application signature $\Theta \times \Theta$ in the monad $\mathsf{Cont}_{\mathbb{N}}$.

Next, given a set $X$ and $k : X \to \mathbb{N}$, define $\hat{k} : X + \{*\} \to \mathbb{N}$ by $\hat{k}(x) = k(x)$ for all

$x \in X$ and $\hat{k}(*) = 0$. This induces a function

$$\begin{array}{rcl} c_{\mathsf{abs}}(X)\colon \mathsf{Cont}'_{\mathbb{N}}(X) & \longrightarrow & \mathsf{Cont}_{\mathbb{N}}(X) \\ t & \mapsto & (k \mapsto 1 + t(\hat{k})) \end{array}$$

which is the desired action of the abstraction signature $\Theta'$.

Altogether, the transformations $c_{\mathsf{app}}$ and $c_{\mathsf{abs}}$ form the desired action of $\Sigma_{\mathsf{LC}}$ in $\mathsf{Cont}_{\mathbb{N}}$ and thus give an initial morphism, i.e., a natural transformation $\iota\colon \mathsf{LC} \to \mathsf{Cont}_{\mathbb{N}}$ which respects the $\Sigma_{\mathsf{LC}}$-model structure. Now let $0_X$ be the function that is constantly zero on $X$. Then the sought "size" map $s : \prod_{X:\mathsf{Set}} \mathsf{LC}(X) \to \mathbb{N}$ is given by $s_X(t) = \iota_X(t, 0_X)$.

### 3.5.4 Example: Counting the number of redexes

We now consider an example of recursive computation: a function $r$ such that $r(t)$ is the number of redexes of the $\lambda$-term $t$ of $\mathsf{LC}(X)$. Informally, the equations defining $r$ are

$$\begin{aligned} r(x) &= 0, \qquad (x \text{ variable}) \\ r(\mathsf{abs}(t)) &= r(t), \\ r(\mathsf{app}(t,u)) &= r(t) + r(u) + \begin{cases} 1 & \text{if } t \text{ is an abstraction} \\ 0 & \text{otherwise.} \end{cases} \end{aligned}$$

In order to compute recursively the number of $\beta$-redexes in a term, we need to keep track, not only of the number of redexes in subterms, but also whether the head construction of subterms is the abstraction; in the affirmative case we use the value $1$ and $0$ otherwise. Hence, we define a $\Sigma_{\mathsf{LC}}$-action on the monad $W := \mathsf{Cont}_{\mathbb{N}\times\{0,1\}}$. We denote by $\pi_1$, $\pi_2$ the projections that access the two components of the product $\mathbb{N} \times \{0, 1\}$.

For any set $X$ and function $k : X \to \mathbb{N} \times \{0, 1\}$, let us denote by $\hat{k} : X + \{*\} \to \mathbb{N} \times \{0, 1\}$ the function which sends $x \in X$ to $k(x)$ and $*$ to $(0, 0)$. Now, consider the function

$$\begin{array}{rcl} c_{\mathsf{abs}}(X)\colon W'(X) & \longrightarrow & W(X) \\ t & \mapsto & (k \mapsto (\pi_1(t(\hat{k})), 1)). \end{array}$$

Then $c_{\mathsf{abs}}$ is an action of the abstraction signature $\Theta'$ in $W$.

Next, we specify an action $c_{\mathsf{app}} : W \times W \to W$ of the application signature $\Theta \times \Theta$:

Given a set $X$, consider the function

$$\begin{array}{rcl} c_{\mathsf{app}}(X)\colon W(X)\times W(X) & \longrightarrow & W(X) \\ (t,u) & \mapsto & (k\mapsto(\pi_1(t(k))+\pi_1(u(k))+\pi_2(t(k)),0)). \end{array}$$

Then $c_{\mathsf{app}}$ is an action of the abstraction signature $\Theta\times\Theta$ in $W$.

Overall we have a $\Sigma_{\mathsf{LC}}$-action from which we get an initial morphism $\iota\colon \mathsf{LC}\to W$. If $0_X$ is the constant function $X\to\mathbb{N}\times\{0,1\}$ returning the pair $(0,0)$, then $\pi_1(\iota(0_X)) : \mathsf{LC}(X)\to\mathbb{N}$ is the desired function $r$.

## 3.6 Presentations of signatures and syntaxes

In this section, we identify a superclass of algebraic parametric modules that are still effective: we call them *presentable* parametric modules, inducing *presentable* signatures.

**Definition 56.** Given a parametric module $\Sigma$, a *presentation*[5] *of* $\Sigma$ is given by an algebraic parametric module $\Upsilon$ and an epimorphism of parametric modules $p:\Upsilon\longrightarrow\Sigma$. In that case, we say that $\Sigma$ *is presented by* $p:\Upsilon\longrightarrow\Sigma$.

A parametric module for which a presentation exists is called *presentable*. The induced signature is then called *presentable*.

Presentations for a signature are not essentially unique; indeed, signatures can have many different presentations.

*Remark.* By definition, any construction which can be encoded through a presentable signature $\Sigma$ can alternatively be encoded through any algebraic signature "presenting" $\Sigma$. The former encoding is finer than the latter in the sense that terms which are different in the latter encoding can be identified by the former. In other words, a certain amount of semantics is integrated into the syntax.

The main desired property of our presentable signatures is that, thanks to the following theorem, they are effective:

**Theorem 57** (`PresentableisEffective`)**.** *Any presentable signature is effective.*

5. In algebra, a presentation of a group $G$ is an epimorphism $F\to G$ where $F$ is free (together with a generating set of relations among the generators).

The proof is discussed in Section 3.7.

Using the axiom of choice, we can prove a stronger statement:

**Theorem 58** (`is_right_adjoint_functor_of_reps_from_pw_epi_choice`). *We assume the axiom of choice. Let $\Sigma$ be a parametric module, and let $p : \Upsilon \longrightarrow \Sigma$ be a presentation of $\Sigma$. Then the functor $p^* : \mathsf{Mon}^{\Sigma} \longrightarrow \mathsf{Mon}^{\Upsilon}$ has a left adjoint.*

In the proof of Theorem 58, the axiom of choice is used to show that endofunctors on Set preserve epimorphisms.

Theorem 57 follows from Theorem 58 since the left adjoint $p^! : \mathsf{Mon}^{\Upsilon} \longrightarrow \mathsf{Mon}^{\Sigma}$ preserves colimits, in particular, initial objects. However, Theorem 57 is proved in Section 3.7 without appealing to the axiom of choice: there, only some specific endofunctor on Set is considered, for which preservation of epimorphisms can be proved without using the axiom of choice.

**Definition 59.** We call a syntax *presentable* if it is generated by a presentable signature.

Next, we give important examples of presentable signatures:

**Theorem 60.** *The following hold:*

1. *Any algebraic signature is presentable.*

2. *Any colimit of presentable parametric modules is presentable.*

3. *The product of two presentable parametric modules is presentable (in the case when one of them is $\Theta$, see `har_ binprodR_ isPresentable`)*

*Proof.* Items 1–2 are easy to prove. For Item 3, if $\Sigma_1$ and $\Sigma_2$ are presented by $\coprod_i \Upsilon_i$ and $\coprod_j \Phi_j$ respectively, then $\Sigma_1 \times \Sigma_2$ is presented by $\coprod_{i,j} \Upsilon_i \times \Phi_j$. □

**Corollary 61.** *Any colimit of algebraic parametric modules is effective.*

*Proof.* A colimit of algebraic parametric modules is presentable, by Theorem 60, hence effective, by Theorem 57 □

## 3.7 Proof of Theorem 57

In this section, we prove Theorem 57. This proof is mechanically checked in our library; the reader may thus prefer to look at the formalised statements in the library.

Note that the proof of Theorem 57 rests on the more technical Lemma 66 below.

**Proposition 62** (`epiSig_equiv_pwEpi_SET`)**.** *Epimorphisms of parametric modules are exactly pointwise epimorphisms.*

*Proof.* In any category, a morphism $f : a \to b$ is an epimorphism if and only if the following diagram is a pushout diagram ([ML98, Exercise III.4.4]) :

$$\begin{array}{ccc} a & \xrightarrow{f} & b \\ {\scriptstyle f}\big\downarrow & & \big\downarrow{\scriptstyle id} \\ b & \xrightarrow[id]{} & b \end{array}$$

Using this characterization of epimorphisms, the proof follows from the fact that colimits are computed pointwise in the category of parametric modules. □

Another important ingredient will be the following quotient construction for monads. Let $R$ be a monad preserving epimorphisms, and let $\sim$ be a "compatible" family of relations on (the functor underlying) $R$, that is, for any $X : \mathsf{Set}_0$, $\sim_X$ is an equivalence relation on $RX$ such that, for any $f : X \to Y$, the function $R(f)$ maps related elements in $RX$ to related elements in $RY$. Taking the pointwise quotient, we obtain a quotient $\pi : R \to \overline{R}$ in the functor category, satisfying the usual universal property. We want to equip $\overline{R}$ with a monad structure that upgrades $\pi : R \to \overline{R}$ into a quotient in the category of monads. In particular, this means that we need to fill in the square

$$\begin{array}{ccc} R \cdot R & \xrightarrow{\quad\mu\quad} & R \\ {\scriptstyle \pi\cdot\pi}\big\downarrow & & \big\downarrow{\scriptstyle \pi} \\ \overline{R} \cdot \overline{R} & \dashrightarrow_{\overline{\mu}} & \overline{R} \end{array}$$

with a suitable $\overline{\mu} : \overline{R} \cdot \overline{R} \longrightarrow \overline{R}$ satisfying the monad laws. But $\pi$ is epi, and hence so is $\pi{\cdot}\pi = \pi\overline{R} \circ R\pi$ since epis are closed under composition and $R$ preserves epimorphisms. Thus, this is possible when any two elements in $RRX$ that are mapped to the same element by $\pi \cdot \pi$ (the left vertical morphism) are also mapped to the same element by $\pi \circ \mu$ (the top-right composition). It turns out that this is the only extra condition needed for the upgrade. We summarize the construction in the following lemma:

**Lemma 63** (`projR_monad`)**.** *Given a monad $R$ preserving epimorphisms, and a compatible relation $\sim$ on $R$ such that for any set $X$ and $x, y \in RRX$, we have that if $(\pi \cdot \pi)_X(x) \sim (\pi \cdot \pi)_X(y)$ then $\pi(\mu(x)) \sim \pi(\mu(y))$. Then we can construct the quotient $\pi : R \to \overline{R}$ in the category of monads, satisfying the usual universal property.*

Note that the axiom of choice implies that epimorphisms have sections, and thus that any endofunctor on Set preserves epimorphisms.

**Definition 64.** An *epi-parametric module* is a parametric module $\Sigma$ that preserves the epimorphicity in the category of endofunctors on Set: for any monad morphism $f : R \longrightarrow S$, if $U(f)$ is an epi of functors, then so is $U(\Sigma(f))$. Here, we denote by $U$ the forgetful functor from monads resp. modules to endofunctors.

**Example 65** (`BindingSigAreEpiSig`)**.** All the algebraic parametric modules are epi-parametric modules.

We are now in a position to state and prove the main technical lemma:

**Lemma 66** (`push_initiality`)**.** *Let $\Upsilon$ be effective, such that both $\hat{\Upsilon}$ and $\Upsilon(\hat{\Upsilon})$ preserve epimorphisms (as noted above, this condition is automatically fulfilled if one assumes the axiom of choice). Let $F : \Upsilon \to \Sigma$ be a morphism of parametric modules. Suppose that $\Upsilon$ is an epi-parametric module and $F$ is an epimorphism. Then $\Sigma$ is effective.*

*Proof sketch.* As before, we denote by $\hat{\Upsilon}$ the initial $\Upsilon$-model, as well as—by abuse of notation—its underlying monad. For each set $X$, we consider the equivalence relation $\sim_X$ on $\hat{\Upsilon}(X)$ defined as follows: for all $x, y \in \hat{\Upsilon}(X)$ we stipulate that $x \sim_X y$ if and only if $i_X(x) = i_X(y)$ for each (initial) morphism of $\Upsilon$-models $i : \hat{\Upsilon} \to F^*S$ with $S$ a $\Sigma$-model and $F^*S$ the $\Upsilon$-model induced by $F : \Upsilon \to \Sigma$.

By Lemma 63, as $\hat{\Upsilon}$ preserves epimorphisms, we obtain the quotient monad, which we call $\hat{\Upsilon}/F$, and the epimorphic projection $\pi : \hat{\Upsilon} \to \hat{\Upsilon}/F$. We now equip $\hat{\Upsilon}/F$ with a $\Sigma$-action, and show that the induced model is initial, in four steps:

(i) We equip $\hat{\Upsilon}/F$ with a $\Sigma$-action, i.e., with a morphism of $\hat{\Upsilon}/F$-modules $m_{\hat{\Upsilon}/F} : \Sigma(\hat{\Upsilon}/F) \to \hat{\Upsilon}/F$. We define $u : \Upsilon(\hat{\Upsilon}) \to \Sigma(\hat{\Upsilon}/F)$ as $u = F_{\hat{\Upsilon}/F} \circ \Upsilon(\pi)$. Then $u$ is epimorphic, by composition of epimorphisms and by using Proposition 62.

Let $m_{\hat{\Upsilon}} : \Upsilon(\hat{\Upsilon}) \to \hat{\Upsilon}$ be the action of the initial model of $\Upsilon$. We define $m_{\hat{\Upsilon}/F}$ as the unique morphism making the following diagram commute in the category of endofunctors on Set:

$$\begin{array}{ccc} \Upsilon(\hat{\Upsilon}) & \xrightarrow{m_{\hat{\Upsilon}}} & \hat{\Upsilon} \\ {\scriptstyle u}\downarrow & & \downarrow{\scriptstyle \pi} \\ \Sigma(\hat{\Upsilon}/F) & \underset{m_{\hat{\Upsilon}/F}}{\dashrightarrow} & \hat{\Upsilon}/F \end{array}$$

Uniqueness follows from pointwise surjectivity of $u$. Existence follows from the compatibility of $m_{\hat{\Upsilon}}$ with the congruence $\sim_X$. The diagram necessary to turn $m_{\hat{\Upsilon}/F}$ into a module morphism on $\hat{\Upsilon}/F$ is proved by pre-composing it with the epimorphism $(\Sigma(\pi) \circ F_{\hat{\Upsilon}}) \cdot \pi : \Upsilon(\hat{\Upsilon}) \cdot \hat{\Upsilon} \to \Sigma(\hat{\Upsilon}/F) \cdot \hat{\Upsilon}/F$ (this is where the preservation of epimorphims by $\Upsilon(\hat{\Upsilon})$ is required) and unfolding the definitions.

(ii) Now, $\pi$ can be seen as a morphism of $\Upsilon$-models between $\hat{\Upsilon}$ and $F^*\hat{\Upsilon}/F$, by naturality of $F$ and using the previous diagram.

It remains to show that $(\hat{\Upsilon}/F, m_{\hat{\Upsilon}/F})$ is initial in the category of $\Sigma$-models.

(iii) Given a $\Sigma$-model $(S, m_s)$, the initial morphism of $\Upsilon$-models $i_S : \hat{\Upsilon} \to F^*S$ induces a monad morphism $\iota_S : \hat{\Upsilon}/F \to S$. We need to show that the morphism $\iota$ is a morphism of $\Sigma$-models. Pre-composing the involved diagram by the epimorphism $\Sigma(\pi) \circ F_{\hat{\Upsilon}} : \Upsilon(\hat{\Upsilon}) \to \Sigma(\hat{\Upsilon}/F)$ and unfolding the definitions shows that $\iota_S : \hat{\Upsilon}/F \to S$ is a morphism of $\Sigma$-models.

(iv) We show that $\iota_S$ is the only morphism $\hat{\Upsilon}/F \to S$. Let $g$ be such a morphism. Then $g \circ \pi : \hat{\Upsilon} \to S$ defines a morphism in the category of $\Upsilon$-models. Uniqueness of $i_S$ yields $g \circ \pi = i_S$, and by uniqueness of the diagram defining $\iota_S$ it follows that $g = i'_S$. □

**Lemma 67** (`algebraic_model_Epi` and `BindingSig_on_model_isEpi`). *Let $\Sigma$ be an algebraic parametric module. Then $\hat{\Sigma}$ and $\Sigma(\hat{\Sigma})$ preserve epimorphisms.*

*Proof.* The initial model of an algebraic parametric module $\Sigma$ is obtained as the initial chain of the endofunctor $R \mapsto \mathsf{Id} + \Sigma(R)$, where $\Sigma$ denotes (by abuse of notation) the endofunctor on endofunctors on Set corresponding to the parametric module $\Sigma$.

Then the proof follows from the fact that this endofunctor preserves preservation of epimorphisms. □

*Proof of Theorem 57.* Let $p : \Upsilon \to \Sigma$ be a presentation of $\Sigma$. We need to construct an initial model for $\Sigma$.

As the parametric module $\Upsilon$ is algebraic, it is effective (by Theorem 53) and is an epi-parametric module (by Example 65). We can thus instantiate Lemma 66 to see that $\Sigma$ is effective, thanks to Lemma 67. □

## 3.8 Constructions of presentable signatures

Complex signatures are naturally built as the sum of basic components, generally referred as "arities" (which in our settings can be captured with a single parametric module, see remark after Example 38). Thanks to Theorem 60, Item 2, direct sums (or, indeed, any colimit) of presentable parametric modules are presentable, hence effective by Theorem 57.

In this section, we show that, besides algebraic signatures, there are other interesting examples of signatures which are presentable, and which hence can be *safely* added to any presentable signature. *Safely* here means that the resulting signature is still presentable.

### 3.8.1 Post-composition with a presentable functor

A functor $F : \mathsf{Set} \to \mathsf{Set}$ is *polynomial* if it is of the form $FX = \coprod_{n\in\mathbb{N}} a_n \times X^n$ for some sequence $(a_n)_{n\in\mathbb{N}}$ of sets. Note that if $F$ is polynomial, then the signature $F \cdot \Theta$ is algebraic.

**Definition 68.** Let $G : \mathsf{Set} \to \mathsf{Set}$ be a functor. A *presentation of* $G$ is a pair consisting of a polynomial functor $F : \mathsf{Set} \to \mathsf{Set}$ and an epimorphism $p : F \to G$. The functor $G$ is called *presentable* if there is a presentation of $G$.

**Proposition 69.** *Given a presentable functor $G$, the signature $G \cdot \Theta$ is presentable.*

*Proof.* Let $p : F \to G$ be a presentation of $G$; then a presentation of $G \cdot \Theta$ is given by the induced epimorphism $F \cdot \Theta \to G \cdot \Theta$. □

**Proposition 70.** *Here we assume the axiom of excluded middle. An endofunctor on* Set *is presentable if and only if it is finitary (i.e., it preserves filtered colimits).*

*Proof.* This is a corollary of Proposition 5.2 of [AP04], since $\omega$-accessible functors are exactly the finitary ones. □

We now give several examples of presentable signatures obtained from presentable functors.

### Example: Adding a syntactic commutative binary operator, e.g., parallel-or

Consider the functor $\mathsf{square} : \mathsf{Set} \to \mathsf{Set}$ mapping a set $X$ to $X \times X$; it is polynomial. The associated signature $\mathsf{square} \cdot \Theta$ encodes a binary operator, such as the application of the lambda calculus.

Sometimes such binary operators are asked to be *commutative*; a simple example of such a commutative binary operator is standard integer addition.

Another example, more specific to formal computer languages, is a "concurrency" operator $P \mid Q$ of a process calculus, such as the $\pi$-calculus, for which it is natural to require commutativity as a structural congruence relation: $P \mid Q \equiv Q \mid P$.

Such a commutative binary operator can be specified via the following presentable signature: we denote by $\mathcal{S}_2 : \mathsf{Set} \to \mathsf{Set}$ the endofunctor that assigns, to each set $X$, the set $(X \times X)/(x, y) \sim (y, x)$ of unordered pairs of elements of $X$. This functor is presented by the obvious projection $\mathsf{square} \to \mathcal{S}_2$. By Proposition 69, the signature $\mathcal{S}_2 \cdot \Theta$ is presentable; it encodes a commutative binary operator.

### Example: Adding a maximum operator

Let $\mathsf{list} : \mathsf{Set} \to \mathsf{Set}$ be the functor associating, to any set $X$, the set $\mathsf{list}(X)$ of (finite) lists with entries in $X$; specifically, it is given on objects as $X \mapsto \coprod_{n \in \mathbb{N}} X^n$.

We now consider the syntax of a "maximum" operator, acting, e.g., on a list of natural numbers:

$$\mathsf{max} : \mathsf{list}(\mathbb{N}) \to \mathbb{N}$$

It can be specified via the algebraic signature $\mathsf{list} \cdot \Theta$.

However, this signature is "rough" in the sense that it does not take into account some semantic aspects of a maximum operator, such as invariance under repetition or permutation of elements in a list.

For a finer encoding, consider the functor $\mathcal{P}_{\text{fin}} : \mathsf{Set} \to \mathsf{Set}$ associating, to a set $X$, the set $\mathcal{P}_{\text{fin}}(X)$ of its finite subsets. This functor is presented by the epimorphism $\mathsf{list} \to \mathcal{P}_{\text{fin}}$.

By Proposition 69, the signature $\mathcal{P}_{\text{fin}} \cdot \Theta$ is presentable; it encodes the syntax of a "maximum" operator accounting for invariance under repetition or permutation of elements in a list.

#### Example: Adding an application à la Differential LC

Let $R$ be a commutative (semi)ring. To any set $S$, we can associate the *free $R$-module* $R\langle S\rangle$; its elements are formal linear combinations $\sum_{s\in S} a_s s$ of elements of $S$ with coefficients $a_s$ from $R$; with $a_s = 0$ almost everywhere. Ignoring the $R$-module structure on $R\langle S\rangle$, this assignment induces a functor $R\langle_\rangle : \mathsf{Set} \to \mathsf{Set}$ with the obvious action on morphisms. For simplicity, we restrict our attention to the semiring $(\mathbb{N}, +, \times)$.

This functor is presentable: a presentation is given by the polynomial functor $\mathsf{list} : \mathsf{Set} \to \mathsf{Set}$, and the epimorphism

$$p : \mathsf{list} \longrightarrow \mathbb{N}\langle_\rangle$$
$$p_X\left([x_1, \ldots, x_n]\right) := x_1 + \ldots + x_n \ .$$

By Proposition 69, this yields a presentable signature, which we call $\mathbb{N}\langle\Theta\rangle$.

The Differential Lambda Calculus (DLC) [ER03b] of Ehrhard and Regnier is a lambda calculus with operations suitable to express differential constructions. The calculus is parametrized by a semiring $R$; again we restrict to $R = (\mathbb{N}, +, \times)$.

DLC has a binary "application" operator, written $(s)t$, where $s \in T$ is an element of the inductively defined set $T$ of terms and $t \in \mathbb{N}\langle T\rangle$ is an element of the free $(\mathbb{N}, +, \times)$-module. This operator is thus specified by the presentable signature $\Theta \times \mathbb{N}\langle\Theta\rangle$.

### 3.8.2 Example: Adding a syntactic closure operator

Given a quantification construction (e.g., abstraction, universal or existential quantification), it is often useful to take the associated closure operation. One well-known example is the universal closure of a logic formula. Such a closure is invariant under permutation of the fresh variables. A closure can be syntactically encoded in a rough way by iterating the closure with respect to one variable at a time. Here our framework allows a refined syntactic encoding which we explain below.

Let us start with binding a fixed number $k$ of fresh variables. The elementary signature $\Theta^{(k)}$ already specifies an operation that binds $k$ variables. However, this encoding does not reflect invariance under variable permutation. To enforce this invariance, it suffices to quotient the signature $\Theta^{(k)}$ with respect to the action of the group $\mathfrak{S}_k$ of permutations of the set $k$, that is, to consider the colimit of the following one-object diagram:

$$\Theta^{(k)} \circlearrowleft \Theta^{(\sigma)}$$

where $\sigma$ ranges over the elements of $\mathfrak{S}_k$. We denote by $\mathcal{S}^{(k)}\Theta$ the resulting signature presented by the projection $\Theta^{(k)} \to \mathcal{S}^{(k)}\Theta$. By universal property of the quotient, a model of it consists of a monad $R$ with an action $m : R^{(k)} \to R$ that satisfies the required invariance.

Now, we want to specify an operation which binds an arbitrary number of fresh variables, as expected from a closure operator. One rough solution is to consider the coproduct $\coprod_k \mathcal{S}^{(k)}\Theta$. However, we encounter a similar inconvenience as for $\Theta^{(k)}$. Indeed, for each $k' > k$, each term already encoded by the signature $\mathcal{S}^{(k)}\Theta$ may be considered again, encoded (differently) through $\mathcal{S}^{(k')}\Theta$.

Fortunately, a finer encoding is provided by the following simple colimit of presentable parametric modules. The crucial point here is that, for each $k$, all natural injections from $\Theta^{(k)}$ to $\Theta^{(k+1)}$ induce the same canonical injection from $\mathcal{S}^{(k)}\Theta$ to $\mathcal{S}^{(k+1)}\Theta$. We thus have a natural colimit for the sequence $k \mapsto \mathcal{S}^{(k)}\Theta$ and thus a signature $\operatorname{colim}_k \mathcal{S}^{(k)}\Theta$ which, as a colimit of presentable parametric modules, is presentable (Theorem 60, Item 2).

Accordingly, we define a total closure on a monad $R$ to be an action of the signature $\operatorname{colim}_k \mathcal{S}^{(k)}\Theta$ in $R$. It can easily be checked that a model of this signature is a monad $R$ together with a family of module morphisms $(e_k : R^{(k)} \to R)_{k \in \mathbb{N}}$ compatible in the sense that for each injection $i : k \to k'$ the following diagram commutes:

$$\begin{array}{ccc} R^{(k)} & \xrightarrow{R^{(i)}} & R^{(k')} \\ & \searrow^{e_k} & \downarrow e_{k'} \\ & & R \end{array}$$

### 3.8.3 Example: Adding an explicit substitution

*Explicit substitution* was introduced by Abadi et al. [Aba+90] as a theoretical device to study the theory of substitution and to describe concrete implementations of substitution algorithms. In this section, we explain how we can extend any presentable signature with an explicit substitution construction, and we offer some refinements from a purely syntactic point of view. In fact, we will show three solutions, differing in the amount of "coherence" which is handled at the syntactic level (e.g., invariance under permutation and weakening). We follow the approach initiated by Ghani, Uustalu, and Hamana in [GUH06].

Let $R$ be a monad. We have already considered (see Lemma 33) the (unary) substitution $\sigma_R : R' \times R \to R$. More generally, we have the sequence of substitution operations

$$\mathsf{subst}_p : R^{(p)} \times R^p \longrightarrow R. \tag{3.3}$$

We say that $\mathsf{subst}_p$ is the $p$-substitution in $R$; it simultaneously replaces the $p$ extra variables in its first argument with the $p$ other arguments, respectively. (Note that $\mathsf{subst}_1$ is the original $\sigma_R$.)

We observe that, for fixed $p$, the group $\mathfrak{S}_p$ of permutations on $p$ elements has a natural action on $R^{(p)} \times R^p$, and that $\mathsf{subst}_p$ is invariant under this action.

Thus, if we fix an integer $p$, there are two ways to internalise $\mathsf{subst}_p$ in the syntax: we can choose the elementary signature $\Theta^{(p)} \times \Theta^p$, which is rough in the sense that the above invariance is not reflected; and, alternatively, if we want to reflect the permutation invariance syntactically, we can choose the quotient $Q_p$ of the above parametric module by the action of $\mathfrak{S}_p$.

By universal property of the quotient, a model of our quotient $Q_p$ is given by a monad $R$ with an action $m : R^{(p)} \times R^p \to R$ satisfying the desired invariance.

Before turning to the encoding of the entire series $(\mathsf{subst}_p)_{p \in \mathbb{N}}$, we recall how, as noticed already in [GUH06], this series enjoys further coherence. In order to explain this coherence, we start with two natural numbers $p$ and $q$ and the module $R^{(p)} \times R^q$. Pairs in this module are almost ready for substitution: what is missing is a map $u : I_p \longrightarrow I_q$, where $I_n$ denotes the set $\{1, \ldots, p\}$. But such a map can be used in two ways: letting $u$ act covariantly on the first factor leads us into $R^{(q)} \times R^q$ where we can apply $\mathsf{subst}_q$; while letting $u$ act contravariantly on the second factor leads us into $R^{(p)} \times R^p$ where we can apply $\mathsf{subst}_p$. The good news is that we obtain the same result. More precisely, the following diagram is commutative:

$$\begin{array}{ccc} R^{(p)} \times R^{q} & \xrightarrow{R^{(p)} \times R^{u}} & R^{(p)} \times R^{p} \\ {\scriptstyle R^{(u)} \times R^{q}} \downarrow & & \downarrow {\scriptstyle \mathsf{subst}_p} \\ R^{(q)} \times R^{q} & \xrightarrow[\mathsf{subst}_q]{} & R \end{array} \tag{3.4}$$

Note that in the case where $p$ equals $q$ and $u$ is a permutation, we recover exactly the invariance by permutation considered earlier.

Abstracting over the numbers $p, q$ and the map $u$, this exactly means that our series factors through the coend $\int^{p:\mathbb{F}} R^{(\underline{p})} \times R^{\overline{p}}$, where covariant (resp. contravariant) occurrences of the bifunctor have been underlined (resp. overlined), and the category $\mathbb{F}$ is the full subcategory of Set whose objects are natural numbers. Thus we have a canonical morphism

$$\mathsf{isubst}_R : \int^{p:\mathbb{F}} R^{(\underline{p})} \times R^{\overline{p}} \longrightarrow R.$$

Abstracting over $R$, we obtain the following:

**Definition 71.** *Integrated substitution*

$$\mathsf{isubst} : \int^{p:\mathbb{F}} \Theta^{(\underline{p})} \times \Theta^{\overline{p}} \longrightarrow \Theta$$

is the parametric module morphism obtained by abstracting over $R$ the linear morphisms $\mathsf{isubst}_R$.

Thus, if we want to internalise the whole sequence $(\mathsf{subst}_p)_{p:\mathbb{N}}$ in the syntax, we have at least three solutions: we can choose the algebraic signature

$$\coprod_{p:\mathbb{N}} \Theta^{(p)} \times \Theta^{p}$$

which is rough in the sense that the above invariance and coherence is not reflected; we can choose the presentable signature

$$\coprod_{p:\mathbb{N}} Q_p,$$

which reflects the invariance by permutation, but not more; and finally, if we want to reflect the whole coherence syntactically, we can choose the presentable signature

$$\int^{p:\mathbb{F}} \Theta^{(\underline{p})} \times \Theta^{\overline{p}}.$$

Thus, whenever we have a presentable signature, we can safely extend it by adding one or the other of the three above signatures, for a (more or less coherent) explicit substitution.

Ghani, Uustalu, and Hamana already studied this problem in [GUH06]. Our solution proposed here does not require the consideration of a *strength*.

### 3.8.4 Example: Adding a coherent fixed-point operator

In the same spirit as in the previous section, we define, in this section,

- for each $n \in \mathbb{N}$, a notion of *$n$-ary fixed-point operator* in a monad;
- a notion of *coherent fixed-point operator* in a monad, which assigns, in a "coherent" way, to each $n \in \mathbb{N}$, an $n$-ary fixed-point operator.

We furthermore explain how to safely extend any presentable syntax with a syntactic coherent fixed-point operator.

There is one fundamental difference between the integrated substitution of the previous section and our coherent fixed points: while every monad has a canonical integrated substitution, this is not the case for coherent fixed-point operators.

Let us start with the unary case.

**Definition 72.** A *unary fixed-point operator for a monad $R$* is a module morphism $f$ from $R'$ to $R$ that makes the following diagram commute,

$$\begin{array}{ccc} R' & \xrightarrow{(id_{R'},f)} & R' \times R \\ & {}_{f}\searrow \qquad \swarrow_{\sigma} & \\ & R & \end{array}$$

where $\sigma$ is the substitution morphism defined in Lemma 33.

Accordingly, the signature for a syntactic unary fixpoint operator is $\Theta'$, ignoring the commutation requirement (which we address later in Section 4.4.4, after extending signatures induced by parametric modules with equations).

Let us digress here and examine what the unary fixpoint operators are for the lambda calculus, more precisely, for the monad $\mathsf{LC}_{\beta\eta}$ of the lambda-calculus modulo $\beta$- and $\eta$-equivalence. How can we relate the above notion to the classical notion of fixed-point combinator? Terms are built out of two constructions, $\mathsf{app} : \mathsf{LC}_{\beta\eta} \times \mathsf{LC}_{\beta\eta} \to \mathsf{LC}_{\beta\eta}$ and $\mathsf{abs} : \mathsf{LC}'_{\beta\eta} \to \mathsf{LC}_{\beta\eta}$. A fixed-point combinator is a term $Y$ satisfying, for any (possibly open) term $t$, the equation

$$\mathsf{app}(t, \mathsf{app}(Y, t)) = \mathsf{app}(Y, t).$$

Given such a combinator $Y$, we define a module morphism $\hat{Y} : \mathsf{LC}'_{\beta\eta} \to \mathsf{LC}_{\beta\eta}$. It associates, to any term $t$ depending on an additional variable $*$, the term $\hat{Y}(t) := \mathsf{app}(Y, \mathsf{abs}\ t)$. This term satisfies $t\{* := \hat{Y}(t)\} = \hat{Y}(t)$, which is precisely the diagram of Definition 72 for a unary fixed-point operator. Thus, $\hat{Y}$ is a unary fixed-point operator for the monad $\mathsf{LC}_{\beta\eta}$. Conversely, we have:

**Proposition 73.** *Any fixed-point combinator in* $\mathsf{LC}_{\beta\eta}$ *comes from a unique fixed-point operator.*

*Proof.* We construct a bijection between the subset of $\mathsf{LC}_{\beta\eta}\emptyset$ consisting of (closed) fixed-point combinator on the one hand and the set of module morphisms from $\mathsf{LC}'_{\beta\eta}$ to $\mathsf{LC}_{\beta\eta}$ satisfying the fixed-point property on the other hand.

A closed lambda term $t$ is mapped to the morphism $u \mapsto \hat{t}\, u := \mathsf{app}(t, \mathsf{abs}\ u)$. We have already seen that if $t$ is a fixed-point combinator, then $\hat{t}$ is a fixed-point operator.

For the inverse function, note that a module morphism $f$ from $\mathsf{LC}'_{\beta\eta}$ to $\mathsf{LC}_{\beta\eta}$ induces a closed term $Y_f := \mathsf{abs}(f_1(\mathsf{app}(*, **)))$ where $f_1 : \mathsf{LC}_{\beta\eta}(\{*, **\}) \to \mathsf{LC}_{\beta\eta}(\{*\})$.

A small calculation shows that $Y \mapsto \hat{Y}$ and $f \mapsto Y_f$ are inverse to each other.

It remains to be proved that if $f$ is a fixed-point operator, then $Y_f$ satisfies the fixed-point combinator equation. Let $t \in \mathsf{LC}_{\beta\eta}X$, then we have

$$\begin{aligned} \mathsf{app}(Y_f, t) &= \mathsf{app}(\mathsf{abs}\ f_X(\mathsf{app}(*, **)), t) && (3.5)\\ &= f_X(\mathsf{app}(t, **)) && (3.6)\\ &= \mathsf{app}(t, f_X(\mathsf{app}(t, **))) && (3.7)\\ &= \mathsf{app}(t, \mathsf{app}(Y_f, t)) && (3.8) \end{aligned}$$

where (3.5) comes from the definition of $Y_f$ (and naturality of $f$). Equality (3.6) follows from $\beta$-reduction, Equality 3.7 from the definition of a fixed-point operator. Finally,

Equality 3.8 comes from the equality $\mathsf{app}(Y_f, t) = f_X(\mathsf{app}(t, **))$, which is obtained by chaining the equalities from (3.5) to (3.6). This concludes the construction of the bijection. $\square$

After this digression, we now turn to the $n$-ary case.

**Definition 74.**

- A *rough $n$-ary fixed-point operator* for a monad $R$ is a module morphism $f : (R^{(n)})^n \to R^n$ making the following diagram commute:

$$\begin{array}{ccc} (R^{(n)})^n & \xrightarrow{id_{(R^{(n)})^n}, f, .., f} & (R^{(n)})^n \times (R^n)^n \\ {\scriptstyle f}\downarrow & & \wr\parallel \\ R^n & \xleftarrow[(\mathsf{subst}_n)^n]{} & (R^{(n)} \times R^n)^n \end{array}$$

  where $\mathsf{subst}_n$ is the $n$-substitution as in Section 3.8.3.

- An *$n$-ary fixed-point operator* is just a rough $n$-ary fixed-point operator which is furthermore invariant under the natural action of the permutation group $\mathfrak{S}_n$.

The type of $f$ above is canonically isomorphic to

$$(R^{(n)})^n + (R^{(n)})^n + \ldots + (R^{(n)})^n \to R,$$

which we abbreviate to[6] $n \times (R^{(n)})^n \to R$.

Accordingly, a natural signature for encoding a *syntactic*[7] rough $n$-ary fixpoint operator is $n \times (\Theta^{(n)})^n$.

Similarly, a natural signature for encoding a syntactic $n$-ary fixpoint operator is $(n \times (\Theta^{(n)})^n)/\mathfrak{S}_n$ obtained by quotienting the previous parametric module by the action of $\mathfrak{S}_n$.

Now we let $n$ vary and say that a *total fixed-point operator* on a given monad $R$ assigns to each $n \in \mathbb{N}$ an $n$-ary fixpoint operator on $R$. Obviously, the natural signature for the encoding of a syntactic total fixed-point operator is $\coprod_n (\Theta^{(n)})^n/\mathfrak{S}_n$. Alternatively, we may wish to discard those total fixed-point operators that do not satisfy some coherence conditions analogous to what we encountered in Section 3.8.3, which we now introduce.

6. In the following, we similarly write $n$ instead of $I_n$ in order to make equations more readable.
7. The adjective *syntactic* means here that we do not deal with the equation.

Let $R$ be a monad with a sequence of module morphisms $\mathsf{fix}_n : n \times (R^{(n)})^n \to R$. We call this family *coherent* if, for any $p, q \in \mathbb{N}$ and $u : p \to q$, the following diagram commutes:

$$
\begin{array}{ccc}
p \times (R^{(p)})^q & \xrightarrow{\;p \times (R^{(p)})^u\;} & p \times (R^{(p)})^p \\
{\scriptstyle u \times (R^{(u)})^q}\big\downarrow & & \big\downarrow{\scriptstyle \mathsf{fix}_p} \\
q \times (R^{(q)})^q & \xrightarrow[\;\mathsf{fix}_q\;]{} & R
\end{array}
\tag{3.9}
$$

These conditions have an interpretation in terms of a coend, just as we already encountered in Section 3.8.3. This leads us to the following

**Definition 75.** Given a monad $R$, we define a *coherent fixed-point operator on $R$* to be a module morphism from $\int^{n:\mathbb{F}} \underline{n} \times (R^{(\underline{n})})^{\overline{n}}$ to $R$ where, for every $n \in \mathbb{N}$, the $n$-th component is a (rough)[8] $n$-ary fixpoint operator.

Now, the natural signature for a syntactic coherent fixed-point operator is $\int^{n:\mathbb{F}} \underline{n} \times (\Theta^{(\underline{n})})^{\overline{n}}$. Thus, given a presentable signature $\Sigma$, we can safely extend it with a syntactic coherent fixed-point operator by adding the presentable signature

$$\int^{n:\mathbb{F}} \underline{n} \times (\Theta^{(\underline{n})})^{\overline{n}}$$

to $\Sigma$.

8. As in Section 3.8.3, invariance follows from coherence.

CHAPTER 4

# ALGEBRAIC 2-SIGNATURES FOR MONADS

This chapter is adapted from [Ahr+19b].

The presentable signatures of Chapter 3 allow to specify syntaxes satisfying some equations by considering colimits of algebraic parametric modules. However, it seems quite limited: for example, we don't know how to specify an associative operation by a presentable signature. This motivates the work of the present chapter: we identify the class of algebraic 2-signatures which are effective: they are particular signatures over the category of monads consists of extensions of signatures induced by parametric modules with a family of equational arities.

It is not clear if any syntax generated by a presentable signature can also be generated by an algebraic 2-signature, although we do not know of any counter-example. Conversely, algebraic 2-signatures take into account operations that we do not know how to specify using a presentable signature, such as an associative operation. Signatures induced by parametric modules and models in the sense of Chapter 3 are referred to as 1-signatures and 1-models in this chapter.

## 4.1 Introduction

There is a well-established theory of presentations of monads through generating (first-order) operations equipped with relations among the corresponding derived operations. Algebraic 1-signatures can be considered as generating monads by *binding operations*. Various algebraic structures generated by binding operations have been considered by many, going back at least to Fiore, Plotkin, and Turi [FPT99], Gabbay and Pitts [GP99], and Hofmann [Hof99].

If $p : \hat{\Sigma} \to R$ is a monad epimorphism, we understand that $R$ is generated by a family of operations whose binding arities are given by $\Sigma$, subject to suitable identifications. In particular, for $\Sigma := \Theta \times \Theta + \Theta'$, $\hat{\Sigma}$ may be understood as the monad $\mathsf{LC}$ of syntactic terms of the lambda calculus (see Section 3.2), and we have an obvious epimorphism $p : \hat{\Sigma} \to \mathsf{LC}_{\beta\eta}$, where $\mathsf{LC}_{\beta\eta}$ is the monad of lambda-terms modulo $\beta$ and $\eta$. In order to cover such equations, the approach in the first-order case suggests to identify $p$ as the coequalizer of a pair of parallel arrows from $T$ to $\hat{\Sigma}$ where $T$ is again a "free" monad. Let us see what comes out when we attempt to find such an encoding for the $\beta$-equation of the monad $\mathsf{LC}_{\beta\eta}$. It should say that for each set $X$, the following two maps from $\hat{\Sigma}(X + \{*\}) \times \hat{\Sigma}(X)$ to $\hat{\Sigma}(X)$,

- $(t, u) \mapsto \mathsf{app}(\mathsf{abs}(t), u)$
- $(t, u) \mapsto t\{* \mapsto u\}$

are equal. Here a problem occurs, namely that the above collections of maps, which can be understood as mere natural transformations, cannot be understood as morphisms of monads. Notably, they do not send variables to variables.

On the other hand, we observe that the members of our equations, which are not morphisms of monads, commute with substitution, and hence are more than natural transformations: indeed they are morphisms of *modules over* $\hat{\Sigma}$. Accordingly, a (second-order) presentation for a monad $R$ could be a diagram

$$T \overset{f}{\rightrightarrows} \hat{\Sigma} \xrightarrow{p} R \tag{4.1}$$

where $\Sigma$ is an algebraic signature, $\hat{\Sigma}$ is the associated free monad, $T$ is a module over $\hat{\Sigma}$, $f$ is a pair of morphisms of modules over $\hat{\Sigma}$, and $p$ is a monad epimorphism. And now we are faced with the task of finding a condition meaning something like "$p$ is the coequalizer of $f$"[1]. To this end, recall that we introduced the category $\mathsf{Mon}^{\Sigma}$ "of models of $\Sigma$", whose objects are monads "equipped with an action of $\Sigma$" (Definition 45). Of course $\hat{\Sigma}$ is equipped with such an action which turns it into the initial object. Now, we define the full subcategory of models satisfying the equation $f$, and require $R$ to be the initial object therein. Our definition is suited to the case where the equation $f$ is parametric in the model: this means that now $T$ and $f$ are functions of the model $S$, and $f(S) = (u(S), v(S))$ is a pair of $S$-module morphisms from $T(S)$ to $S$. We say that

1. This cannot be the case stricto sensu since $f$ is a pair of module morphisms while $p$ is a monad morphism.

$S$ satisfies the equation $f$ if $u(S) = v(S)$. Generalizing the case of one equation to the case of a family of equations yields our notion of 2-signature, which is similar to that introduced by Ahrens [Ahr16] in a slightly different context and is a particular case of signature over the category of monads.

Now we are ready to formulate our main problem: given a 2-signature $(\Sigma, E)$, where $E$ is a family of parametric equations as above, does the subcategory of models of $\Sigma$ satisfying the family of equations $E$ admit an initial object?

We answer positively for a large subclass of 2-signatures which we call *algebraic* 2-signatures (see Theorem 107).

This provides a construction of a monad from an algebraic 2-signature, and we prove furthermore (see Theorem 102) that this construction is *modular*, in the sense that merging two extensions of 2-signatures corresponds to building an amalgamated sum of initial models. This is analogous to Theorem 54 for 1-signatures.

As expected, our initiality property generates a recursion principle which is a recipe allowing us to specify a morphism from the presented monad to any given other monad.

We give various examples of monads arising "in nature" that can be specified via an algebraic 2-signature (see Section 4.4), and we also show through a simple example how our recursion principle applies (see Section 4.5).

**Computer-checked formalization** A summary of our formalization regarding 2-signatures is available at `https://initialsemantics.github.io/doc/50fd617/Modules.SoftEquations.Summary.html`.

## 4.2 2-Signatures and their models

In this section we study *2-signatures* and *models of 2-signatures*. A 2-signature is a signature over the category of monads consisting of a 1-signature $\Sigma$ and a family of particular equational $\Sigma$-arities. We first explain which equational $\Sigma$-arities are involved.

### 4.2.1 Equations

Our equations are analogous to those considered by Ahrens in [Ahr16]: they are parallel module morphisms parametrized by the models of the underlying 1-signature. They are particular equational arities in the sense of Chapter 2. The underlying notion of 1-model is essentially the same as in [Ahr16], even if, there, such equations are interpreted instead as *inequalities*.

Throughout this subsection, we fix an arbitray signature $\Sigma$ over the category of monads, that we instantiate in the examples with particular 1-signatures.

We first give a direct definition as an equational arity (Definition 78), and then rephrase it to give a more intuitive one as a parallel pair of morphisms (Remark 82).

**Definition 76.** We define a **$\Sigma$-module** to be a functor $T$ from the category of models of $\Sigma$ to the category $\int \mathsf{Mod}$ commuting with the forgetful functors to the category $\mathsf{Mon}$ of monads,

$$\begin{array}{ccc} \mathsf{Mon}^{\Sigma} & \xrightarrow{T} & \int \mathsf{Mod} \\ & \searrow \quad \swarrow & \\ & \mathsf{Mon} & \end{array}$$

**Example 77.** To each parametric module $\Psi$ is associated, by precomposition with the projection from $\mathsf{Mon}^{\Sigma}$ to $\mathsf{Mon}$, a $\Sigma$-module still denoted $\Psi$. All the $\Sigma$-modules occurring in this work arise in this way from 1-signatures; in other words, they do not depend on the action of the 1-model. In particular, we have the **tautological $\Sigma$-module** $\Theta$, and, more generally, for any natural number $n \in \mathbb{N}$, a $\Sigma$-module $\Theta^{(n)}$. Also we have another fundamental $\Sigma$-module (arising in this way from) $\Sigma$ itself, if $\Sigma$ is induced by a paremtric module.

**Definition 78.** A **$\Sigma$-equation** is a $\Sigma$-arity $(\mathsf{D}, a, u, v)$ in the sense of Chapter 2, where

- $\mathsf{D}$ is the category determined by two $\Sigma$-modules $\Psi$ and $\Phi$ as follows:
  - objects are models $R$ of $\Sigma$ equipped with a $R$-module morphism $h : \Psi(R) \to \Phi(R)$,
  - morphisms between $(R, h)$ and $(S, i)$ are model morphisms $f : R \to S$ such that the following diagram commutes in the category of functors

$$\begin{array}{ccc} \Psi(R) & \xrightarrow{h} & \Phi(R) \\ {\scriptstyle \Psi(f)}\downarrow & & \downarrow{\scriptstyle \Phi(f)} \\ \Psi(S) & \xrightarrow[i]{} & \Phi(S) \end{array} \quad ;$$

- $a : \mathsf{D} \to \mathsf{Mon}^{\Sigma}$ is the canonical forgetful functor;

Note that such an arity is always equational (Definition 7).

In the following, we provide an alternative description of a $\Sigma$-equation as a parallel pair of $\Sigma$-module morphisms (Remark 82).

**Definition 79.** Let $S$ and $T$ be $\Sigma$-modules. We define a **morphism of $\Sigma$-modules** from $S$ to $T$ to be a natural transformation from $S$ to $T$ which becomes the identity when postcomposed with the forgetful functor $\int \mathsf{Mod} \to \mathsf{Mon}$.

**Example 80.** Each parametric module morphism $\Psi \to \Phi$ upgrades into a morphism of $\Sigma$-modules. Further in that vein, if $\Sigma$ is a parametric module, there is a morphism of $\Sigma$-modules $\tau^{\Sigma} : \Sigma \to \Theta$. It is given, on a model $(R, m)$ of $\Sigma$, by $m : \Sigma(R) \to R$. (Note that it does not arise from a morphism of parametric modules.) When the context is clear, we write simply $\tau$ for this morphism, and call it the **tautological morphism of $\Sigma$-modules**.

**Proposition 81.** *Our $\Sigma$-modules and their morphisms, with the obvious composition and identity, form a category.*

**Remark 82.** A $\Sigma$-equation $(\mathsf{D}, a, u, v)$ is uniquely determined by a pair of parallel morphisms of $\Sigma$-modules $e_1, e_2 : \Phi \to \Psi$:

- $\mathsf{D}$ is the category determined by the $\Sigma$-modules $\Phi$ and $\Psi$, as in Definition 78;
- $a : \mathsf{D} \to \mathsf{Mon}^{\Sigma}$ is the canonical forgetful functor;
- $u$ maps a model $R$ to $(R, e_{1,R})$;
- $v$ maps a model $R$ to $(R, e_{2,R})$.

**Notation 83.** *Thanks to this remark, we allow ourselves to identify $\Sigma$-equations with pairs of parallel $\Sigma$-module morphisms $e_1, e_2 : \Phi \to \Psi$ in the following, that we denote $e_1 = e_2$.*

**Example 84** (Commutativity of a binary operation)**.** Here we instantiate our fixed 1-signature as follows: $\Sigma := \Theta \times \Theta$. In this case, we say that $\tau$ is the (tautological) binary operation. Now we can formulate the usual law of commutativity for this binary operation.

We consider the morphism of 1-signatures $\mathsf{swap} : \Theta^2 \longrightarrow \Theta^2$ that exchanges the two components of the direct product. Again by Example 80, we have an induced morphism of $\Sigma$-modules, still denoted $\mathsf{swap}$.

Then, the $\Sigma$-equation for commutativity is given by the two morphisms of $\Sigma$-modules

$$\begin{array}{l} \Theta^2 \xrightarrow{\mathsf{swap}} \Theta^2 \xrightarrow{\tau} \Theta\ . \\ \Theta^2 \xrightarrow[\tau]{\qquad\qquad} \Theta \end{array}$$

See also Section 4.4.1 where we explain in detail the case of monoids.

For the example of the lambda calculus with $\beta$- and $\eta$-equality (given in Example 86), we need to introduce *currying*:

**Definition 85.** By abstracting over the base monad $R$ the adjunction in the category of $R$-modules of Proposition 34, we can perform **currying** of morphisms of parametric modules: given a morphism of parametric modules $\Sigma_1 \times \Theta \to \Sigma_2$ it produces a new morphism $\Sigma_1 \to \Sigma_2'$. By Example 77, currying acts also on morphisms of $\Sigma$-modules.

Conversely, given a morphism of parametric modules (resp. $\Sigma$-modules) $\Sigma_1 \to \Sigma_2'$, we can define the **uncurryied** map $\Sigma_1 \times \Theta \to \Sigma_2$.

**Example 86** ($\beta$- and $\eta$-conversions)**.** Here we instantiate our fixed 1-signature as follows: $\Sigma_{\mathsf{LC}} := \Theta \times \Theta + \Theta'$. This is the 1-signature of the lambda calculus. We break the tautological $\Sigma$-module morphism into its two pieces, namely $\mathsf{app} := \tau \circ \mathsf{inl} : \Theta \times \Theta \longrightarrow \Theta$ and $\mathsf{abs} := \tau \circ \mathsf{inr} : \Theta' \longrightarrow \Theta$. Applying currying to $\mathsf{app}$ yields the morphism $\mathsf{app}_1 : \Theta \longrightarrow \Theta'$ of $\Sigma_{\mathsf{LC}}$-modules. The usual $\beta$ and $\eta$ relations are implemented in our formalism by two $\Sigma_{\mathsf{LC}}$-equations that we call $e_\beta$ and $e_\eta$ respectively:

$$e_\beta : \begin{array}{l} \Theta' \xrightarrow{\mathsf{abs}} \Theta \xrightarrow{\mathsf{app}_1} \Theta' \\ \Theta' \xrightarrow[1]{\qquad\qquad} \Theta' \end{array} \quad \text{and} \quad e_\eta : \begin{array}{l} \Theta \xrightarrow{\mathsf{app}_1} \Theta' \xrightarrow{\mathsf{abs}} \Theta \\ \Theta \xrightarrow[1]{\qquad\qquad} \Theta \end{array} .$$

### 4.2.2 2-signatures and their models

**Definition 87.** A **2-parametric module** $(\Sigma, E)$ is a pair of a parametric module $\Sigma$ and a family $E$ of $\Sigma$-equations. Such a 2-parametric module induces a signature over the category of monads that we call a **2-signature** also denoted $(\Sigma, E)$: it consists of the extension of the signature induced by $\Sigma$ with the family $E$ of equational arities.

**Example 88.** The 2-signature for a commutative binary operation is $(\Theta^2, \tau \circ \mathsf{swap} = \tau)$ (cf. Example 84).

**Example 89.** The 2-signature of the lambda calculus modulo $\beta$- and $\eta$-equality is $\Upsilon_{\mathsf{LC}_{\beta\eta}} = (\Theta \times \Theta + \Theta', \{e_\beta, e_\eta\})$, where $e_\beta, e_\eta$ are the $\Sigma_{\mathsf{LC}}$-equations defined in Example 86.

Now, we unfold Definition 12 of the category of models in the case of a 2-signature (Definition 92).

**Definition 90** (`satisfies_equation`)**.** We say that a model $M$ of a 1-signature $\Sigma$ **satisfies the $\Sigma$-equation** $e = (e_1, e_2)$ if there is an action of $e$ in $M$, that is, if $e_1(M) = e_2(M)$. If $E$ is a family of $\Sigma$-equations, we say that a model $M$ of $\Sigma$ **satisfies** $E$ if $M$ satisfies each $\Sigma$-equation in $E$.

**Remark 91.** Given a monad $R$ and a 2-signature $\Upsilon = (\Sigma, E)$, an **action of $\Upsilon$ in $R$** is an action of $\Sigma$ in $R$ such that the induced 1-model satisfies all the equations in $E$.

**Definition 92** (`category_model_equations`)**.** For a 2-parametric module $(\Sigma, E)$, the **category of models** $\mathsf{Mon}^{(\Sigma,E)}$ **of** $(\Sigma, E)$ is the category of models of the induced 2-signature. More concretely, it is the full subcategory of the category of models of $\Sigma$ whose objects are models of $\Sigma$ satisfying $E$, or equivalently, monads equipped with an action of $(\Sigma, E)$.

**Example 93.** A model of the 2-signature $\Upsilon_{\mathsf{LC}_{\beta\eta}} = (\Theta \times \Theta + \Theta', \{e_\beta, e_\eta\})$ is given by a model $(R, \mathsf{app}^R : R \times R \to R, \mathsf{abs}^R : R' \to R)$ of the 1-signature $\Sigma_{\mathsf{LC}}$ such that $\mathsf{app}_1^R \cdot \mathsf{abs}^R = 1_{R'}$ and $\mathsf{abs}^R \cdot \mathsf{app}_1^R = 1_R$ (see Example 86).

**Definition 94.** A 2-parametric module $(\Sigma, E)$ is said to be **effective** if the induced 2-signature is, that is, if its category of models $\mathsf{Mon}^{(\Sigma,E)}$ has an initial object, denoted $\widehat{(\Sigma, E)}$.

In Section 4.2.4, we aim to find sufficient conditions for a 2-parametric module $(\Sigma, E)$ to be effective.

### 4.2.3 Modularity for 2-signatures

In this section, we define the category $2\mathsf{PMod}$ of 2-parametric modules and the category $2\mathsf{Mdls}$ of models of 2-signatures, together with functors that relate them with the categories PMod and Mdls of parametric modules and 1-models. The situation is summarized in the commutative diagram of functors

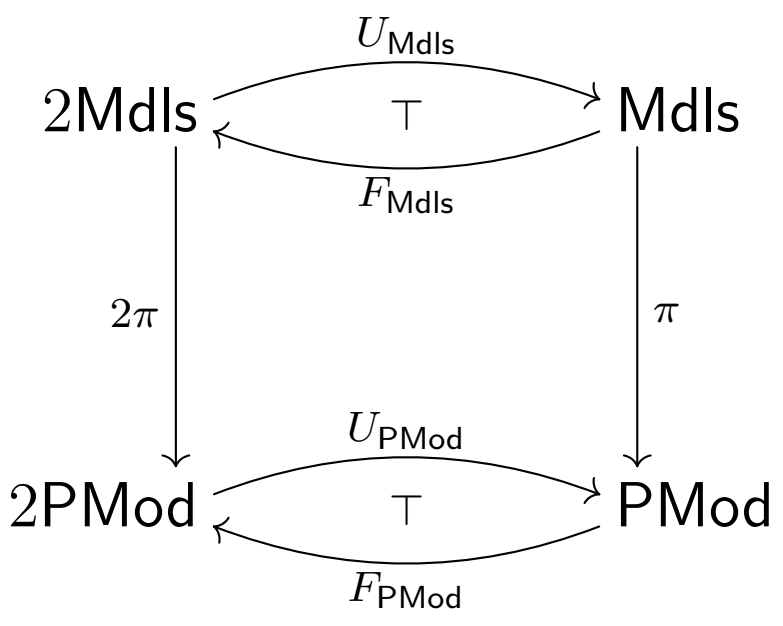

where

- $2\pi$ is a Grothendieck fibration;
- $\pi$ is the Grothendieck fibration defined in Section 3.4.2;
- $U_{\mathsf{PMod}}$ is a coreflection and preserves colimits; and
- $U_{\mathsf{Mdls}}$ is a coreflection.

As a simple consequence of this data, we obtain, in Theorem 102, a *modularity* result in the sense of Ghani, Uustalu, and Hamana [GUH06]: it explains how the initial model of a pushout of 2-parametric modules is the pushout of the initial models of the summands[2].

We start by defining the category $2\mathsf{PMod}$ of 2-parametric modules:

**Definition 95** (`TwoSig_category`)**.** Given 2-parametric modules $(\Sigma_1, E_1)$ and $(\Sigma_2, E_2)$, a **morphism of 2-parametric modules from** $(\Sigma_1, E_1)$ **to** $(\Sigma_2, E_2)$ is a morphism of parametric modules $m : \Sigma_1 \to \Sigma_2$ such that for any model $M$ of $\Sigma_2$ satisfying $E_2$, the $\Sigma_1$-model $m^*M$ satisfies $E_1$.

These morphisms, together with composition and identity inherited from parametric modules, form the category $2\mathsf{PMod}$.

We now study the existence of colimits in $2\mathsf{PMod}$. We know that PMod is cocomplete, and we use this knowledge in our study of $2\mathsf{PMod}$, by relating the two categories:

2. This definition of “modularity” does not seem related to the specific meaning it has in the rewriting community (see, for example, [Gra12]).

Let $F_{\mathsf{PMod}} : \mathsf{PMod} \to 2\mathsf{PMod}$ be the functor which associates to any parametric module $\Sigma$ the empty family of equations, $F_{\mathsf{PMod}}(\Sigma) := (\Sigma, \emptyset)$. Call $U_{\mathsf{PMod}} : 2\mathsf{PMod} \to \mathsf{PMod}$ the forgetful functor defined on objects as $U_{\mathsf{PMod}}(\Sigma, E) := \Sigma$.

**Lemma 96** (`TwoSig_OneSig_is_right_adjoint`, `OneSig_TwoSig_fully_faithful`). *We have $F_{\mathsf{PMod}} \dashv U_{\mathsf{PMod}}$. Furthermore, $U_{\mathsf{PMod}}$ is a coreflection.*

We are interested in specifying new languages by "gluing together" simpler ones. On the level of 2-parametric modules, this is done by taking the coproduct, or, more generally, the pushout of 2-parametric modules:

**Theorem 97** (`TwoSig_PushoutsSET`). *The category $2\mathsf{PMod}$ has pushouts.*

Coproducts are computed by taking the union of the equations and the coproducts of the underlying parametric modules. Coequalizers are computed by keeping the equations of the codomain and taking the coequalizer of the underlying parametric modules. Thus, by decomposing any colimit into coequalizers and coproducts, we have this more general result:

**Proposition 98.** *The category $2\mathsf{PMod}$ is cocomplete and $U_{\mathsf{PMod}}$ preserves colimits.*

We now turn to our modularity result, which states that the initial model of a coproduct of two 2-parametric modules is the coproduct of the initial models of the summands. More generally, the two languages can be amalgamated along a common "core language", by considering a pushout rather than a coproduct.

For a precise statement of that result, we define a "total category of models of 2-signatures":

**Definition 99.** The category $\int^{(\Sigma,E)} \mathsf{Mon}^{(\Sigma,E)}$, or $2\mathsf{Mdls}$ for short, has, as objects, pairs $((\Sigma, E), M)$ of a 2-parametric modules $(\Sigma, E)$ and a model $M$ of $(\Sigma, E)$.

A morphism from $((\Sigma_1, E_1), M_1)$ to $((\Sigma_2, E_2), M_2)$ is a pair $(m, f)$ consisting of a morphism $m : (\Sigma_1, E_1) \to (\Sigma_2, E_2)$ of 2-parametric modules and a morphism $f : M_1 \to m^* M_2$ of $(\Sigma_1, E_1)$-models (or, equivalently, of $\Sigma_1$-models).

This category of models of 2-signatures contains the models of 1-signatures as a coreflective subcategory. Let $F_{\mathsf{Mdls}} : \mathsf{Mdls} \to 2\mathsf{Mdls}$ be the functor which associates to any 1-model $(\Sigma, M)$ the empty family of equations, $F_{\mathsf{Mdls}}(\Sigma, M) := (F_{\mathsf{PMod}}(\Sigma), M)$. Conversely, the forgetful functor $U_{\mathsf{Mdls}} : 2\mathsf{Mdls} \to \mathsf{Mdls}$ maps $((\Sigma, E), M)$ to $(\Sigma, M)$.

**Lemma 100** (`TwoMod_OneMod_is_right_adjoint`, `OneMod_TwoMod_fully_faithful`). *We have $F_{\mathsf{Mdls}} \dashv U_{\mathsf{Mdls}}$. Furthermore, $U_{\mathsf{Mdls}}$ is a coreflection.*

The modularity result is a consequence of the following technical result:

**Proposition 101** (`TwoMod_cleaving`). *The forgetful functor $2\pi$ from $2\mathsf{Mdls}$ to $2\mathsf{PMod}$ is a Grothendieck fibration.*

The *modularity result* below is analogous to the modularity result for 1-signatures (Theorem 54):

**Theorem 102** (Modularity for 2-signatures, `pushout_in_big_rep`). *Suppose we have a pushout diagram of effective 2-parametric modules, as on the left below. This pushout gives rise to a commutative square of morphisms of models in $2\mathsf{Mdls}$ as on the right below, where we only write the second components, omitting the (morphisms of) parametric modules. This square is a pushout square.*

$$\begin{array}{ccc} \Upsilon_0 & \longrightarrow & \Upsilon_1 \\ \downarrow & & \ulcorner\downarrow \\ \Upsilon_2 & \longrightarrow & \Upsilon \end{array} \qquad \begin{array}{ccc} \widehat{\Upsilon}_0 & \longrightarrow & \widehat{\Upsilon}_1 \\ \downarrow & & \ulcorner\downarrow \\ \widehat{\Upsilon}_2 & \longrightarrow & \widehat{\Upsilon} \end{array}$$

Intuitively, the 2-signatures $\Upsilon_1$ and $\Upsilon_2$ specify two extensions of the 2-signature $\Upsilon_0$, and $\Upsilon$ is the smallest extension containing both these extensions. By Theorem 102 the initial model of $\Upsilon$ is the "smallest model containing both the languages generated by $\Upsilon_1$ and $\Upsilon_2$".

### 4.2.4 Initial Semantics for 2-Signatures

We now turn to the problem of constructing the initial model of a 2-signature $(\Sigma, E)$. More specifically, we identify sufficient conditions for $(\Sigma, E)$ to be effective. Our approach is very straightforward: we seek to construct the initial object $\widehat{(\Sigma, E)}$ by applying a suitable quotient construction to the initial object $\hat{\Sigma}$ of $\mathsf{Mon}^{\Sigma}$.

This leads immediately to our first requirement on $(\Sigma, E)$, which is that $\Sigma$ must be an effective 1-signature. (For instance, we can assume that $\Sigma$ is an algebraic 1-signature, see Theorem 53.) This is a very natural hypothesis, since in the case where $E$ is the empty family of $\Sigma$-equations, it is obviously a necessary and sufficient condition.

Some $\Sigma$-equations are never satisfied. In that case, the category $\mathsf{Mon}^{(\Sigma,E)}$ is empty. For example, given any 1-signature $\Sigma$, consider the $\Sigma$-equation $\mathsf{inl}, \mathsf{inr} : \Theta \rightrightarrows \Theta + \Theta$ given by the left and right inclusion. This is obviously an unsatisfiable $\Sigma$-equation. We have to find suitable hypotheses to rule out such unsatisfiable $\Sigma$-equations. This motivates the notion of *elementary* equations.

**Definition 103.** Given a 1-signature $\Sigma$, a $\Sigma$-module $S$ is **nice** if $S$ sends pointwise epimorphic $\Sigma$-model morphisms to pointwise epimorphic module morphisms.

**Definition 104** (`elementary_equation`)**.** Given a 1-signature $\Sigma$, an **elementary $\Sigma$-equation** is a $\Sigma$-equation such that

- the target is a finite derivative of the tautological $\Sigma$-module $\Theta$, i.e., of the form $\Theta^{(n)}$ for some $n \in \mathbb{N}$, and
- the source is a nice $\Sigma$-module.

**Example 105.** Any algebraic parametric module is nice (Example 65). Thus, any $\Sigma$-equation between an algebraic parametric module and $\Theta^{(n)}$, for some natural number $n$, is elementary.

**Definition 106.** A 2-parametric module $(\Sigma, E)$ is said **algebraic** if $\Sigma$ is algebraic and $E$ is a family of elementary equations. The induced 2-signature is then also called **algebraic**.

**Theorem 107** (`elementary_equations_on_alg_preserve_initiality`)**.** *Any algebraic 2-signature is effective.*

The proof of Theorem 107 is given in Section 4.3.

**Example 108.** The 2-signature of lambda calculus modulo $\beta$ and $\eta$ equations given in Example 89 is algebraic. Its initial model is precisely the monad $\mathsf{LC}_{\beta\eta}$ of lambda calculus modulo $\beta\eta$ equations.

The instantiation of the formalized Theorem 107 to this 2-signature is done[3] in `LCBetaEta`.

Let us mention finally that, using the axiom of choice, we can take a similar quotient on all the 1-models of $\Sigma$:

3. An initiality result for this particular case was also previously discussed and proved formally in the Coq proof assistant in [HM10].

**Proposition 109** (`ModEq_Mod_is_right_adjoint`, `ModEq_Mod_fully_faithful`). *Here we assume the axiom of choice. The forgetful functor from the category* $\mathsf{Mon}^{(\Sigma,E)}$ *of 2-models of* $(\Sigma, E)$ *to the category* $\mathsf{Mon}^{\Sigma}$ *of* $\Sigma$*-models has a left adjoint. Moreover, the left adjoint is a reflector.*

## 4.3 Proof of Theorem 107

Our main technical result on effectiveness is the following Lemma 110. In Theorem 107, we give a much simpler criterion that encompasses all the examples we give.

**Lemma 110** (`elementary_equations_preserve_initiality`). *Let* $(\Sigma, E)$ *be a 2-parametric module such that:*

1. $\Sigma$ *sends epimorphic natural transformations to epimorphic natural transformations,*
2. $E$ *is a family of elementary equations,*
3. $\Sigma$ *is effective,*
4. *the initial 1-model of* $\Sigma$ *preserves epimorphisms,*
5. *the image by* $\Sigma$ *of the initial 1-model of* $\Sigma$ *preserves epimorphisms.*

*Then, the 2-signature* $(\Sigma, E)$ *is effective.*

Before tackling the proof of Lemma 110, we discuss how to derive Theorem 107 from it, and we prove some auxiliary results.

The "epimorphism" hypotheses of Lemma 110 are used to transfer structure from the initial model $\hat{\Sigma}$ of the 1-signature $\Sigma$ onto a suitable quotient. There are different ways to prove these hypotheses:

- The axiom of choice implies Conditions 4 and 5 since, in this case, any epimorphism in $\mathsf{Set}$ is split and thus preserved by any functor.
- Condition $5$ is a consequence of Condition 4 if $\Sigma$ sends monads preserving epimorphisms to modules preserving epimorphisms.
- If $\Sigma$ is algebraic, then Conditions 1, 3, 4 and 5 are satisfied (Example 65 and Lemma 66).

From the remarks above, we derive the simpler and weaker statement of Theorem 107 that covers all our examples, which are algebraic.

This section is dedicated to the proof of the main technical result, Lemma 110. The reader inclined to do so may safely skip this section, and rely on the correctness of the machine-checked proof instead.

The proof of Lemma 110 uses some quotient constructions that we present now:

**Proposition 111** (`u_monad_def`). *Given a monad $R$ preserving epimorphisms and a collection of monad morphisms $(f_i : R \to S_i)_{i \in I}$, there exists a* quotient *monad $R/(f_i)$ together with a* projection *$p^R \colon R \longrightarrow R/(f_i)$, which is a morphism of monads such that each $f_i$ factors through $p$.*

*Proof.* The set $R/(f_i)(X)$ is computed as the quotient of $R(X)$ with respect to the relation $x \sim y$ if and only if $f_i(x) = f_i(y)$ for each $i \in I$. This is a straightforward adaptation of Lemma 63. □

Note that epimorphism preservation is implied by the axiom of choice, but can be proven for the monad underlying the initial model $\hat{\Sigma}$ of an algebraic 1-signature $\Sigma$ even without resorting to the axiom of choice.

The above construction can be transported to $\Sigma$-models:

**Proposition 112** (`u_rep_def`). *Let $\Sigma$ be a parametric module sending epimorphic natural transformations to epimorphic natural transformations, and let $R$ be a $\Sigma$-model such that $R$ and $\Sigma(R)$ preserve epimorphisms. Let $(f_i : R \to S_i)_{i \in I}$ be a collection of $\Sigma$-model morphisms. Then the monad $R/(f_i)$ has a natural structure of $\Sigma$-model and the quotient map $p^R \colon R \longrightarrow R/(f_i)$ is a morphism of $\Sigma$-models. Any morphism $f_i$ factors through $p^R$ in the category of $\Sigma$-models.*

The fact that $R$ and $\Sigma(R)$ preserve epimorphisms is implied by the axiom of choice. The proof follows the same line of reasoning as the proof of Proposition 111.

Now we are ready to prove the main technical lemma:

*Proof of Lemma 110.* Let $\Sigma$ be an effective parametric module, and let $E$ be a set of elementary $\Sigma$-equations. The plan of the proof is as follows:

1. Start with the initial model $(\hat{\Sigma}, \sigma)$, with $\sigma : \Sigma(\hat{\Sigma}) \to \hat{\Sigma}$.

2. Construct the quotient model $\hat{\Sigma}/(f_i)$ according to Proposition 112 where $(f_i : \hat{\Sigma} \to S_i)_i$ is the collection of all initial $\Sigma$-morphisms from $\hat{\Sigma}$ to any $\Sigma$-model satisfying the equations. We denote by $\sigma/(f_i) : \Sigma(\hat{\Sigma}/(f_i)) \to \hat{\Sigma}/(f_i)$ the action of the quotient model.

3. Given a model $M$ of the 2-signature $(\Sigma, E)$, we obtain a morphism $i_M : \hat{\Sigma}/(f_i) \to M$ from Proposition 112. Uniqueness of $i_M$ is shown using epimorphicity of the projection $p : \hat{\Sigma} \to \hat{\Sigma}/(f_i)$. For this, it suffices to show uniqueness of the composition $i_M \circ p : \hat{\Sigma} \to M$ in the category of 1-models of $\Sigma$, which follows from initiality of $\hat{\Sigma}$.

4. The verification that $\left(\hat{\Sigma}/(f_i), \sigma/(f_i)\right)$ satisfies the equations is given below. Actually, it follows the same line of reasoning as in the proof of Proposition 111 that $\hat{\Sigma}/(f_i)$ satisfies the monad equations.

Let $e = (e_1, e_2) : U \to \Theta^{(n)}$ be an elementary equation of $E$. We want to prove that the two arrows

$$e_{1,\hat{\Sigma}/(f_i)}, e_{2,\hat{\Sigma}/(f_i)} : U(\hat{\Sigma}/(f_i)) \longrightarrow (\hat{\Sigma}/(f_i))^{(n)}$$

are equal. As $p$ is an epimorphic natural transformation, $U(p)$ also is by definition of an elementary equation. It is thus sufficient to prove that

$$e_{1,\hat{\Sigma}/(f_i)} \circ U(p) = e_{2,\hat{\Sigma}/(f_i)} \circ U(p) \ ,$$

which, by naturality of $e_1$ and $e_2$, is equivalent to $p^{(n)} \circ e_{1,\hat{\Sigma}} = p^{(n)} \circ e_{2,\hat{\Sigma}}$.

Let $x$ be an element of $U(\hat{\Sigma})$ and let us show that $p^{(n)}(e_{1,\hat{\Sigma}}(x)) = p^{(n)}(e_{2,\hat{\Sigma}}(x))$. By definition of $\hat{\Sigma}/(f_i)$ as a pointwise quotient (see Proposition 111), it is enough to show that for any $j$, the equality $f_j^{(n)}(e_{1,\hat{\Sigma}}(x)) = f_j^{(n)}(e_{2,\hat{\Sigma}}(x))$ is satisfied. Now, by naturality of $e_1$ and $e_2$, this equation is equivalent to $e_{1,S_j}(U(f_j)(x))) = e_{2,S_j}(U(f_j)(x)))$ which is true since $S_j$ satisfies the equation $e_1 = e_2$. □

## 4.4 Examples of algebraic 2-signatures

We already illustrated our theory by looking at the paradigmatic case of lambda calculus modulo $\beta$- and $\eta$-equations (Examples 86 and 108). This section collects further examples of application of our results.

In our framework, complex signatures can be built out of simpler ones by taking their coproducts. Note that the class of algebraic 2-parametric modules encompasses the algebraic parametric modules and is closed under arbitrary coproducts: the prototypical examples of algebraic 2-signatures given in this section can be combined with any other algebraic 2-signature, yielding an effective 2-signature thanks to Theorem 107.

### 4.4.1 Monoids

We begin with an example of monad for a first-order syntax with equations. Given a set $X$, we denote by $M(X)$ the free monoid built over $X$. This is a classical example of monad over the category of (small) sets. The monoid structure gives us, for each set $X$, two maps $m_X\colon M(X) \times M(X) \longrightarrow M(X)$ and $e_X\colon 1 \longrightarrow M(X)$ given by the product and the identity respectively. It can be easily verified that $m\colon M^2 \longrightarrow M$ and $e\colon 1 \longrightarrow M$ are $M$-module morphisms. In other words, $(M, \rho) = (M, [m, e])$ is a model of the 1-signature $\Sigma = \Theta \times \Theta + 1$.

We break the tautological morphism of $\Sigma$-modules (cf. Example 80) into constituent pieces, defining $\mathsf{m} := \tau \circ \mathsf{inl} : \Theta \times \Theta \to \Theta$ and $\mathsf{e} := \tau \circ \mathsf{inr} : 1 \to \Theta$.

Over the 1-signature $\Sigma$ we specify equations postulating *associativity* and *left and right unitality* as follows:

$$\begin{array}{ccc} \Theta^3 \xrightarrow{\Theta\times\mathsf{m}} \Theta^2 \xrightarrow{\mathsf{m}} \Theta & \Theta \xrightarrow{\mathsf{e}\times\Theta} \Theta^2 \xrightarrow{\mathsf{m}} \Theta & \Theta \xrightarrow{\Theta\times\mathsf{e}} \Theta^2 \xrightarrow{\mathsf{m}} \Theta \\ \Theta^3 \xrightarrow[\mathsf{m}\times\Theta]{} \Theta^2 \xrightarrow[\mathsf{m}]{} \Theta & \Theta \xrightarrow[1]{} \Theta & \Theta \xrightarrow[1]{} \Theta \end{array}$$

and we denote by $E$ the family consisting of these three $\Sigma$-equations. All are elementary since their codomain is $\Theta$, and their domain a product of $\Theta$s.

One checks easily that $(M, [m, e])$ is the initial model of $(\Sigma, E)$.

Several other classical (equational) algebraic theories, such as groups and rings, can be treated similarly, see Section 4.4.3 below. However, at the present state we cannot model theories with partial construction (e.g., fields).

### 4.4.2 Colimits of algebraic 2-parametric modules

In this section, we argue that our framework encompasses any colimit of algebraic 2-parametric modules.

Actually, the class of algebraic 2-parametric modules is not stable under colimits, as this is not even the case for algebraic parametric modules. However, we can weaken this statement as follows:

**Proposition 113.** *Given any colimit of algebraic 2-parametric modules, there is an algebraic 2-signature yielding an isomorphic category of models.*

*Proof.* As the class of algebraic 2-parametric modules is closed under arbitrary coproducts, using the decomposition of colimits into coproducts and coequalizers, any colimit $\Xi$ of algebraic 2-parametric modules can be expressed as a coequalizer of two morphisms $f, g$ between some algebraic 2-parametric modules $(\Sigma_1, E_1)$ and $(\Sigma_2, E_2)$,

$$(\Sigma_1, E_1) \underset{g}{\overset{f}{\rightrightarrows}} (\Sigma_2, E_2) \xrightarrow{p} \Xi = (\Sigma_3, E_2)\ .$$

where $\Sigma_3$ is the coequalizer of the parametric module morphisms $f$ and $g$. Note that the set of equations of $\Xi$ is $E_2$, by construction of the coequalizer in the category of 2-parametric modules. Now, consider the algebraic 2-signature $\Xi' = (\Sigma_2, E_2 + \text{(4.2)})$ consisting of the 1-signature $\Sigma_2$ and the equations of $E_2$ plus the following elementary equation (see Examples 80 and 105):

$$\begin{array}{c} \Sigma_1 \xrightarrow{f} \Sigma_2 \xrightarrow{\tau^{\Sigma_2}} \Theta\ . \\ \Sigma_1 \xrightarrow[g]{} \Sigma_2 \xrightarrow[\tau^{\Sigma_2}]{} \Theta \end{array} \tag{4.2}$$

We show that $\mathsf{Mon}^{\Xi}$ and $\mathsf{Mon}^{\Xi'}$ are isomorphic. A model of $\Xi'$ is a monad $R$ together with an $R$-module morphism $r : \Sigma_2(R) \to R$ such that $r \circ f_R = r \circ g_R$ and that the equations of $E_2$ are satisfied. By universal property of the coequalizer, this is exactly the same as giving an $R$-module morphism $\Sigma_3(R) \to R$ satisfying the equations of $E_2$, i.e., giving $R$ an action of $\Xi = (\Sigma_3, E_2)$.

It is straightforward to check that this correspondence yields an isomorphism between the category of models of $\Xi$ and the category of models of $\Xi'$. □

This proposition, together with the following corollary, allow us to recover all the examples presented in Chapter 3, as colimits of algebraic parametric modules: syntactic commutative binary operator, maximum operator, application à la differential lambda calculus, syntactic closure operator, integrated substitution operator, coherent fixpoint operator.

**Corollary 114.** *If $F$ is a finitary endofunctor on* Set*, then there is an algebraic 2-signature whose category of models is isomorphic to the category of 1-models of the 1-signature $F \cdot \Theta$.*

*Proof.* It is enough to prove that $F \cdot \Theta$ is a colimit of algebraic parametric modules.

As $F$ is finitary, it is isomorphic to the coend $\int^{n \in \mathbb{F}} F(n) \times _^n$ where $\mathbb{F}$ is the full subcategory of Set of finite ordinals (see, e.g., [VK11, Example 3.19]). As colimits are computed pointwise, the parametric module $F \cdot \Theta$ is the coend $\int^{n \in \mathbb{F}} F(n) \times \Theta^n$, and as such, it is a colimit of algebraic 2-parametric modules. □

However, we do not know whether we can recover Theorem 57 stating that any presentable 1-signature is effective.

### 4.4.3 Algebraic theories

From the categorical point of view, several fundamental algebraic structures in mathematics can be conveniently and elegantly described using finitary monads. For instance, the category of monoids can be seen as the category of Eilenberg–Moore algebras of the monad of lists. Other important examples, like groups and rings, can be treated analogously. A classical reference on the subject is the work of Manes, where such monads are significantly called *finitary algebraic theories* [Man76, Definition 3.17].

We want to show that such "algebraic theories" fit in our framework, in the sense that they can be incorporated into an algebraic 2-signature, with the effect of enriching the initial model with the operations of the algebraic theory, subject to the axioms of the algebraic theory.

For a finitary monad $T$, Corollary 114 says how to encode the 1-signature $T \cdot \Theta$ as an algebraic 2-signature $(\Sigma_T, E_T)$. Models are monads $R$ together with an $R$-linear morphism $r : T \cdot R \to R$.

Now, for any model $(R, m)$ of $T \cdot \Theta$, we would like to enforce the usual $T$-algebra equations on the action $m$. This is done thanks to the following equations, where $\tau$ denotes the tautological morphism of $T \cdot \Theta$-modules:

$$\begin{array}{ccc} \Theta \xrightarrow{\eta_T \cdot \Theta} T \cdot \Theta \xrightarrow{\tau} \Theta & \qquad & T \cdot T \cdot \Theta \xrightarrow{\mu_T \cdot \Theta} T \cdot \Theta \xrightarrow{\tau} \Theta \\ \Theta \xrightarrow[1]{\hspace{4em}} \Theta & \qquad & T \cdot T \cdot \Theta \xrightarrow[T\tau]{} T \cdot \Theta \xrightarrow[\tau]{} \Theta \end{array} \tag{4.3}$$

The first equation is clearly elementary. The second one is elementary thanks to the following lemma:

**Lemma 115.** *Let $F$ be a finitary endofunctor on* $\mathsf{Set}$*. Then $F$ preserves epimorphisms.*

*Proof.* This is a consequence of the axiom of choice, because then any epimorphism in the category of $\mathsf{Set}$ is split, and thus preserved by any functor. Here we provide an alternative proof which does not rely on the axiom of choice. (However, it may require the excluded middle, depending on the chosen definition of finitary functor.)

As $F$ is finitary, it is isomorphic to the coend $\int^{n\in\mathbb{F}} F(n)\times_^n$ [VK11, Example 3.19]. By decomposing it as a coequalizer of coproducts, we get an epimorphism $\alpha : \coprod_{n\in\mathbb{N}} F(n)\times _^n \to F$. Now, let $f : X \to Y$ be a surjective function between two sets. We show that $F(f)$ is epimorphic. By naturality, the following diagram commutes:

$$\begin{array}{ccc} \coprod_{n\in\mathbb{N}} F(n)\times X^n & \xrightarrow{F(n)\times f^n} & \coprod_{n\in\mathbb{N}} F(n)\times Y^n \\ {\scriptstyle \alpha_X}\downarrow & & \downarrow{\scriptstyle \alpha_Y} \\ F(X) & \xrightarrow[F(f)]{} & F(Y) \end{array} .$$

The top-right composite is epimorphic by composition of epimorphisms. Thus, the bottom-left composite is also epimorphic, hence so is $F(f)$ as the last morphism of this composition.

□

In conclusion, we have exhibited the algebraic 2-signature $(\Sigma_T, E'_T)$, where $E'_T$ extends the family $E_T$ with the two elementary equations of Diagram 4.3. This signature allows to enrich any other algebraic 2-signature with the operations of the algebraic theory $T$, subject to the relevant equations.

### 4.4.4 Fixpoint operator

Here, we show the algebraic 2-signature corresponding to a fixpoint operator. In Section 3.8.4, we studied fixpoint operators in the context of 1-signatures. In that setting, we treated a *syntactic* fixpoint operator called *coherent* fixpoint operator, somehow reminiscent of mutual letrec. We were able to impose many natural equations to this operator but we were not able to enforce the fixpoint equation. In this section, we show how a fixpoint operator can be fully specified by an algebraic 2-signature. We restrict our discussion to the unary case; the coherent family of multi-ary fixpoint operators presented in Section 3.8.4, now including the fixpoint equations, can also be specified, in an analogous way, via an algebraic 2-signature.

Let us start by recalling Definition 72: a **unary fixpoint operator for a monad** $R$ is a module morphism $f$ from $R'$ to $R$ that makes the following diagram commute, where $\sigma$ is the substitution morphism defined as the uncurrying (see Definition 85) of the identity morphism on $\Theta'$:

$$\begin{array}{ccc} R' & \xrightarrow{(id_{R'},f)} & R' \times R \\ & {\scriptstyle f}\searrow \quad \swarrow{\scriptstyle \sigma_R} & \\ & R & \end{array}$$

In order to rephrase this definition, we introduce the obviously algebraic 2-signature $\Upsilon_{\mathsf{fix}}$ consisting of the 1-signature $\Sigma_{\mathsf{fix}} = \Theta'$ and the family $E_{\mathsf{fix}}$ consisting of the single following $\Sigma_{\mathsf{fix}}$-equation:

$$e_{\mathsf{fix}} : \begin{array}{l} \Theta' \xrightarrow{\langle 1,\tau \rangle} \Theta' \times \Theta \xrightarrow{\ \sigma\ } \Theta \\ \Theta' \xrightarrow[\tau]{\qquad\qquad\qquad} \Theta \end{array} \tag{4.4}$$

This allows us to rephrase the previous definition as follows: a unary fixpoint operator for a monad $R$ is just an action of the 2-signature $\Upsilon_{\mathsf{fix}}$ in $R$.

## 4.5 Recursion

In this section, we explain how a recursion principle can be derived from our initiality result, and give an example of a morphism—a *translation*—between monads defined via the recursion principle.

### 4.5.1 Principle of recursion

In our context, the recursion principle is a recipe for constructing a morphism from the monad underlying the initial model of a 2-signature $\Upsilon = (\Sigma, E)$ to an arbitrary monad $T$.

**Proposition 116** (Recursion principle)**.** *Let $S$ be the monad underlying the initial model of the 2-signature $\Upsilon$. To any action $a$ of $\Upsilon$ in $T$ is associated a monad morphism $\hat{a}$ :*

$S \to T$.

*Proof.* The action $a$ defines a 2-model $M$ of $\Upsilon$, and $\hat{a}$ is the monad morphism underlying the initial morphism to $M$. □

Hence the recipe consists in the following two steps:

1. give $T$ an action of the 1-signature $\Sigma$;

2. check that all the equations in $E$ are satisfied for the induced model.

In the next section, we illustrate this principle.

### 4.5.2 Translation of lambda calculus with fixpoint to lambda calculus

In this section, we consider the 2-signature $\Upsilon_{\mathsf{LC}_{\beta\eta,\mathsf{fix}}} := \Upsilon_{\mathsf{LC}_{\beta\eta}} + \Upsilon_{\mathsf{fix}}$ where the two components have been introduced above (see Example 93 and Section 4.4.4).

As a coproduct of algebraic 2-parametric modules, $\Upsilon_{\mathsf{LC}_{\beta\eta,\mathsf{fix}}}$ is itself algebraic, and thus the initial model exists. The underlying monad $\mathsf{LC}_{\beta\eta,\mathsf{fix}}$ of the initial model can be understood as the monad of lambda calculus modulo $\beta$ and $\eta$ enriched with an *explicit* fixpoint operator $\mathsf{fix} : \mathsf{LC}'_{\beta\eta,\mathsf{fix}} \longrightarrow \mathsf{LC}_{\beta\eta,\mathsf{fix}}$. Now we build by recursion a monad morphism from this monad to the “bare” monad $\mathsf{LC}_{\beta\eta}$ of lambda calculus modulo $\beta$ and $\eta$.

As explained in Section 4.5.1, we need to define an action of $\Upsilon_{\mathsf{LC}_{\beta\eta,\mathsf{fix}}}$ in $\mathsf{LC}_{\beta\eta}$, that is to say an action of $\Upsilon_{\mathsf{LC}_{\beta\eta}}$ plus an action of $\Upsilon_{\mathsf{fix}}$. For the action of $\Upsilon_{\mathsf{LC}_{\beta\eta}}$, we take the one yielding the initial model.

Now, in order to find an action of $\Upsilon_{\mathsf{fix}}$ in $\mathsf{LC}_{\beta\eta}$, we choose a fixpoint combinator $Y$ (say the one of Curry) and take the action $\hat{Y}$ as defined at the end of Section 4.4.4.

In more concrete terms, our translation is a kind of compilation which replaces each occurrence of the explicit fixpoint operator $\mathsf{fix}(t)$ with $\mathsf{app}(Y, \mathsf{abs}\ t)$.

PART II

# Semantics

CHAPTER 5

# REDUCTION MONADS AND THEIR SIGNATURES

In this chapter, we study *reduction monads*, which are essentially the same as monads relative to the free functor from sets into graphs and signatures over their category. The statements here are not formalized.

Reduction monads abstract two aspects of the lambda calculus: on the one hand, in the monadic viewpoint, it is an object equipped with a well-behaved substitution; on the other hand, in the graphical viewpoint, it is an oriented graph whose vertices are terms and whose edges witness the reductions between two terms.

We study presentations of reduction monads. To this end, we propose a notion of *reduction signature*. They are, in particular, signatures in the sense of Chapter 2 consisting of a signature over the category of monads (for example, a 2-signature, as introduced in Chapter 4) extended with a family of arities specifying reduction rules.

As usual, such a signature plays the role of a virtual presentation, and specifies arities for generating operations—possibly subject to equations—together with arities for generating reduction rules.

The main result of this chapter identifies a class of effective reduction signatures. In particular, the lambda calculus is naturally specified by such a signature.

## 5.1 Introduction

The lambda calculus has been a central object in theoretical computer science for decades. However, the corresponding mathematical structure does not seem to have been identified once and for all. In particular, two complementary viewpoints on the (pure untyped) lambda calculus have been widespread: some consider it as a graph (or a preorder, or a category), while others view it as a monad (on the category of sets).

The first account incorporates the $\beta$-reduction, while the second addresses substitution but incorporates only the $\beta$-equality. Merging these two perspectives led Lüth and Ghani [LG97b] to consider monads on the category of preordered sets, and Ahrens [Ahr16] to consider monads relative to the free functor from sets into preorders. In the present chapter, we propose a variant of their approaches. Here we call *reduction monad* a monad relative to the natural injection of sets in graphs, and of course the lambda-calculus yields such a reduction monad. Our main contribution concerns the generation of reduction monads by syntactic (possibly binding) operations (possibly subject to equations) and reduction rules. As is common in similar contexts, we propose a notion of signature for reduction monads, which we call "reduction signatures". Each reduction signature comes equipped with the category of its models: such a model is a reduction monad "acted upon" by the signature. A reduction signature may be understood as a virtual presentation: when an initial model exists (that is, when the signature is effective), it inherits a kind of presentation given by the action of the signature. We identify a natural criterion for a signature to be effective. As should be expected, we give an effective reduction signature specifying the lambda calculus with its reduction rules, which yields a new, high-level definition of the (pure untyped) lambda calculus.

### 5.1.1 Terminology and notations

In this section, we set up some terminology and notations that we use in this chapter.

**Signature for monads** By **signatures for monads**, we mean signatures over the category of monads. In the examples, they are in particular 2-signatures of Chapter 4, or just 1-signatures of Chapter 3. The letter $\Sigma$ is usually associated with a signature for monads.

**Models of signature for monads** In order to avoid any confusion with the notion of model of operational signatures that we introduce here, we refer to models of a signature $\Sigma$ for monads as **$\Sigma$-monads**.

**Substitution** Let $R$ be a monad, $M$ be a $R$-module, $t$ an element of $M(X)$ and $f$ a function from $X$ to $R(Y)$. Then, we denote by $t\{f\} \in R(Y)$ the substitution of all the variables in $t$ by the corresponding term given by $f$.

**Unary substitution** We abbreviate $t\{x \mapsto \text{if } x = y \text{ then } u \text{ else } x\}$ as $t\{y := u\}$.

### 5.1.2 Plan of the chapter

In Section 5.2, we define reduction monads. In Section 5.3, we present our take on reduction rules. This enables us to define a class of signatures over the category of reduction monads—*reduction signatures*—in Section 5.4. Section 5.5 is devoted to the proof of our theorem of effectivity for these signatures, under some natural assumptions. Then, in Section 5.6, we give a detailed example of a reduction signature specifying the lambda calculus with explicit substitutions of [Kes09]. Finally, in Section 5.7, we explain the recursion principle which, as usual, can be derived from initiality in our categories of models.

## 5.2 Reduction monads

Here below, we define *the category of reduction monads* in Section 5.2.1. We also consider some examples of reduction monads, in Section 5.2.2.

### 5.2.1 The category of reduction monads

**Definition 117.** A **reduction monad** $R$ is given by:

1. a monad on sets (the monad of *terms*), that we still denote by $R$, or by $\underline{R}$ when we want to be explicit;

2. an $R$-module $\mathsf{Red}(R)$ (the module of *reductions*);

3. a morphism of $R$-modules $\mathsf{red}_R : \mathsf{Red}(R) \to R \times R$ (*source* and *target* of rules).

We set $\mathsf{source}_R := \pi_1 \circ \mathsf{red}_R : \mathsf{Red}(R) \to R$, and $\mathsf{target}_R := \pi_2 \circ \mathsf{red}_R : \mathsf{Red}(R) \to R$

**Notation 118.** *For a reduction monad $R$, a set $X$, and elements $s, t \in R(X)$, we think of the fiber $\mathsf{red}_R(X)^{-1}(s,t)$ as the set of "reductions from $s$ to $t$". We sometimes write $m : s \blacktriangleright t : R(X)$, or even $m : s \blacktriangleright t$ when there is no ambiguity, instead of $m \in \mathsf{red}_R(X)^{-1}(s,t)$.*

**Remark 119.** Note that for a given reduction monad $R$, set $X$, and $s, t : R(X)$, there can be multiple reductions from $s$ to $t$, that is, the fibre $s \blacktriangleright t$ is not necessarily a subsingleton.

**Remark 120.** Let $R$ be a reduction monad, $X$ and $Y$ two sets, $f : X \to R(Y)$ a substitution, and $u$ and $v$ two elements of $R(X)$ related by $m : u \blacktriangleright v$. The module structure on $\mathsf{Red}(R)$ yields a reduction denoted $m\{f\}$ between $u\{f\}$ and $v\{f\}$.

However, if we are given two substitutions $f$ and $g$, and for all $x \in X$, a reduction $m_x : f(x) \blacktriangleright g(x)$, then it does not follow that there is a reduction between $u\{f\}$ and $u\{g\}$. This leaves the door open for non-congruent reductions.

Our main examples of reduction monads are given by variants of the lambda calculus. We have collected these examples in Section 5.2.2.

**Definition 121.** A **morphism of reduction monads** from $R$ to $S$ is given by a pair $(f, \alpha)$ of

1. a monad morphism $f : R \to S$, and
2. a natural transformation $\alpha : \mathsf{Red}(R) \to \mathsf{Red}(S)$

satisfying the following two conditions:

3. $\alpha$ is an $R$-module morphism between $\mathsf{Red}(R)$ and the pullback module $f^*\mathsf{Red}(S)$ of $\mathsf{Red}(S)$ along $f$ (see Definition 26), and
4. the square

$$\begin{array}{ccc} \mathsf{Red}(R) & \xrightarrow{\alpha} & \mathsf{Red}(S) \\ {\scriptstyle \mathsf{red}_R}\downarrow & & \downarrow{\scriptstyle \mathsf{red}_S} \\ R \times R & \xrightarrow[f \times f]{} & S \times S \end{array}$$

commutes in the category of functors and natural transformations.

In Section 5.7 we specify morphisms of reduction monads via a recursion principle.

Intuitively, a morphism $(f, \alpha)$ as above maps terms of $R$ to terms of $S$ via $f$, and reductions of $R$ to reductions of $S$ via $\alpha$. Condition 3 states compatibility of the map of reductions with substitution. Condition 4 states preservation of source and target by the map of reductions: a reduction $m : u \blacktriangleright v$ is mapped by $\alpha$ to a reduction $\alpha(m) : f(u) \blacktriangleright f(v)$.

**Proposition 122** (Category of reduction monads)**.** *Reduction monads and their morphisms, with the obvious composition and identity, form a category* $\mathsf{RedMon}$*, equipped with a forgetful functor to the category of monads.*

It turns out that reduction monads are the same as monads relative to the free functor from sets to graphs (for the definition of relative monads, see [ACU15, Definition 2.1]):

**Theorem 123.** *The category of reduction monads is isomorphic to the category of monads relative to the functor mapping a set to its discrete graph.*

*Proof.* This is obvious after unfolding the definitions. □

**Remark 124.** In the previous theorem, the involved category of graphs has pairs of parallel functions $E \underset{t}{\overset{s}{\rightrightarrows}} V$ as objects: $E$ is the set of edges, $V$ the set of vertices, $s$ and $t$ maps respectively any edge to its source and target. Adopting the *indexed point of view* rather than this fibrational approach, we obtain an equivalent (but not isomorphic) category: a graph is a set $V$ of vertices, and for each pair of vertices there is a set of edges between them. Our notion of reduction monads could also be rephrased in this way, yielding an equivalent (but not isomorphic) category.

### 5.2.2 Examples of reduction monads

We are interested in reduction monads with underlying monad $\mathsf{LC}$, the monad of syntactic lambda terms specified by the 1-signature $\Sigma_{\mathsf{LC}} = \Theta \times \Theta + \Theta'$.

**Example 125** (Lambda calculus with head-$\beta$-reduction)**.** Consider the reduction monad $\mathsf{LC}_{\mathsf{head}\text{-}\beta}$ given as follows:

1. the underlying monad is $\mathsf{LC}$;
2. $\mathsf{Red}(R)$ is the module $\mathsf{LC}' \times \mathsf{LC}$;
3. $\mathsf{red}_R(X)$ is the morphism $(u, v) \mapsto \big(\mathsf{app}(\mathsf{abs}(u), v), u\{* := v\}\big)$.

This reduction monad deserves to be called *the reduction monad of the lambda calculus with head-$\beta$-reduction*.

**Example 126** (Lambda calculus with head-$\eta$-expansion)**.** Consider the reduction monad $\mathsf{LC}_{\mathsf{head}\text{-}\eta}$ given as follows:

1. the underlying monad is $\mathsf{LC}$;
2. $\mathsf{Red}(R)$ is the module $\mathsf{LC}$;

3. $\mathsf{red}_R(X)$ is the morphism $t \mapsto \big(t, \mathsf{abs}(\mathsf{app}(\iota t, *))\big)$, where $\iota : \mathsf{LC}(X) \to \mathsf{LC}'(X)$ is the natural injection (weakening).

This reduction monad deserves to be called *the reduction monad of the lambda calculus with head-$\eta$-expansion*.

Obviously, to get the analogous reduction rule for the $\eta$-contraction, it is enough to swap, in the previous example, the components of the output of the morphism $\mathsf{red}_R$.

Given two reduction monads with the same underlying monad on sets, we define the *amalgamation* of the reduction monads as follows:

**Definition 127.** Given reduction monads $R$ and $S$ with the same underlying monad on sets, we define the reduction monad $R \amalg S$ as follows:

1. the underlying monad on sets is still $R$ (or, equivalently, $S$);
2. the $R$-module $\mathsf{Red}(R \amalg S)$ is the coproduct $\mathsf{Red}(R) \amalg \mathsf{Red}(S)$.
3. the module morphism $\mathsf{red}_{R \amalg S}$ is induced by $\mathsf{red}_R$ and $\mathsf{red}_S$.

This is the pushout, in the category of reduction monads, of $uR \to R$ and $uR \to S$, with the reduction monad $uR := (\underline{R}, 0, !)$.

**Example 128.** The reduction monad $\mathsf{LC}_{\mathsf{head}\text{-}\beta/\eta} := \mathsf{LC}_{\mathsf{head}\text{-}\beta} \amalg \mathsf{LC}_{\mathsf{head}\text{-}\eta}$ has, as reductions, $\beta$-reductions and $\eta$-expansions at the head of lambda terms.

So far we have only considered reductions at the root of a lambda term. The following construction allows us to propagate reductions into subterms.

**Definition 129.** Let $R$ be a reduction monad over the monad LC on sets. We define the reduction monad $R^{\mathsf{cong}}$ as follows:

1. the underlying monad is LC;
2. $\mathsf{Red}(R^{\mathsf{cong}})(X)$ is generated by the following constructions:
   (a) for $m : u \blacktriangleright v$ in $\mathsf{Red}(R)$, $m$ is also in $\mathsf{Red}(R^{\mathsf{cong}})$
   (b) for $m : u \blacktriangleright v : \mathsf{LC}(X)$ in $\mathsf{Red}(R^{\mathsf{cong}})$ and $t \in \mathsf{LC}(X)$, we have $\mathsf{app\text{-}cong}_1(m, t) : \mathsf{app}(u, t) \blacktriangleright \mathsf{app}(v, t)$
   (c) for $m : u \blacktriangleright v : \mathsf{LC}(X)$ in $\mathsf{Red}(R^{\mathsf{cong}})$ and $t \in \mathsf{LC}(X)$, we have $\mathsf{app\text{-}cong}_2(t, m) : \mathsf{app}(t, u) \blacktriangleright \mathsf{app}(t, v)$

(d) for $m : u \blacktriangleright v : \mathsf{LC}'(X)$ in $\mathsf{Red}(R^{\mathsf{cong}})$ we have $\mathsf{abs\text{-}cong}(m) : \mathsf{abs}(u) \blacktriangleright \mathsf{abs}(v)$

3. $\mathsf{red}_{R^{\mathsf{cong}}}$ is obvious.

**Example 130.** A reduction in the reduction monad $\mathsf{LC}_{\beta/\eta} := (\mathsf{LC}_{\mathsf{head}\text{-}\beta/\eta})^{\mathsf{cong}}$ is "one step" of $\beta$-reduction or $\eta$-expansion, anywhere in the source term.

The *closure under identity and composition of reductions* of a reduction monad is defined as follows:

**Definition 131.** Given a reduction monad $R$, we define the reduction monad $R^*$ as follows:

1. the underlying monad on sets is still $R$;
2. the $R$-module $\mathsf{Red}(R^*)$ is defined as follows. For $n \in \mathbb{N}$ we define the module $\mathsf{Red}(R)^n$ of "$n$ composable reductions", namely as the limit of the diagram

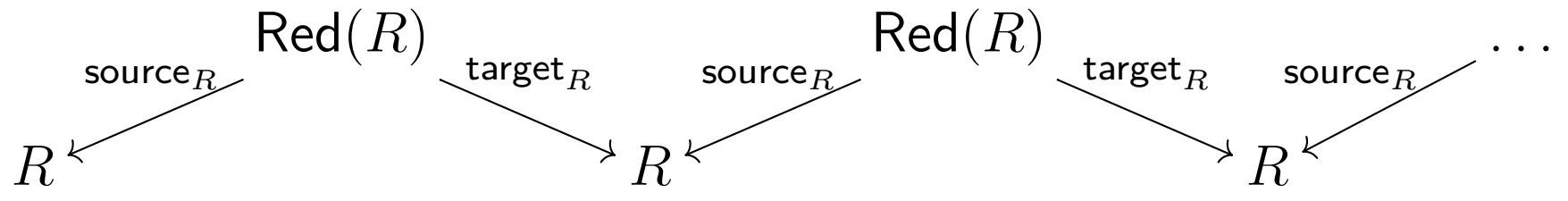

   with $n$ copies of $\mathsf{Red}(R)$ (and hence $n+1$ copies of $R$). We obtain $n+1$ projections $\pi_i : \mathsf{Red}(R)^n \to R$, and we call $p_n := (\pi_0, \pi_n) : \mathsf{Red}(R)^n \to R \times R$. We set $\mathsf{Red}(R^*) := \coprod_n \mathsf{Red}(R)^n$.

3. the module morphism is $\mathsf{red}_{R^*} := [p_n]_{n\in\mathbb{N}} : \coprod_n \mathsf{Red}(R)^n \to R \times R$ the universal morphism induced by the family $(p_n)_{n\in N}$.

**Example 132** (The reduction monad of the lambda calculus)**.** The *reduction monad of the lambda calculus* is defined to be the reduction monad $\mathsf{LC}^*_{\beta/\eta}$.

In Section 5.4 we introduce signatures that allow for the specification of reduction monads. The signature specifying the reduction monad of lambda calculus of Example 130 is given in Example 152.

## 5.3 Reduction rules

In this section, we define an abstract notion of reduction rule over a signature for monads $\Sigma$ (Section 5.3.2). We first focus, in Section 5.3.1, on the example of the congruence rule for the application construction in the signature $\Sigma_{\mathsf{LC}}$ for the monad of the lambda calculus, in order to motivate the definitions. The purpose of a reduction rule over $\Sigma$ is to be "validated" in a reduction monad equipped with an action of $\Sigma$ (this is what we will call a reduction $\Sigma$-monad in Section 5.3.3). We make this notion of validation precise in Section 5.3.4, as an action of the reduction rule in the reduction $\Sigma$-monad. Finally, we give a protocol for specifying reduction rules in Section 5.3.5 that we apply in Section 5.3.6 to some examples.

### 5.3.1 Example: congruence rule for application

We give some intuitions of the definition of reduction rule with the example of the congruence rule for application, given, e.g., in Selinger's lecture notes [Sel08], as follows:

$$\frac{T \rightsquigarrow T' \qquad U \rightsquigarrow U'}{\mathsf{app}(T, U) \rightsquigarrow \mathsf{app}(T', U')}$$

This rule is parameterized by four *metavariables*: $T$, $T'$, $U$, and $U'$. The conclusion and the hypotheses are given by pairs of terms built out of these metavariables.

We formalize this rule as follows: for any monad $R$ equipped with an application operation $\mathsf{app} : R \times R \to R$, we associate a module of metavariables $\mathcal{V}(R) = R \times R \times R \times R$, one factor for each of the metavariables $T$, $T'$, $U$, and $U'$. Each hypothesis or conclusion is described by a parallel pair of morphisms from $\mathcal{V}(R)$ to $R$: for example, the conclusion $c_R : \mathcal{V}(R) \to R$ maps a set $X$ and a quadruple $(T, T', U, U')$ to the pair $(\mathsf{app}(T, U), \mathsf{app}(T', U'))$. These assignments are actually functorial in $R$, and abstracting over $R$ yields our notion of *term-pair* over the $\Sigma$-module $\mathcal{V}$, as morphisms from $\mathcal{V}$ to $\Theta \times \Theta$, where $\Sigma$ is any signature including a single first-order binary operation $\mathsf{app}$ (for example, $\Sigma_{\mathsf{LC}}$). The three term-pairs, one for each hypothesis and one for the conclusion, define the desired reduction rule.

Now, we explain in which sense such a rule can be validated in a reduction monad $R$: intuitively, it means that for any set $X$, any quadrupe $(T, T', U, U') \in \mathcal{V}(R)$, any reductions $s : T \blacktriangleright T'$ and $t : U \blacktriangleright U'$, there is a reduction $\mathsf{app\text{-}cong}(s, t) : \mathsf{app}(T, U) \blacktriangleright \mathsf{app}(T', U')$. Of course, this only makes sense if the monad $\underline{R}$ underlying the reduction monad is equipped with an application operation, that is, with an operation of $\Sigma_{\mathsf{app}}$. We call such a structure a *reduction* $\Sigma_{\mathsf{app}}$*-monad*.

### 5.3.2 Definition of reduction rules

In this subsection, $\Sigma$ is a signature for monads. We present our notion of *reduction rule over* $\Sigma$, from which we build *reduction signatures* in Section 5.4.

We begin with the definition of *term-pair*, alluded to already in Section 5.3.1:

**Definition 133.** Given a $\Sigma$-module $\mathcal{V}$, a **term-pair from** $\mathcal{V}$ is a pair $(n, p)$ of a natural number $n$ and a morphism of $\Sigma$-modules $p : \mathcal{V} \to \Theta^{(n)} \times \Theta^{(n)}$.

Many term-pairs are of a particularly simple form, namely a pair of projections, which intuitively picks two among the available metavariables. Because of their ubiquity, we introduce the following notation:

**Definition 134.** Let $n_1$, ..., $n_p$ be a list of natural numbers. For $i, j \in \{1, \ldots p\}$, we define the pair projection $\pi_{i,j}$ and the projection $\pi_i$ as the following $\Sigma$-module morphisms, for any signature $\Sigma$:

$$\begin{aligned} \pi_{i,j} &: \Theta^{(n_1)} \times \ldots \Theta^{(n_p)} \to \Theta^{(n_i)} \times \Theta^{(n_j)} & \pi_{i,j,R,X}(T_1, \ldots, T_p) &= (T_i, T_j) \\ \pi_i &: \Theta^{(n_1)} \times \ldots \Theta^{(n_p)} \to \Theta^{(n_i)} & \pi_{i,R,X}(T_1, \ldots, T_p) &= T_i \end{aligned}$$

Some term-pairs, such as the conclusions of the congruence rules for application and abstraction, are more complicated: intuitively, they are constructed from term constructions applied to metavariables.

**Example 135** (term-pair of the conclusion of the congruence for application)**.** The term-pair corresponding to the conclusion $\mathsf{app}(T, U) \rightsquigarrow \mathsf{app}(T', U')$ of congruence for application (Section 5.3.1) is given by $(0, c)$, on the $\Sigma_{\mathsf{LC}}$-module $\Theta^4$. Here, we have

$$\begin{aligned} &c : \mathcal{V} \to \Theta \times \Theta \\ &c_{R,X}(T, T', U, U') := \Big(\mathsf{app}_{R,X}(T, U), \mathsf{app}_{R,X}(T', U')\Big) \end{aligned}$$

More schematically:

$$c : \Theta^4 \xrightarrow{\mathsf{app} \circ \pi_{1,3}, \mathsf{app} \circ \pi_{2,4}} \Theta \times \Theta$$

We now give our definition of *reduction rule*, making precise the intuition developed in Section 5.3.1.

**Definition 136.** A **reduction rule** $\mathcal{A} = (\mathcal{V}, (n_i, h_i)_{i \in I}, (n, c))$ **over** $\Sigma$ is given by:

- Metavariables: a $\Sigma$-module $\mathcal{V}$ of metavariables, that we sometimes denote by $\mathsf{MVar}_{\mathcal{A}}$;
- Hypotheses: a finite family of term-pairs $(n_i, h_i)_{i \in I}$ from $\mathcal{V}$;
- Conclusion: a term-pair $(n, c)$ from $\mathcal{V}$.

**Example 137** (Reduction rule for congruence of application)**.** The reduction rule $\mathcal{A}_{\mathsf{app\text{-}cong}}$ for congruence of application (Section 5.3.1) is defined as follows:

- Metavariables: $\mathcal{V} = \Theta^4$ for the four metavariables $T$, $T'$, $U$, and $U'$;
- Hypotheses: Given by two term-pairs $(0, h_1)$ and $(0, h_2)$:

$$h_1 : \Theta^4 \xrightarrow{\pi_{1,2}} \Theta \times \Theta \qquad h_2 : \Theta^4 \xrightarrow{\pi_{3,4}} \Theta \times \Theta$$

- Conclusion: Given by the term-pair $(0, c)$ of Example 135.

More examples of reduction rules are given in Section 5.3.6.

### 5.3.3 Reduction $\Sigma$-monads

As already said, the purpose of a reduction rule is to be validated in a reduction monad $R$. However, as the hypotheses or the conclusion of the reduction rule may refer to some operations specified by a signature $\Sigma$ for monads, this reduction monad $R$ must be equipped with an action of $\Sigma$, hence the following definition:

**Definition 138.** Let $\Sigma$ be a signature for monads. The **category** $\mathsf{RedMon}^{\Sigma}$ **of reduction $\Sigma$-monads** is the category of models of the pullback of $\Sigma$ along the forgetful functor $U : \mathsf{RedMon} \to \mathsf{Mon}$ (see Chapter 2).

In other words, it is defined as the following pullback:

$$\begin{array}{ccc} \mathsf{RedMon}^{\Sigma} & \longrightarrow & \mathsf{RedMon} \\ \downarrow & \lrcorner & \downarrow \\ \mathsf{Mon}^{\Sigma} & \longrightarrow & \mathsf{Mon} \end{array}$$

More concretely,

- a **reduction $\Sigma$-monad** is a reduction monad $R$ equipped with an **action** $\rho$ of $\Sigma$ in $R$, thus inducing a $\Sigma$-monad that we denote also by $R$, or by $\underline{R}$ when we want to be explicit;

- a **morphism of reduction $\Sigma$-monads** $R \to S$ is a morphism $f : R \to S$ of reduction monads compatible with the action of $\Sigma$, i.e, whose underlying monad morphism is a $\Sigma$-monad morphism.

### 5.3.4 Action of a reduction rule

Let $\Sigma$ be a signature for monads. In this section, we introduce the notion of *action of a reduction rule over $\Sigma$ in a reduction $\Sigma$-monad*. Intuitively, such an action is a "map from the hypotheses to the conclusion" of the reduction rule. To make this precise, we need to first take the product of the hypotheses; this product is, more correctly, a *fibred* product.

**Definition 139.** Let $(n, p)$ be a term-pair from a $\Sigma$-module $\mathcal{V}$, and $R$ be a reduction $\Sigma$-monad. We denote by $p^*(\mathsf{Red}(R)^{(n)})$ the pullback of $\mathsf{red}_R^{(n)} : \mathsf{Red}(R)^{(n)} \to R^{(n)} \times R^{(n)}$ along $p_R : \mathcal{V}(R) \to R^{(n)} \times R^{(n)}$:

$$\begin{array}{ccc} p^*(\mathsf{Red}(R)^{(n)}) & \longrightarrow & \mathsf{Red}(R)^{(n)} \\ \big\downarrow & \lrcorner & \big\downarrow \mathsf{red}_R^{(n)} \\ \mathcal{V}(R) & \xrightarrow[p_R]{} & R^{(n)} \times R^{(n)} \end{array}$$

We denote by $p^*(\mathsf{red}_R^{(n)}) : p^*(\mathsf{Red}(R)^{(n)}) \to \mathcal{V}(R)$ the projection morphism on the left.

**Definition 140.** Let $\mathcal{A} = (\mathcal{V}, (n_i, h_i)_{i \in I}, (n, c))$ be a reduction rule, and $R$ be a reduction $\Sigma$-monad. The **$R$-module $\mathsf{Hyp}_{\mathcal{A}}(R)$ of hypotheses of $\mathcal{A}$** is $\prod_{i \in I / \mathcal{V}(R)} h_i^* \mathsf{Red}(R)^{(n_i)}$, i.e., the fiber product of all the $R$-modules $h_i^* \mathsf{Red}(R)^{(n_i)}$ along their projection to $\mathcal{V}(R)$. It thus comes with a projection $\mathsf{hyp}_{\mathcal{A}}(R) : \mathsf{Hyp}_{\mathcal{A}}(R) \to \mathcal{V}(R)$

The **$R$-module $\mathsf{Con}_{\mathcal{A}}(R)$ of conclusion of $\mathcal{A}$** is $c^* \mathsf{Red}(R)^{(n)}$, and comes with a projection $\mathsf{con}_{\mathcal{A}}(R) : \mathsf{Con}_{\mathcal{A}}(R) \to \mathcal{V}(R)$.

**Example 141.** Let $R$ be a reduction $\Sigma_{\mathsf{LC}}$-monad. The $R$-module of conclusion of the congruence reduction rule for application (Example 137) maps a set $X$ to the set of quintuples $(T, T', U, U', m)$ where $(T, T', U, U') \in R^4(X)$ and $m$ is a reduction $m : \mathsf{app}(T, U) \blacktriangleright \mathsf{app}(T', U')$.

The $R$-module of hypotheses of this reduction rule maps a set $X$ to the set of sextuples $(T, T', U, U', m, n)$ where $(T, T', U, U') \in R^4(X)$, $m : T \blacktriangleright T'$, and $n : U \blacktriangleright U'$.

**Definition 142.** Let $\mathcal{A}$ be a reduction rule over $\Sigma$. An **action of $\mathcal{A}$ in a reduction $\Sigma$-monad** $R$ is a morphism in the slice category $\mathsf{Mod}(R)/\mathsf{MVar}_{\mathcal{A}}(R)$ between $\mathsf{hyp}_{\mathcal{A}}(R)$ and $\mathsf{con}_{\mathcal{A}}(R)$, that is, a morphism of $R$-modules

$$\tau : \mathsf{Hyp}_{\mathcal{A}}(R) \to \mathsf{Con}_{\mathcal{A}}(R)$$

making the following diagram commute:

$$\begin{array}{ccc} \mathsf{Hyp}_{\mathcal{A}}(R) & \xrightarrow{\quad\tau\quad} & \mathsf{Con}_{\mathcal{A}}(R) \\ & \searrow \quad \swarrow & \\ & \mathsf{MVar}_{\mathcal{A}}(R) & \end{array} \tag{5.1}$$

**Example 143** (Action of the congruence rule for application)**.** Consider the reduction rule of the congruence for application of Example 137. Let $R$ be a reduction $\Sigma_{\mathsf{LC}}$-monad $R$. An action in $R$ is a $R$-module morphism mapping, for each set $X$, a sextuple $(T, T', U, U', r, s)$ with $r : T \blacktriangleright T'$ and $s : U \blacktriangleright U'$ to a quintuple $(A, A', B, B', m)$ with $m : \mathsf{app}(A, B) \blacktriangleright \mathsf{app}(A', B')$. The commutation of the triangle (5.1) ensures that $(A, A', B, B') = (T, T', U, U')$.

Alternatively (as justified formally by Lemma 157), an action is a morphism mapping the same sextuple to a reduction $m : \mathsf{app}(T, U) \blacktriangleright \mathsf{app}(T', U')$.

**Remark 144.** Any reduction rule $\mathcal{A}$ over $\Sigma$ induces an arity $(\mathsf{D}, a, u, v)$ over $\mathsf{RedMon}^{\Sigma}$ defined as follows:

- $a : \mathsf{D} \to \mathsf{RedMon}^{\Sigma}$ is the Grothendieck fibration corresponding (through the Grothendieck construction) to the functor mapping a reduction $\Sigma$-monad $R$ to the category $\mathsf{Mod}(R)/\mathsf{MVar}_{\mathcal{A}}(R)$;
- $u$ maps a reduction $\Sigma$-monad $R$ to $\mathsf{hyp}_{\mathcal{A}}(R) : \mathsf{Hyp}_{\mathcal{A}}(R) \to \mathsf{MVar}_{\mathcal{A}}(R)$
- $v$ maps a reduction $\Sigma$-monad $R$ to $\mathsf{con}_{\mathcal{A}}(R) : \mathsf{Con}_{\mathcal{A}}(R) \to \mathsf{MVar}_{\mathcal{A}}(R)$

Then, the notion of action of this induced arity (see Chapter 2) coincides with the one introduced in this section.

### 5.3.5 Protocol for specifying reduction rules

In Section 5.3.6, we adopt the following schematic presentation of a reduction rule over a signature $\Sigma$:

$$\frac{s_1(T_1,\ldots,T_q) \rightsquigarrow t_1(T_1,\ldots,T_q) \qquad \ldots \qquad s_n(T_1,\ldots,T_q) \rightsquigarrow t_n(T_1,\ldots,T_q)}{s_0(T_1,\ldots,T_q) \rightsquigarrow t_0(T_1,\ldots,T_q)}$$

where $s_i$ and $t_i$ are expressions depending on metavariables $T_1$, ..., $T_q$. Each pair $(s_i, t_i)$ defines a term-pair as follows:

$$p_i : M_1 \times \cdots \times M_q \to \Theta^{(m_i)} \times \Theta^{(m_i)}$$
$$p_{i,R,X}(T_1,\ldots,T_q) := (s_i(T_1,\ldots,T_q), t_i(T_1,\ldots,T_q)) \qquad (5.2)$$

where the $\Sigma$-modules $M_1$, ..., $M_q$, and the natural numbers $m_0$, ..., $m_n$ are inferred for Equation (5.2) to be well defined for all $i \in \{0,\ldots,n\}$.

The induced reduction rule is:

- Metavariables: the $\Sigma$-module of metavariables is $\mathcal{V} = M_1 \times \cdots \times M_q$;
- Hypotheses: the hypotheses are the term-pairs $(m_i, p_i)_{i\in\{1,\ldots,n\}}$;
- Conclusion: the conclusion is the term-pair $(m_0, p_0)$.

Typically, $M_i = \Theta^{(n_i)}$ for some natural number $n_i$, as in the examples that we consider in this section. In practice, there are several choices for building the reduction rule out of such a schematic presentation, depending on the order in which the metavariables are picked. This order is irrelevant: the different possible versions of reduction rules are all equivalent, in the sense that taking one or the other as part of a reduction signature yields isomorphic category of models.

### 5.3.6 Examples of reduction rules

This section collects a list of motivating examples of reduction rules.

For the rest of this section, we assume that we have fixed a signature for monads $\Sigma$. Figure 5.1 shows some notable examples of reduction rules. In order, they are: reflexivity, transitivity, congruence for $\mathsf{abs}$, $\beta$-reduction, $\eta$-expansion, and expansion of the fixpoint operator.

$$\frac{}{T \rightsquigarrow T}\ \mathsf{Refl} \qquad \frac{T \rightsquigarrow U \quad U \rightsquigarrow W}{T \rightsquigarrow W}\ \mathsf{Trans} \qquad \frac{T \rightsquigarrow U}{\mathsf{abs}(T) \rightsquigarrow \mathsf{abs}(U)}\ \mathsf{abs\text{-}Cong}$$

$$\frac{}{\mathsf{app}(\mathsf{abs}(T), U) \rightsquigarrow T\{* := U\}}\ \beta\text{-}\mathsf{Red} \qquad \frac{}{T \rightsquigarrow \mathsf{abs}(\iota(T))}\ \eta\text{-}\mathsf{Exp} \qquad \frac{}{\mathsf{fix}(T) \rightsquigarrow T\{* := \mathsf{fix}(T)\}}\ \mathsf{fix\text{-}Exp}$$

Figure 5.1: Examples of reduction rules.

| Rule | Signature | Metavariables | Hypothesis | Conclusion |
|---|---|---|---|---|
| Refl | any | $\Theta$ | | $(0, (\mathtt{id}, \mathtt{id}))$ |
| Trans | any | $\Theta \times \Theta \times \Theta$ for $(T, U, W)$ | $(0, \pi_{1,2})$, $(0, \pi_{2,3})$ | $(0, \pi_{1,3})$ |
| abs-Cong | $\Sigma_{\mathsf{LC}}$ | $\Theta' \times \Theta'$ for $(T, U)$ | $(1, \mathtt{id})$ | $(0, \mathsf{abs} \times \mathsf{abs})$ |
| $\beta$-Red | $\Sigma_{\mathsf{LC}}$ | $\Theta' \times \Theta$ for $(T, U)$ | | $(0, c_{\beta\text{-Red}})$ |
| $\eta$-Exp | $\Sigma_{\mathsf{LC}}$ | $\Theta$ | | $(0, (\mathtt{id}, \mathsf{abs} \circ \iota))$ |
| fix-Exp | $\Sigma_{\mathsf{fix}}$ | $\Theta'$ | | $(0, c_{\mathsf{fix\text{-}Exp}})$ |

$$c_{\beta\text{-Red},R,X}(T, U) = (\mathsf{app}(\mathsf{abs}(T), U), T\{* := U\})$$
$$c_{\mathsf{fix\text{-}Exp},R,X}(T) = (\mathsf{fix}(T), T\{* := \mathsf{fix}(T)\})$$

Figure 5.2: Modules and term pairs relative to the reduction rules of Figure 5.1.

For the example of the fixpoint operator (rule fix-Exp), we consider the 1-signature $\Sigma_{\mathsf{fix}}$, as described in Section 4.4.4. (without enforcing the fixpoint equation, which is replaced here by the reduction rule under consideration). A $\Sigma_{\mathsf{fix}}$-monad is a monad $R$ equipped with an $R$-module morphism $\mathsf{fix} : R' \to R$.

Figure 5.2 lists the modules and term pairs for hypothesis and conclusion of each of these reduction rules. There, $\pi_{i,j}$ designates the pair projection described in Definition 134.

## 5.4 Signatures for reduction monads and Initiality

In this section, we define *reduction signatures* which are particular signatures over the category of reduction monads (see Section 5.4.1). They consist of a signature for monads $\Sigma$ (more precisely, a pullback of such a signature along the forgetful functor from reduction monads to monads) and a family of arities induced by reduction rules over $\Sigma$ (as explained in Remark 144). Our main result, Theorem 150 (see Section 5.4.3), states that a reduction signature is effective as soon as its underlying signature for monads is effective.

### 5.4.1 Reduction signatures

We define here *reduction signatures* and their *models*.

**Definition 145.** A **reduction signature** is a signature over the category of reduction monads, consisting of:

- the pullback of a signature for monads $\Sigma$ along the forgetful functor from reduction monads to monads;
- a family of arities induced by reduction rules over $\Sigma$ (as explained in Remark 144).

A reduction signature is thus determined by a signature for monads $\Sigma$ and a family of reduction rules $\mathfrak{R}$ over $\Sigma$. We denote $(\Sigma, \mathfrak{R})$ the induced reduction signature.

We unfold the definition of model for reduction signatures:

**Remark 146.** Given a reduction monad $R$ and a reduction signature $\mathcal{S} = (\Sigma, \mathfrak{R})$, an **action of $\mathcal{S}$ in $R$** consists of an action of $\Sigma$ in its underlying monad $\underline{R}$ and an action of each reduction rule of $\mathfrak{R}$ in $R$. A **model of $\mathcal{S}$** is a reduction monad equipped with an action of $\mathcal{S}$, or equivalently, a reduction $\Sigma$-monad equipped with an action of each reduction rule of $\mathfrak{R}$.

### 5.4.2 Morphisms of models

Here we unfold the definition of morphism between models of a reduction signature: it relies on the functoriality of the assignments $R \mapsto \mathsf{Hyp}_{\mathcal{A}}(R)$ and $R \mapsto \mathsf{Con}_{\mathcal{A}}(R)$, for a given reduction rule $\mathcal{A}$ on a signature $\Sigma$ for monads.

**Definition 147.** Let $\Sigma$ be a signature for monads, and $\mathcal{A}$ be a reduction rule over $\Sigma$. Definition 140 assigns to each $\Sigma$-monad $R$ the $R$-modules $\mathsf{Hyp}_{\mathcal{A}}(R)$ and $\mathsf{Con}_{\mathcal{A}}(R)$. These assignments extend to functors $\mathsf{Hyp}_{\mathcal{A}}, \mathsf{Con}_{\mathcal{A}} : \mathsf{RedMon}^{\Sigma} \to \int \mathsf{Mod}$.

**Proposition 148.** *Given the same data, the functors* $\mathsf{Hyp}_{\mathcal{A}}$ *and* $\mathsf{Con}_{\mathcal{A}}$ *commute with the forgetful functors to* $\mathsf{Mon}$*:*

$$\begin{array}{ccc} \mathsf{RedMon}^{\Sigma} & \overset{\mathsf{Hyp}_{\mathcal{A}}}{\underset{\mathsf{Con}_{\mathcal{A}}}{\rightrightarrows}} & \int \mathsf{Mod} \\ & \searrow \quad \swarrow & \\ & \mathsf{Mon} & \end{array}$$

Let $\mathcal{S} = (\Sigma, \mathfrak{R})$ be a reduction signature. As usual, we denote $\mathsf{RedMon}^{\mathcal{S}}$ the category of models of $\mathcal{S}$. More concretely, a morphism between models $R$ and $T$ of $\mathcal{S}$ is a morphism $f$ of reduction $\Sigma$-monads commuting with the action of any reduction rule, in the sense that for any reduction rule $\mathcal{A} \in \mathfrak{R}$, the following diagram of natural transformations commutes:

$$\begin{array}{ccc} \mathsf{Hyp}_{\mathcal{A}}(R) & \longrightarrow & \mathsf{Con}_{\mathcal{A}}(R) \\ {\scriptstyle \mathsf{Hyp}_{\mathcal{A}}(f)}\downarrow & & \downarrow{\scriptstyle \mathsf{Con}_{\mathcal{A}}(f)} \\ \mathsf{Hyp}_{\mathcal{A}}(T) & \longrightarrow & \mathsf{Con}_{\mathcal{A}}(T) \end{array}$$

**Example 149** (Example 143 continued)**.** Consider the reduction signature consisting of the signature $\Sigma_{\mathsf{app}}$ of a binary operation $\mathsf{app}$ and a single reduction rule of congruence for application (Example 137).

Let $R$ and $T$ be models for this signature: they are reduction $\Sigma_{\mathsf{app}}$-monads equipped with an action $\rho$ and $\tau$, in the alternative sense of Example 143. A $\Sigma_{\mathsf{app}}$-monad morphism $(f, \alpha)$ between $R$ and $T$ is a model morphism if, for any set $X$, any sextuple $(A, A', B, B', m, n)$ where $(A, A', B, B') \in R^4(X)$, $m : A \blacktriangleright A'$, and $n : B \blacktriangleright B'$, the reduction $\rho(A, A', B, B') : \mathsf{app}(A, B) \blacktriangleright \mathsf{app}(A', B')$ is mapped to the reduction $\tau(f(A), f(A'), f(B), f(B'))$ by $\alpha : \mathsf{Red}(R) \to \mathsf{Red}(T)$.

### 5.4.3 The main result

We state our main result, Theorem 150, which gives a sufficient condition for $\mathcal{S}$ to be effective.

**Theorem 150.** *Let* $(\Sigma, \mathfrak{R})$ *be a reduction signature. If* $\Sigma$ *is effective, then so is* $(\Sigma, \mathfrak{R})$*.*

The proof of this theorem is given in Section 5.5.

Theorem 107 entails the following corollary:

**Corollary 151.** *Let $(\Sigma, \mathfrak{R})$ be a reduction signature. If $\Sigma$ is an algebraic 2-signature (Definition 106), then $(\Sigma, \mathfrak{R})$ is effective.*

All the examples of reduction signatures considered here satisfy the condition of Corollary 151.

**Example 152** (Reduction signature for Example 130)**.** The reduction monad of Example 130 is generated by the reduction signature $\mathcal{S}_{\mathsf{LC}_{\beta/\eta}}$ that is given by the signature $\Sigma_{\mathsf{LC}}$ together with the following reduction rules (see Section 5.3.6):

- the reduction rule for $\beta$-reduction;
- the reduction rule for $\eta$-expansion;
- the congruence rule for abstraction;
- two unary congruence rules for application:

$$\frac{T \rightsquigarrow T'}{\mathsf{app}(T,U) \rightsquigarrow \mathsf{app}(T',U)} \qquad \frac{U \rightsquigarrow U'}{\mathsf{app}(T,U) \rightsquigarrow \mathsf{app}(T,U')}$$

**Remark 153** (Continuation of Remark 119)**.** Just as our reduction monads are "proof-relevant" (cf. Remark 119), our notion of reduction signature allows for the specification of multiple reductions between terms. As a trivial example , duplicating the $\beta$-rule in the signature $\mathcal{S}_{\mathsf{LC}_{\beta/\eta}}$ yields two distinct $\beta$-reductions in the initial model.

**Example 154** (Reduction signature of lambda calculus with a fixpoint operator)**.** The signature $\mathcal{S}_{\mathsf{LC}_{\mathsf{fix}}}$ specifying the reduction monad $\mathsf{LC}_{\mathsf{fix}}$ of the lambda calculus with a fixpoint operator extends the signature $\mathcal{S}_{\mathsf{LC}_{\beta/\eta}}$ of Example 152 with:

- a new operation $\mathsf{fix} : \Theta' \to \Theta$ (thus extending the signature for monads $\Sigma_{\mathsf{LC}}$);
- the reduction rule for the fixpoint reduction (cf. Section 5.3.6);
- a congruence rule for fix:

$$\frac{T \rightsquigarrow T'}{\mathsf{fix}(T) \rightsquigarrow \mathsf{fix}(T')}.$$

## 5.5 Proof of Theorem 150

This section details the proof of Theorem 150.

Let $\mathcal{S} = (\Sigma, (\mathcal{A}_i)_{i\in I}))$ be a reduction signature. We denote by $\mathcal{U}^\Sigma$ the forgetful functor from the category of reduction $\Sigma$-monads to the category of $\Sigma$-monads.

In Section 5.5.1, we first reduce to the case of reduction rules $(\mathcal{V}, (n_j, h_j)_{j\in J}, (n, c))$ for which $n = 0$, that we call *normalized*. Then, in Section 5.5.2, we give an alternative definition of the category of models that we make use of in the proof of effectivity, in Section 5.5.3.

### 5.5.1 Normalizing reduction rules

**Definition 155.** A reduction rule $(\mathcal{V}, (n_j, h_j)_{j\in J}, (n, c))$ is said to be **normalized** if $n = 0$.

**Lemma 156.** *Let $\mathcal{A} = (\mathcal{V}, (n_j, h_j)_{j\in J}, (n, c))$ be a reduction rule over $\Sigma$. Then there exists a normalized reduction rule $\mathcal{A}'$ over $\Sigma$ such that the induced notion of action is equivalent, in the sense that:*

- *for a reduction $\Sigma$-monad $R$, there is a bijection between actions of $\mathcal{A}$ in $R$ and actions of $\mathcal{A}'$ in $R$;*
- *a morphism between reduction $\Sigma$-monads equipped with an action of $\mathcal{A}$ preserves the action (in the sense of Section 5.4.2) if and only if it preserves the corresponding action of $\mathcal{A}'$ through the bijection.*

Before tackling the proof, we give an alternative definition of action and model morphism:

**Lemma 157.** *Let $\mathcal{A} = (\mathcal{V}, (n_i, h_i)_{i\in I}, (n, c))$ be a reduction rule over $\Sigma$. By universal property of the pullback $\mathsf{Con}_{\mathcal{A}}(R) = c^*\mathsf{Red}(R)^{(n)}$, an action can be alternatively be defined as an $R$-module morphism $\sigma : \mathsf{Hyp}_{\mathcal{A}}(R) \to \mathsf{Red}(R)^{(n)}$ making the following diagram commute*

$$\begin{array}{ccc} \mathsf{Hyp}_{\mathcal{A}}(R) & \xrightarrow{\sigma} & \mathsf{Red}(R)^{(n)} \\ \downarrow & & \downarrow {\mathsf{red}_R}^{(n)} \\ \mathcal{V}(R) & \xrightarrow[c]{} & R^{(n)} \times R^{(n)} \end{array} \quad . \tag{5.3}$$

**Lemma 158.** *Using this alternative definition of action, a morphism between models $R$ and $T$ of a reduction signature $\mathcal{S} = (\Sigma, \mathfrak{R})$ is a morphism $f$ of reduction $\Sigma$-monads making the following diagram commute, for any reduction rule $\mathcal{A} = (\mathcal{V}, (n_i, l_i, r_i)_{i\in I}, (n, l, r))$ of $\mathfrak{R}$:*

$$\begin{array}{ccc} \mathsf{Hyp}_{\mathcal{A}}(R) & \longrightarrow & \mathsf{Red}(R)^{(n)} \\ {\scriptstyle \mathsf{Hyp}_{\mathcal{A}}(f)}\downarrow & & \downarrow{\scriptstyle \mathsf{Red}(f)^{(n)}} \\ \mathsf{Hyp}_{\mathcal{A}}(T) & \longrightarrow & \mathsf{Red}(T)^{(n)} \end{array}$$

We now prove Lemma 156 using these alternative definitions:

*Proof of Lemma 156.* The reduction rule $\mathcal{A}' = (\mathcal{V}', (n_j, h'_j)_{j\in J}, (0, c'))$ is defined as follows:

- Metavariables: $\mathcal{V}' = \mathcal{V} \times \Theta^n$
- Hypotheses: For each $j \in J$, $h'_j : \mathcal{V}' \to \Theta^{(n_j)} \times \Theta^{(n_j)}$ is defined as the composition of $\pi_1 : \mathcal{V} \times \Theta^n \to \mathcal{V}$ with $h_j : \mathcal{V} \to \Theta^{(n_j)} \times \Theta^{(n_j)}$.
- Conclusion: The morphism $c' : \mathcal{V} \times \Theta^n \to \Theta \times \Theta$ is the $n^{th}$ uncurrying (see Definition 85) of $c : \mathcal{V} \to \Theta^{(n)} \times \Theta^{(n)}$.

Now, consider an action for the reduction rule $\mathcal{A}$ in a reduction $\Sigma$-monad $R$: it is an $R$-module morphism $\tau : \mathsf{Hyp}_{\mathcal{A}}(R) \to \mathsf{Red}(R)^{(n)}$ such that the following square commutes:

$$\begin{array}{ccc} \mathsf{Hyp}_{\mathcal{A}}(R) & \xrightarrow{\tau} & \mathsf{Red}(R)^{(n)} \\ \downarrow & & \downarrow{\scriptstyle \mathsf{red}_R^{(n)}} \\ \mathcal{V}(R) & \xrightarrow[c]{} & R^{(n)} \times R^{(n)} \end{array} .$$

Equivalently, through the adjunction mentioned above, it is given by an $R$-module morphism $\tau^* : \mathsf{Hyp}_R \times R^m \to M$ such that the following diagram commutes:

$$\begin{array}{ccc} \mathsf{Hyp}_{\mathcal{A}}(R) \times R^n & \xrightarrow{\tau^*} & \mathsf{Red}(R) \\ \downarrow & & \downarrow{\scriptstyle \mathsf{red}_R} \\ \mathcal{V}(R) \times R^n & \xrightarrow[c^*]{} & R \times R \end{array}$$

This is exactly the definition of an action of $\mathcal{A}'$. It is then straightforward to check that one action is preserved by a reduction monad morphism if and only if the other one is. $\square$

**Corollary 159.** *For each reduction signature, there exists a reduction signature yielding an isomorphic category of models and whose underlying reduction rules are all normalized.*

*Proof.* Just replace each reduction rule with the one given by Lemma 156. □

Thanks to this lemma, we assume in the following that all the reduction rules of the given signature $\mathcal{S}$ are normalized.

### 5.5.2 Models as vertical algebras

In this section, we give an alternative definition for the category of models of $\mathcal{S}$ that is convenient in the proof of effectivity.

First we rephrase the notion of action of a reduction rule as an algebra structure for a suitably chosen endofunctor. Indeed, an action of a normalized reduction rule $\mathcal{A} = (\mathcal{V}, (n_j, h_j)_{j\in J}, (0, c))$ in a reduction $\Sigma$-monad $R$ is given by a $R$-module morphism $\tau : \mathsf{Hyp}_{\mathcal{A}}(R) \to \mathsf{Red}(R)$ such that the following square commutes:

$$\begin{array}{ccc}
\mathsf{Hyp}_{\mathcal{A}}(R) & \xrightarrow{\tau} & \mathsf{Red}(R) \\
\downarrow & & \downarrow{\scriptstyle \mathsf{red}_R} \\
\mathcal{V}(R) & \xrightarrow[p]{} & R \times R
\end{array}$$

We can rephrase this commutation by stating that this morphism $\tau$ is a morphism in the slice category $\mathsf{Mod}(\underline{R})/\underline{R}^2$ from an object that we denote by $F_{\mathcal{A}|\underline{R}}(\mathsf{Red}(R), \mathsf{red}_R)$, to $(\mathsf{Red}(R), \mathsf{red}_R)$. Here, we use the notation $\underline{R}$ to refer explicitly to the underlying monad of $R$. Actually, the domain is functorial in its argument, and thus the action $\tau$ can be thought of as an algebra structure on $(\mathsf{Red}(R), \mathsf{red}_R)$:

**Lemma 160.** *Given any $\Sigma$-monad $R$, the assignment $(M, p : M \to R \times R) \mapsto F_{\mathcal{A}|R}(M, c)$ yields an endofunctor $F_{\mathcal{A}|R}$ on $\mathsf{Mod}(R)/R^2$. An action of $\mathcal{A}$ in a reduction $\Sigma$-monad $R$ is exactly the same as an algebra structure for this endofunctor on $(\mathsf{Red}(R), \mathsf{red}_R) \in \mathsf{Mod}(R)/R^2$.*

*Furthermore, the assignment $R \mapsto F_{\mathcal{A}|\underline{R}}(\mathsf{Red}(R), \mathsf{red}_R)$ yields an endofunctor $F_{\mathcal{A}}$ on the category of reduction $\Sigma$-monads. This functor preserves the underlying $\Sigma$-monad, in the sense that $\mathcal{U}^\Sigma \cdot F_{\mathcal{A}} = \mathcal{U}^\Sigma$.*

*Proof.* This is a consequence of the functoriality of $\mathsf{Hyp}_{\mathcal{A}}$, as noticed in Section 5.4.2. □

Now, we give our alternative definition of the category of models:

**Proposition 161.** *Let $F_{\mathcal{S}} : \mathsf{RedMon}^{\Sigma} \to \mathsf{RedMon}^{\Sigma}$ be the coproduct $\coprod_i F_{\mathcal{A}_i}$. Then, the category of models of $\mathcal{S}$ is isomorphic to the* ***category of vertical algebras*** *of $F_{\mathcal{S}}$ defined as follows:*

- *an object is an algebra $r : F_{\mathcal{S}}(R) \to R$ such that $r$ is mapped to the identity by $\mathcal{U}^{\Sigma}$*
- *morphisms are the usual $F_{\mathcal{S}}$-algebra morphisms.*

We adopt this definition in the following. We show now a property of the category of models that will prove useful in the proof of effectivity:

**Lemma 162.** *The forgetful functor from the category of models of $\mathcal{S}$ to the category of $\Sigma$-monads is a fibration.*

The proof relies on some additional lemmas, in particular the following one, that we will specialize by taking $p = \mathcal{U}^{\Sigma}$ (requiring to show that $\mathcal{U}^{\Sigma} : \mathsf{RedMon}^{\Sigma} \to \mathsf{Mon}^{\Sigma}$ is a fibration) and $F = F_{\mathcal{S}}$:

**Lemma 163.** *Let $p : E \to B$ be a fibration and $F$ an endofunctor on $E$ satisfying $p \cdot F = p$. Then the category of vertical algebras of $F$ is fibered over $B$.*

*Proof.* Let $r : F(R) \to R$ be an algebra over $X \in B$. Let $a : Y \to X$ be a morphism in $B$. Let $\overline{a} : a^*R \to R$ be the associated cartesian morphism in $E$. We define the reindexing of $r$ along $a$ as follows: the base object is $a^*R$, and the algebra structure $\rho : F(a^*R) \to R$ is given by the unique morphism which factors $F(a^*R) \xrightarrow{F(\overline{a})} F(R) \xrightarrow{r} R$ through the cartesian morphism $\overline{a} : a^*R \to R$. Thus, the square

$$\begin{array}{ccc} F(a^*R) & \xrightarrow{F(\overline{a})} & F(R) \\ {\scriptstyle\rho}\downarrow & & \downarrow{\scriptstyle r} \\ a^*R & \xrightarrow[\overline{a}]{} & R \end{array}$$

commutes, so $\overline{a}$ is a morphism of algebras between $\rho$ and $r$. Next, we prove that it is a cartesian morphism: let $s : F(S) \to S$ be a vertical algebra over an object $Z$ of $B$, and $v : s \to r$ be a morphism of algebras such that there exists $b : Z \to Y$ such that $p(v) = Z \xrightarrow{b} Y \xrightarrow{a} X$ . We need to show that there exists a unique algebra morphism $w : s \to \rho$ such that $v = \overline{a} \circ w$ and $p(w) = b$.

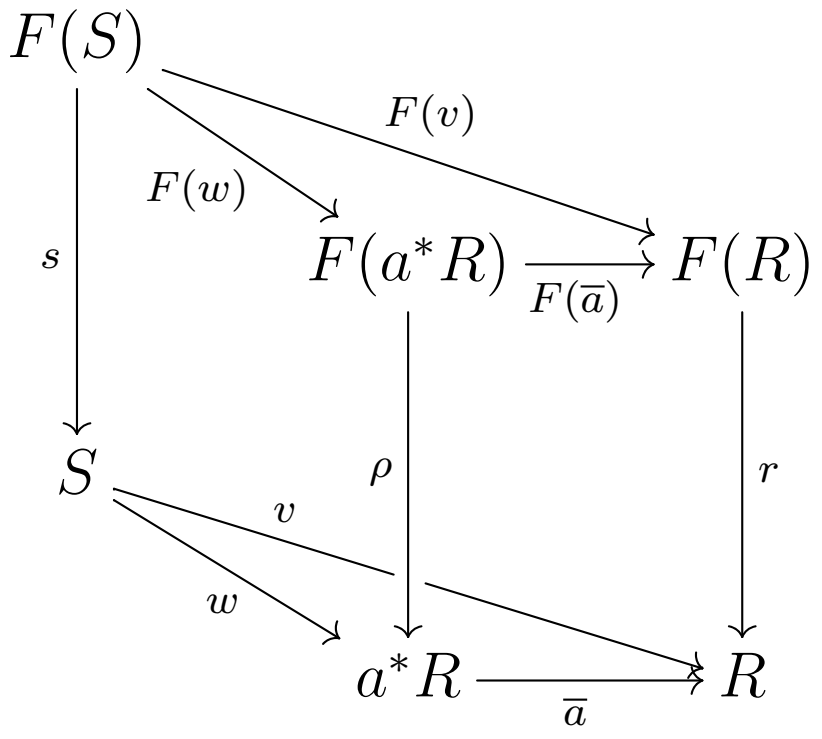

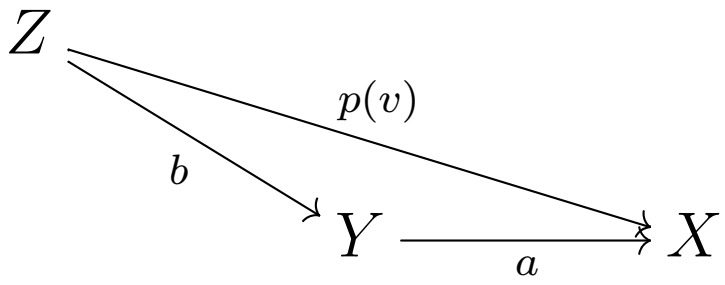

Uniqueness follows from the fact that $\overline{a}$ is cartesian for the fibration $p : E \to B$. Moreover, as $\overline{a}$ is cartesian, we get a morphism $w : S \to a^*R$. We turn it into an algebra morphism by showing that the following square commutes:

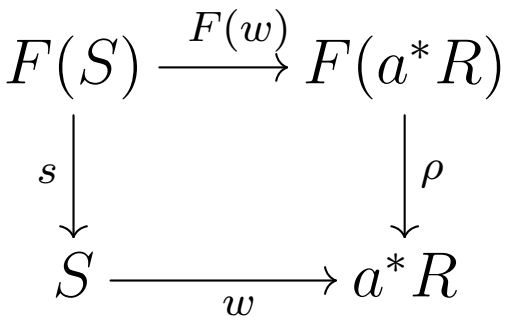

As $\overline{a}$ is cartesian and both $w$ and $F(w)$ are sent to $b$ by $p$, it is enough to show equalities of both morphisms after postcomposing with $\overline{a}$. The fact that $v$ is an algebra morphism allows us to conclude. □

We want to apply this lemma for proving Lemma 162. We thus need to show that $\mathcal{U}^\Sigma : \mathsf{RedMon}^\Sigma \to \mathsf{Mon}^\Sigma$ is a fibration:

**Lemma 164.** *The forgetful functors* $\mathsf{RedMon} \to \mathsf{Mon}$ *and* $\mathcal{U}^\Sigma : \mathsf{RedMon}^\Sigma \to \mathsf{Mon}^\Sigma$ *are fibrations.*

*Proof.* We have the two following pullbacks:

$$\begin{array}{ccccc}
\mathsf{RedMon}^{\Sigma} & \longrightarrow & \mathsf{RedMon} & \longrightarrow & V(\int \mathsf{Mod}) \\
{\scriptstyle \mathcal{U}^{\Sigma}}\big\downarrow & \lrcorner & \big\downarrow & \lrcorner & \big\downarrow{\scriptstyle \mathrm{cod}} \\
\mathsf{Mon}^{\Sigma} & \longrightarrow & \mathsf{Mon} & \xrightarrow[\Theta\times\Theta]{} & \int \mathsf{Mod}
\end{array}$$

where $V(\int \mathsf{Mod})$ is the full subcategory of arrows of $\int \mathsf{Mod}$ which are vertical (that is, they are mapped to the identity monad morphism by the functor from $\int \mathsf{Mod}$ to $\mathsf{Mon}$), and $\mathrm{cod}$ maps such an arrow to its codomain. By Propositions 25 and 29, the category $\int \mathsf{Mod}$ has fibred finite limits , so that $\mathrm{cod}$ is a fibration ([See00, Exercise 9.4.2 (i)]).

Now, Proposition 8.1.15 of [Bor94] states that a pullback of a fibration is a fibration. Thus, the middle functor $\mathsf{RedMon} \to \mathsf{Mon}$ is a fibration, and then, $\mathcal{U}^{\Sigma} : \mathsf{RedMon}^{\Sigma} \to \mathsf{Mon}^{\Sigma}$ also is. □

Finally, gathering all these lemmas yields a proof that the category of models of $\mathcal{S}$ is indeed fibered over the category of $\Sigma$-monads:

*Proof of Lemma 162.* Apply Lemma 163 with the fibration $p = \mathcal{U}^{\Sigma}$ (Lemma 164) and $F = F_{\mathcal{S}}$. □

### 5.5.3 Effectivity

In this section, we prove that $\mathcal{S}$ has an initial model, provided that there exists an initial $\Sigma$-monad. The category of models of $\mathcal{S}$ is fibered over the category of $\Sigma$-monads. A promising candidate for the initial model is the initial object, if it exists, in the fiber category over the initial $\Sigma$-monad:

**Lemma 165.** *Let $p : E \to B$ be a fibration, $b_0$ be an initial object in $B$ and $e_0$ be an object over $b_0$ that is initial in the fiber category over $b_0$. Then $e_0$ is initial in $E$.*

In the following, we thus construct the initial object in a fiber category over a given $\Sigma$-monad $R$. This fiber category can be characterized as a category of algebras:

**Lemma 166.** *The fiber category over a given $\Sigma$-monad $R$ through the fibration from models of $\mathcal{S}$ (Lemma 162) is the category of algebras of the endofunctor $F_{\mathcal{S}|R} = \coprod_i F_{\mathcal{A}_i|R}$ on the slice category $\mathsf{Mod}(R)/R^2$.*

Thus, our task is to construct the initial algebra of some specific endofunctor.

Adámek's theorem [Adá74] provides a sufficient condition for the existence of an initial algebra:

**Lemma 167** (Adámek)**.** *Let $F$ be a finitary endofunctor on a cocomplete category $C$. Then the category of algebras of $F$ has an initial object.*

This initial object can be computed as a colimit of a chain, but we do not rely here on the exact underlying construction.

The first requirement to apply this lemma is that the base category is cocomplete, and this is indeed the case:

**Lemma 168.** *The category $\mathsf{Mod}(R)/R^2$ is cocomplete for any monad $R$.*

*Proof.* The category of modules $\mathsf{Mod}(R)$ over a given monad $R$ is cocomplete by Proposition 25, so any of its slice categories is, by the dual of [ML98, Exercise V.1.1], in particular $\mathsf{Mod}(R)/R^2$. □

Let us show that the finitarity requirement of Lemma 167 is also satisfied for the case of a signature with a single reduction rule:

**Lemma 169.** *Let $\mathcal{A} = (\mathcal{V}, (n_i, h_i)_{i \in I}, (0, c))$ be a normalized reduction rule over $\Sigma$, and $R$ be a $\Sigma$-monad. Then, $F_{\mathcal{A}|R}$ is finitary.*

*Proof.* In this proof, we denote by $F$ the endofunctor $F_{\mathcal{A}|R}$ on $\mathsf{Mod}(R)/R^2$, by $\pi : D/d \to D$ the projection for a general slice category, and by $\alpha : \pi \to d$ the natural transformation from $\pi$ to the functor constant at $d$ induced by the underlying morphism of a slice object: $\alpha_p : \pi(p) \to d$. Note that $\pi$ creates colimits, by the dual of [ML98, Exercise V.1.1].

Given a filtered diagram we want to show that the image by $F$ of the colimiting cocone is colimiting. As $\pi$ creates colimits, this is enough to show that the image by $\pi \cdot F$ of the colimiting cocone is colimiting. Thus, it is enough to prove that $\pi \cdot F : \mathsf{Mod}(R)/R^2 \to \mathsf{Mod}(R)$ is finitary.

Given any $q \in \mathsf{Mod}(R)/R^2$ the module $\pi(F(q))$ is $\mathsf{Hyp}_{\mathcal{A}}(R)$, which can be computed as the limit of the following finite diagram:

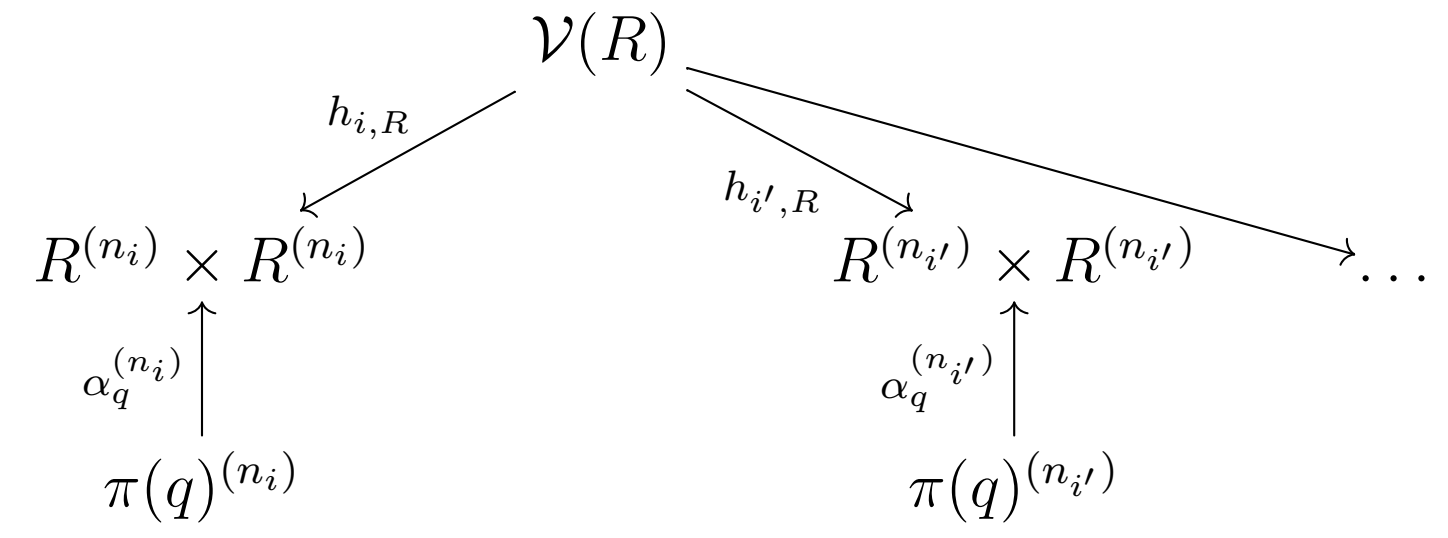

Let $J : \mathsf{C} \to \mathsf{Mod}(R)/R^2$ be a filtered diagram. As $\pi$ preserves colimits (since it creates them), $\pi(F(\operatorname{colim} J))$ is the limit of the following diagram:

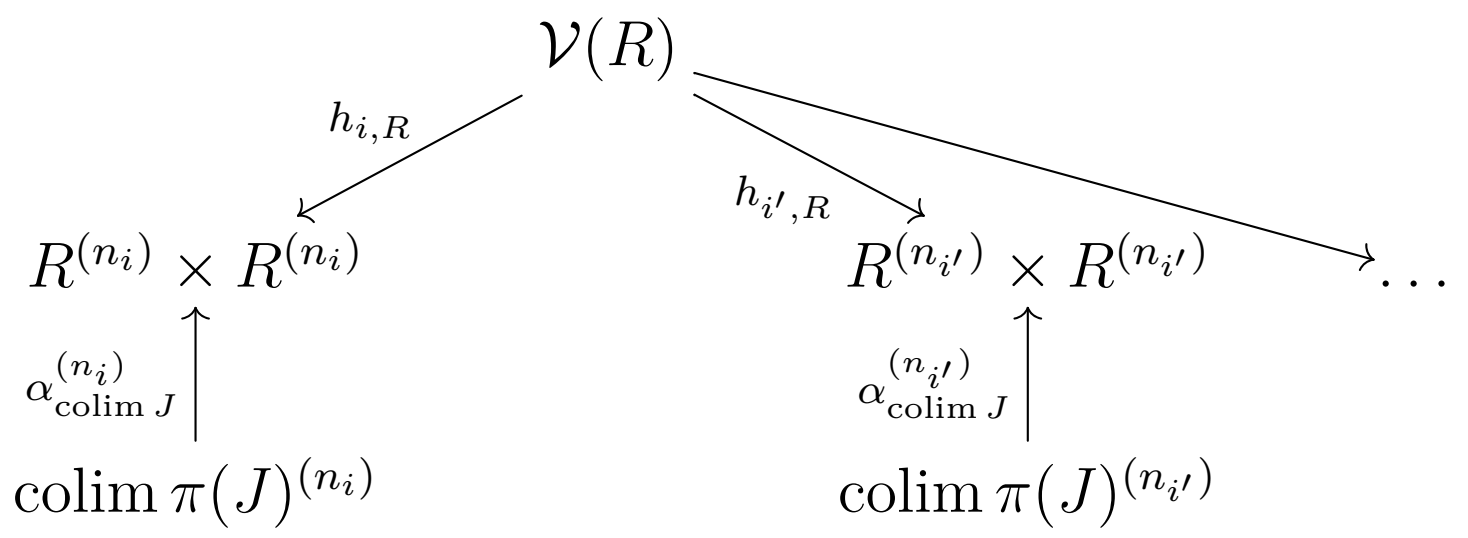

Now, as limits and colimits are computed pointwise in the category of modules, and as finite limits commute with filtered colimits in $\mathsf{Set}$ ([ML98, Section IX.2, Theorem 1]), we have that $\pi(F(\operatorname{colim} J))$, as the limit of such a diagram, is canonically isomorphic to the colimit of $\pi \cdot F \cdot J$.

□

Now, consider a signature $\mathcal{S}$ with a family of reduction rules $(\mathcal{A}_i)_i$. The functor that we are concerned with is $F_{\mathcal{S}|R} = \coprod_i F_{\mathcal{A}_i|R}$, for a given $\Sigma$-monad $R$:

**Lemma 170.** *For any $\Sigma$-monad $R$, the functor $F_{\mathcal{S}|R} = \coprod_i F_{\mathcal{A}_i|R}$ is finitary.*

*Proof.* This is a coproduct of finitary functors (by Lemma 169), and so is finitary as colimits commute with colimits, by [ML98, Equation V.2.2]. □

Now we are ready to tackle the proof of our main result:

*Proof of Theorem 150.* We assume that $\Sigma$ is effective; let $R$ be the initial $\Sigma$-monad. We want to show that $\mathcal{S}$ has an initial model. We apply Lemma 165 with $p$ the fibration from models to $\Sigma$-monads (Lemma 162): we are left with providing an initial object in the fiber category over $R$. By Lemma 166, this boils down to constructing an initial algebra for the endofunctor $F_{\mathcal{S}|R}$ on the category $\mathsf{Mod}(R)/R^2$. We apply Lemma 167: $\mathsf{Mod}(R)/R^2$ is indeed cocomplete by Lemma 168, and $F_{\mathcal{S}|R}$ is finitary by Lemma 170). □

## 5.6 Example: Lambda calculus with explicit substitutions

Here, we give a signature specifying the reduction monad of the lambda calculus with explicit substitutions as described in [Kes09]. One feature of this example is that it involves operations subject to some equations, and on top of this syntax with equations, a "graph of reductions".

In Section 5.6.1, we present the underlying signature for monads, and in Section 5.6.2, we list the reduction rules of the signature.

### 5.6.1 Signature for the monad of the lambda calculus with explicit substitutions

We give here the signature for the monad of the lambda calculus with explicit substitutions: first the syntactic operations, and then the equation that the explicit substitution must satisfy.

#### Operations

The lambda calculus with explicit substitutions extends the lambda calculus with an explicit unary substitution operator $t[x/u]$. Here, the variable $x$ is assumed not to occur freely in $u$. In our setting, it is specified as an operation $\mathsf{esubst}_X : \mathsf{LC}'(X) \times \mathsf{LC}(X) \to \mathsf{LC}(X)$. It is thus specified by the signature induced by the parametric module $\Theta' \times \Theta$. An action of this signature in a monad $R$ yields a map $\mathsf{esubst}_X : R(X + \{*\}) \times R(X) \to R(X)$ for each set $X$, where $\mathsf{esubst}_X(t, u)$ is meant to model the explicit substitution $t[*/u]$.

**Definition 171.** The signature $\Upsilon_{\mathsf{LC}_{\mathsf{ex}}}$ for the monad of the lambda calculus with explicit substitutions without equations is the coproduct of $\Theta' \times \Theta$ and $\Sigma_{\mathsf{LC}}$.

### Equation

The syntax of lambda calculus with explicit substitutions of [Kes09] is subject to the equation (see [Kes09, Figure 1, "Equations"])

$$t[x/u][y/v] = t[y/v][x/u] \quad \text{if } y \notin \mathsf{fv}(u) \text{ and } x \notin \mathsf{fv}(v) \ . \tag{5.4}$$

We rephrase it as an equality between two parallel $\Upsilon_{\mathsf{LC_{ex}}}$-module morphisms from $\Theta'' \times \Theta \times \Theta$, modelling the metavariables $t$, $u$, and $v$, to $\Theta$:

$$\begin{array}{l} \Theta'' \times \Theta \times \Theta \xrightarrow{\Theta'' \times \iota \times \Theta} \Theta'' \times \Theta' \times \Theta \xrightarrow{\mathsf{esubst}' \times \Theta} \Theta' \times \Theta \xrightarrow{\mathsf{esubst}} \Theta \\ \Theta'' \times \Theta \times \Theta \xrightarrow[\Theta'' \times \Theta \times \iota]{} \Theta'' \times \Theta \times \Theta' \xrightarrow[\langle \mathsf{esubst}^{\vee} \circ \pi_{1,3}, \pi_2 \rangle]{} \Theta' \times \Theta \xrightarrow[\mathsf{esubst}]{} \Theta \end{array} \tag{5.5}$$

Here, $\iota$ denotes the inclusion $\Theta \to \Theta'$, and $\mathsf{esubst}^{\vee}$ is the composition of $\mathsf{esubst}'$ with $\mathsf{swap} : \Theta'' \to \Theta''$ swapping the two fresh variables.

Now we are ready to define the signature of the lambda calculus monad with explicit substitutions:

**Definition 172.** The signature $\Sigma_{\mathsf{LC_{ex}}}$ of the lambda calculus monad with explicit substitutions consists of $\Upsilon_{\mathsf{LC_{ex}}}$ and the single $\Sigma_{\mathsf{LC_{ex}}}$-equation stating the equality between the two morphisms of Equation 5.5.

**Lemma 173.** *The signature $\Sigma_{\mathsf{LC_{ex}}}$ for monads is effective*

*Proof.* This is a direct consequence of Corollary 151. □

## 5.6.2 Reduction rules for lambda calculus with explicit substitutions

The reduction signature for the lambda calculus with explicit substitutions consists of two components: the first one is the signature for monads $\Sigma_{\mathsf{LC_{ex}}}$ of Definition 172; the second one is the list of reduction rules that we enumerate here, taken from [Kes09, Figure 1, "Rules"]. Except for congruence, none of them involve hypotheses.

First, let us state the congruence rules (that are implicit in [Kes09]):

$$\frac{T \rightsquigarrow T'}{\mathsf{app}(T,U) \rightsquigarrow \mathsf{app}(T',U)} \qquad \frac{U \rightsquigarrow U'}{\mathsf{app}(T,U) \rightsquigarrow \mathsf{app}(T,U')} \qquad \frac{T \rightsquigarrow T'}{\mathsf{abs}(T) \rightsquigarrow \mathsf{abs}(T')}$$

$$\frac{}{(\lambda x.t)u \rightsquigarrow t[x/u]}\ \beta\text{-red} \qquad \frac{x \notin \mathsf{fv}(t)}{t[x/u] \rightsquigarrow t}\ \mathsf{Gc} \qquad \frac{}{x[x/u] \rightsquigarrow u}\ \mathsf{var}[]$$

$$\frac{}{(t\,u)[x/v] \rightsquigarrow t[x/v]\,u[x/v]}\ \mathsf{app}[] \qquad \frac{}{(\lambda y.t)[x/v] \rightsquigarrow \lambda y.t[x/v]}\ \mathsf{abs}[]$$

$$\frac{x \notin \mathsf{fv}(v) \qquad y \in \mathsf{fv}(u)}{t[x/u][y/v] \rightsquigarrow t[y/v][x/u[y/v]]}\ [][]$$

Figure 5.3: Reduction rules of lambda calculus with explicit substitutions.

$$\frac{T \rightsquigarrow T'}{\mathsf{esubst}(T,U) \rightsquigarrow \mathsf{esubst}(T',U)} \qquad \frac{U \rightsquigarrow U'}{\mathsf{esubst}(T,U) \rightsquigarrow \mathsf{esubst}(T,U')}$$

They are translated into reduction rules through the protocol described in Section 5.3.5.

Figure 5.3 gives Kesner's rules. Five out of six of Kesner's rules translate straightforwardly, see Figure 5.4. Note how the explicit weakening $\iota : \Theta \to \Theta'$ accounts for the side condition $x \notin \mathsf{fv}(t)$ of the Gc-rule in Figure 5.3.

Expressing the side condition $y \in \mathsf{fv}(u)$ of the [][]-rule of Figure 5.3 requires the definition of the $\Sigma_{\mathsf{LC_{ex}}}$-module $\Theta_*$ such that $\mathsf{LC_{ex*}}$ is the submodule of $\mathsf{LC_{ex}}'$ of terms that really depend on the fresh variable.

We propose an approach based on the informal intuitive idea of defining inductively the submodule $R_*$ of $R'$ depending on the fresh variable $*$ as follows, for a given $\Sigma_{\mathsf{LC_{ex}}}$-monad $R$:

- $\eta(*) \in R_*(X)$, for any set $X$;
- (application)
  - if $t \in R(X)$ and $u \in R_*(X)$, then $\mathsf{app}(\iota(t), u) \in R_*(X)$
  - if $t \in R_*(X)$ and $u \in R(X)$, then $\mathsf{app}(t, \iota(u)) \in R_*(X)$
  - if $t \in R_*(X)$ and $u \in R_*(X)$, then $\mathsf{app}(t, u) \in R_*(X)$
- if $t \in R_*(X + \{x\})$, then $\lambda x.t \in R_*(X)$;
- (explicit substitution)
  - if $t \in R(X + \{x\})$ and $u \in R_*(X)$, then $\iota(t)[x/u] \in R_*(X)$;
  - if $t \in R_*(X + \{x\})$ and $u \in R(X)$, then $t[x/\iota(u)] \in R_*(X)$;

$$\frac{}{\mathsf{app}(\mathsf{abs}(T), U) \rightsquigarrow \mathsf{esubst}(T, U)}\ \beta\text{-red} \qquad \frac{}{\mathsf{esubst}(\iota(T), U) \rightsquigarrow T}\ \mathsf{Gc}$$

$$\frac{}{\mathsf{esubst}(\mathsf{app}(T, U), V) \rightsquigarrow \mathsf{app}(\mathsf{esubst}(T, V), \mathsf{esubst}(U, V))}\ \mathsf{app}[]$$

$$\frac{}{\mathsf{esubst}(*, T) \rightsquigarrow T}\ \mathsf{var}[] \qquad \frac{}{\mathsf{esubst}(\mathsf{abs}'(T), V) \rightsquigarrow \mathsf{abs}(\mathsf{esubst}^{\vee}(T, \iota(V)))}\ \mathsf{abs}[]$$

$$\frac{}{\mathsf{esubst}(\mathsf{esubst}'(T, \kappa(U)), V) \rightsquigarrow \mathsf{esubst}(\mathsf{esubst}^{\vee}(T, \iota(V)), \mathsf{esubst}(\kappa(U), V))}\ [][]$$

$$\kappa : \Theta_* \to \Theta'$$
$$\mathsf{esubst}^{\vee} : \Theta'' \times \Theta' \to \Theta'$$

where $\Theta_*$ is the 1-hole context $\Sigma_{\mathsf{LC_{ex}}}$-submodule of $\Theta'$ (Definition 174), and $\mathsf{esubst}^{\vee}$ is defined as the composition

$$\Theta'' \times \Theta' \xrightarrow{\mathsf{swap} \times \Theta'} \Theta'' \times \Theta' \xrightarrow{\mathsf{esubst}} \Theta'$$

Here, swap exchanges the fresh variables:

$$\mathsf{swap}_{X,R} : R((X + \{*_1\}) + \{*_2\}) \to R((X + \{*_1\}) + \{*_2\})(t)$$
$$\mathsf{swap}_{X,R} : t \mapsto t\{*_1 := *_2; *_2 := *_1\}$$

Figure 5.4: Reduction rules of Figure 5.3 reformulated in our setting.

– if $t \in R_*(X + \{x\})$ and $u \in R_*(X)$, then $t[x/u] \in R_*(X)$.

Guided by this intuition, we now formally define a $\Sigma_{\mathsf{LC_{ex}}}$-module $\Theta_*$ equipped with a morphism $\kappa : \Theta_* \to \Theta'$.

The previous informal inductive definition is translated as an initial algebra for an endofunctor on the category of $\Sigma_{\mathsf{LC_{ex}}}$-modules, which is cocomplete (colimits are computed pointwise). This endofunctor maps a $\Sigma_{\mathsf{LC_{ex}}}$-module $M$ to the coproduct of the following $\Sigma_{\mathsf{LC_{ex}}}$-modules:

- the terminal $\Sigma_{\mathsf{LC_{ex}}}$-module $1$, playing the rôle of the fresh variable;
- the coproduct $M \times \Theta + \Theta \times M + M \times M$, one summand for each case of the application;
- the derived module $M'$ for abstraction;
- the coproduct $M' \times \Theta + \Theta' \times M + M' \times M$, one summand for each case of the explicit substitution.

This functor is finitary, so the initial algebra exists thanks to Adámek's theorem (already cited, as Theorem 167). Unfortunately, the resulting $\Sigma_{\mathsf{LC_{ex}}}$-module does not yield the module that we are expecting in the case of the monad $\mathsf{LC_{ex}}$: it does not satisfy Equation 5.5, and thus contains more terms than necessary. To obtain the desired $\Sigma_{\mathsf{LC_{ex}}}$-module, we equip $\Theta'$ with its canonical algebra structure, inducing a morphism from the initial algebra, and we define $\Theta_*$ as the image of this morphism, thus equipped with an inclusion $\kappa : \Theta_* \to \Theta'$.

**Definition 174.** We define the $\Sigma_{\mathsf{LC_{ex}}}$**-module of "one-hole contexts"** to be $\Theta_*$, equipped with an inclusion $\kappa : \Theta_* \to \Theta'$.

**Remark 175.** Such a definition can be worked out for any algebraic signature for monads.

Now we define the signature of the reduction monad of lambda calculus with explicit substitutions:

**Definition 176.** The reduction signature $\mathcal{S}_{\mathsf{LC_{ex}}}$ of the lambda calculus reduction monad with explicit substitutions consists of the signature $\Sigma_{\mathsf{LC_{ex}}}$ of Definition 172 and all the reduction rules specified in this section.

**Lemma 177.** *The reduction signature $\mathcal{S}_{\mathsf{LC_{ex}}}$ is effective.*

*Proof.* Apply Theorem 150. The underlying signature for monads is effective by Lemma 173. □

# 5.7 Recursion

In this section, we derive, for any effective reduction signature $\mathcal{S}$, a recursion principle from initiality. In Section 5.7.1, we state this recursion principle, then we give an example of application in Section 5.7.2, by translating lambda calculus with a fixpoint operator to lambda calculus. In Section 5.7.2, we apply this principle to translate lambda calculus with explicit substitutions into lambda calculus with *unary congruent substitution*. Then, in Section 5.7.4, we translate this latter variant of lambda calculus into lambda calculus closed under identity and composition of reductions.

## 5.7.1 Recursion principle for effective signatures

The recursion principle associated to an effective signature provides a way to construct a morphism from the reduction monad underlying the initial model of that signature to a given reduction monad.

**Proposition 178** (Recursion principle)**.** *Let $\mathcal{S}$ be an effective reduction signature, and $R$ be the reduction monad underlying the initial model. Let $T$ be a reduction monad. Any action $\tau$ of $\mathcal{S}$ in $T$ induces a reduction monad morphism $\hat{\tau} : R \to T$.*

*Proof.* The action $\tau$ defines a model $M$ of $\mathcal{S}$. By initiality, there is a unique model morphism from the initial model to $M$, and $\hat{\tau}$ is the reduction monad morphism underlying it. □

In the next sections, we illustrate this principle.

## 5.7.2 Translation of lambda calculus with fixpoint to lambda calculus

In this section, we consider the signature $\mathcal{S}_{\mathsf{LC_{fix}}}$ of Example 154 for the lambda calculus with an explicit fixpoint operator. We build, by recursion, a reduction monad morphism from the initial model $\mathsf{LC}_{\mathsf{fix}}$ of this signature to $\mathsf{LC}^*_{\beta/\eta}$, the "closure under identity and composition of reductions" (Definition 131) of the initial model $\mathsf{LC}_{\beta/\eta}$ of the signature $\mathcal{S}_{\mathsf{LC}_{\beta/\eta}}$ (Example 152).

As explained in Section 5.7.1, we need to define an action of $\mathcal{S}_{\mathsf{LC}_{\mathsf{fix}}}$ in $\mathsf{LC}^*_{\beta/\eta}$. Note that $\mathcal{S}_{\mathsf{LC}_{\mathsf{fix}}}$ is an extension of $\mathcal{S}_{\mathsf{LC}_{\beta/\eta}}$ (Example 154). First, we focus on the core $\mathcal{S}_{\mathsf{LC}_{\beta/\eta}}$ part: we show that the reduction monad $\mathsf{LC}^*_{\beta/\eta}$ inherits the canonical action of $\mathcal{S}_{\mathsf{LC}_{\beta/\eta}}$ in $\mathsf{LC}_{\beta/\eta}$.

**Lemma 179.** *There is an action*[1] *of $\mathcal{S}_{\mathsf{LC}_{\beta/\eta}}$ in $\mathsf{LC}^*_{\beta/\eta}$.*

*Proof.* The challenge is to give an action of reduction rules with hypotheses: now the input reductions of the rule may be actually sequences of reductions. This concerns congruence for application and abstraction. We take the example of abstraction: suppose we have a sequence of reductions $r_1 \ldots r_n$ going from $t_0$ to $t_n$. We want to provide a reduction between $\mathsf{abs}(t_0)$ and $\mathsf{abs}(t_n)$. For each $i$, we have a reduction between $\mathsf{abs}(t_{i-1})$ and $\mathsf{abs}(t_i)$. By composing the corresponding sequence, we obtain the desired reduction. □

The action for the extra parts of $\mathcal{S}_{\mathsf{LC}_{\mathsf{fix}}}$ requires the following:

- an operation $\mathsf{fix} : {\mathsf{LC}^*_{\beta/\eta}}' \to \mathsf{LC}^*_{\beta/\eta}$: for this, we choose a fixpoint combinator $Y$ and set $\mathsf{fix}_X(t) = \mathsf{app}(Y, \mathsf{abs}(t))$, in accordance with Section 3.8.4;
- an action of the reduction rule

$$\frac{}{\mathsf{fix}(T) \rightsquigarrow T\{* := \mathsf{fix}(T)\}}$$

  A fixpoint combinator $Y$ is a closed term with the property that for any other term $t$, the term $\mathsf{app}(Y, t)$ $\beta$-reduces in some steps to $\mathsf{app}(t, \mathsf{app}(Y, t))$. We denote by $r \in \mathsf{Red}(\mathsf{LC}^*_{\beta/\eta})(\{*\})$ a reduction between $\mathsf{app}(Y, *)$ and $\mathsf{app}(*, \mathsf{app}(Y, *))$. Then, $r$ induces an $\underline{\mathsf{LC}_{\beta/\eta}}$-module morphism $\hat{r} : \underline{\mathsf{LC}_{\beta/\eta}} \to \mathsf{Red}(\mathsf{LC}^*_{\beta/\eta})$ by mapping an element $t \in \mathsf{LC}_{\beta/\eta}(X)$ to $r\{* := t\}$. We define the action of this reduction rule as the composition of the following reductions:

$$\mathsf{app}(Y, \mathsf{abs}(t)) \rightsquigarrow_{\hat{r}(\mathsf{abs}(t))} \mathsf{app}(\mathsf{abs}(t), \mathsf{app}(Y, \mathsf{abs}(t))) \rightsquigarrow_\beta t\{* := \mathsf{app}(Y, \mathsf{abs}(t))\}$$

1. A formalized proof in Agda of this statement is available at `https://github.com/amblafont/unary-subst-LCstar/blob/master/fiberlambda.agda`.

- an action of the congruence rule

$$\frac{T \rightsquigarrow T'}{\mathsf{fix}(T) \rightsquigarrow \mathsf{fix}(T')}$$

  that can be defined in the obvious way using the congruences of application and abstraction.

In more concrete terms, our translation is a kind of compilation which replaces each occurrence of the explicit fixpoint operator $\mathsf{fix}(t)$ with $\mathsf{app}(Y, \mathsf{abs}(t))$, and each fixpoint reduction with a composite of $\beta$-reductions.

### 5.7.3 Translation of lambda calculus with explicit substitutions into lambda calculus with congruent unary substitution

Here, we consider the reduction signature $\mathcal{S}_{\mathsf{LC_{ex}}} = (\Sigma_{\mathsf{LC_{ex}}}, \mathfrak{R}_{\mathsf{LC_{ex}}})$ introduced in Definition 176. The underlying monad of the initial model $\mathsf{LC_{ex}}$ is the monad of lambda calculus with an application and abstraction operation, and an explicit substitution operator $\mathsf{LC_{ex}}' \times \mathsf{LC_{ex}} \to \mathsf{LC_{ex}}$ satisfying Equation 5.5, for $R = \mathsf{LC_{ex}}$. The associated reduction monad has all the rules specified in Section 5.6.

We build, by recursion, a reduction monad morphism from the initial model $\mathsf{LC_{ex}}$ of this signature to $\mathsf{LC_{1\text{-}cong}}$, a variant of the lambda calculus specified by the signature $\mathcal{S}_{\mathsf{LC}_{\beta/\eta}}$ (Example 152) extended with the *congruence for unary substitution*:

$$\frac{T \rightsquigarrow T'}{U\{* := T\} \rightsquigarrow U\{* := T'\}}$$

Note that this reduction rule accounts for the reflexivity rule, and makes congruences for application (but not congruence for abstraction) redundant:

**Reflexivity** Any $U \in \mathsf{LC}(X)$ can be weakened into $\iota(U) \in \mathsf{LC}'(X)$. Then, consider any reduction $m : T \blacktriangleright T'$ . The action of the reduction rule above yields a reduction between $\iota(U)\{* := T\} = U$ and $\iota(U)\{* := T'\} = U$. By choosing adequatly $m$ (for example, take the beta-reduction between $\mathsf{app}(\mathsf{abs}(*), \mathsf{abs}(*))$ and $\mathsf{abs}(*)$), this yields an action of the reflexivity reduction rule.

**Congruence** Consider the left congruence rule (the cases of the right one and the congruence for abstraction are similar): from any reduction $m : T \blacktriangleright T'$, we want a reduction between $\mathsf{app}(T, U)$ and $\mathsf{app}(T', U)$, for $T, T', U \in \mathsf{LC}(X)$. We get it by applying the action of unary congruent substitution to $m$ for the term $\mathsf{app}(*, U)$. One checks that this indeed defines an action for the left congruence reduction rule of application.

As explained in Section 5.7.1, we need to define an action of $\mathcal{S}_{\mathsf{LC}_{\mathsf{ex}}}$ in $\mathsf{LC}_{\mathsf{1\text{-}cong}}$:

- the operations of application and abstraction are those of $\mathsf{LC}_{\mathsf{1\text{-}cong}}$ as the initial $\Sigma_{\mathsf{LC}}$-monad (recall that the underlying monad of $\mathsf{LC}_{\mathsf{1\text{-}cong}}$ is just LC);
- the explicit substitution operation $\mathsf{LC}_{\mathsf{1\text{-}cong}}{}' \times \mathsf{LC}_{\mathsf{1\text{-}cong}} \to \mathsf{LC}_{\mathsf{1\text{-}cong}}$ is defined using the monadic substitution, mapping a pair $(t, u) \in \mathsf{LC}_{\mathsf{1\text{-}cong}}(X + \{*\}) \times \mathsf{LC}_{\mathsf{1\text{-}cong}}(X)$ to the monadic substitution $t\{* := u\}$;
- Equation 5.5 for the underlying monad is satisfied thanks to the usual monadic equations;
- the action of the congruence rules for application and abstraction are induced by the action of the congruence rule for unary substitution, as explained above;
- $\mathsf{LC}_{\mathsf{1\text{-}cong}}$ has already an action for $\beta$-reduction;
- all the actions for the remaining reduction rules involving explicit substitution (except the congruences for explicit substitution that are discussed below) are given by an action of the reflexivity reduction rule;
- the non-obvious actions are the ones of the congruence rules for explicit substitution:

  $$\frac{T \rightsquigarrow T'}{T[x/U] \rightsquigarrow T'[x/U]} \qquad \frac{U \rightsquigarrow U'}{T[x/U] \rightsquigarrow T[x/U']}$$

  The left one is obtained from the substitution of the module of reductions (see Remark 120). The right one is exactly given by the action of the congruence rule for unary substitution.

Finally, by the recursion principle, we get a reduction monad morphism from $\mathsf{LC}_{\mathsf{ex}}$ to $\mathsf{LC}_{\mathsf{1\text{-}cong}}$. This translation replaces the explicit substitution operator $t[x/u]$ with the corresponding monadic substitution $t\{x := u\}$, and all the reductions are translated to reflexivity except for the ones for the $\beta$-reduction and congruences.

### 5.7.4 Translation of lambda calculus with congruent unary substitution into lambda calculus

In the previous section, we translated lambda calculus with explicit substitution into lambda calculus with congruent unary substitution. In this section, we translate this variant of lambda calculus into $\mathsf{LC}^*_{\beta/\eta}$ (introduced in Section 5.7.2), the closure under identity and composition of reductions the initial model $\mathsf{LC}_{\beta/\eta}$ of the signature $\mathcal{S}_{\mathsf{LC}_{\beta/\eta}}$ (Example 152).

As per Section 5.7.1, we need to define an action in $\mathsf{LC}^*_{\beta/\eta}$ of the signature $\mathcal{S}_{\mathsf{LC}_{\beta/\eta}}$ extended with the reduction rule:

$$\frac{T \rightsquigarrow T'}{U\{* := T\} \rightsquigarrow U\{* := T'\}} \tag{5.6}$$

Thanks to Lemma 179, we have an action of $\mathcal{S}_{\mathsf{LC}_{\beta\eta}}$ in $\mathsf{LC}^*_{\beta/\eta}$. Thus, the main challenge consists in equipping $\mathsf{LC}^*_{\beta/\eta}$ with an action of the rule (5.6).

**Proposition 180.** *The reduction monad* $\mathsf{LC}^*_{\beta/\eta}$ *can be equipped*[2] *with an action of the rule* (5.6).

*Proof.* We will write $r : R \to \mathsf{LC} \times \mathsf{LC}$ in place of $\mathsf{red}_{\mathsf{LC}^*_{\beta/\eta}} : \mathsf{Red}(\mathsf{LC}^*_{\beta/\eta}) \to \mathsf{LC} \times \mathsf{LC}$.

Such an action is equivalently given (see Lemma 160) by a morphism $\alpha : \mathsf{LC}' \times R \to R$ such that the following diagram commutes, where $q_X(t, m) = \big(t\{* := \mathsf{source}(m)\}, t\{* := \mathsf{target}(m)\}\big)$.

$$\begin{array}{ccc} \mathsf{LC}' \times R & \overset{\alpha}{\dashrightarrow} & R \\ & {\scriptstyle q}\searrow \quad \swarrow{\scriptstyle r} & \\ & \mathsf{LC} \times \mathsf{LC} & \end{array} \tag{5.7}$$

We first construct the collection of functions $(\alpha_X)_X$ with $\alpha_X : \mathsf{LC}'(X) \times R(X) \to R(X)$ and then shows the two required properties, i.e., that it commutes with substitution (thus inducing a $\mathsf{LC}$-module morphism), and that it satisfies Equation 5.7.

The construction of the collection of functions (without naturality conditions) is worked out in the functor category $[\mathsf{Set}_0, \mathsf{Set}]$, where $\mathsf{Set}_0$ is the discretized category of sets (this base category allows us to get rid of naturality conditions). This is done by recursion on the first argument. More formally, we exploit some initiality property of $\mathsf{LC}' \cdot j$, where $j : \mathsf{Set}_0 \to \mathsf{Set}$ is the inclusion of the discretized category of sets into sets. Indeed, $\mathsf{LC}' \cdot j$

2. A formalized proof in Agda of this statement is available at `https://github.com/amblafont/unary-subst-LCstar/blob/master/fiberlambda.agda`.

is the initial algebra of the endofunctor

$$\Psi : [\mathsf{Set}_0, \mathsf{Set}] \to [\mathsf{Set}_0, \mathsf{Set}] \qquad F \mapsto j + 1 + F \times F + F'$$

where $F'$ is the functor mapping a set $X$ to $F(X + 1)$.

The two properties that we want to show about the collection of functions $(\alpha_X)_X$ are then done by *induction* on the first argument, again exploiting initiality of $\mathsf{LC}' \cdot j$, as we explain below. The proof goes as follows:

1. construct (by initiality) a morphism from $\mathsf{LC}' \cdot j$ to the exponential of $R \cdot j$ with itself, that is, to the functor $R^R : \mathsf{Set}_0 \to \mathsf{Set}$ defined on objects by $X \mapsto R(X)^{R(X)}$;

2. show that the induced morphism from $\mathsf{LC}' \cdot j \times R \cdot j \to R \cdot j$ yields a $\mathsf{LC}$-module morphism $\alpha : \mathsf{LC}' \times R \to R$;

3. show the commutation required by Equation 5.7.

Note how working in the functor category $[\mathsf{Set}_0, \mathsf{Set}]$ allows us to define the functor $R^R$ as above, without worrying about the functorial action on morphisms. Below we sometimes omit the explicit precomposition with $j$ in order to simplify the notation. Now we perform the steps explained above.

1. As we argued before, $\mathsf{LC}' \cdot j$ is the initial algebra of $\Psi$, so our task consists in equipping $R^R$ with an algebra structure for $\Psi$, that we split into the following four components, using the universal properties of the coproduct and the exponential in the category $[\mathsf{Set}_0, \mathsf{Set}]$:

   (a) The morphism $j \times R \to R$ corresponds to the case of variables. We expect that the resulting module morphism $\alpha : \mathsf{LC}' \times R \to R$ satisfies $\alpha_X(\eta(x), m) = \mathsf{refl}_X(x)$ for any $x \in X$, where $\mathsf{refl} : \mathsf{LC} \to R$ maps a term to the reflexive reduction on itself and $\eta : \mathsf{Id} \to \mathsf{LC}'$ is the unit of the monad $\mathsf{LC}$. Accordingly, we define the morphism $j \times R \to R$ as mapping a pair $(x, m) \in X \times R(X)$ to $\mathsf{refl}_X(\eta(x))$.

   (b) The morphism $R \to R$ corresponds to the case of the fresh variable $*$. We expect that $\alpha_X(*, m) = m$. Accordingly, the required morphism is taken as the identity on $R$.

(c) The morphism $(R \times R)^R \to R^R$ corresponds to the case of an application. We expect that $\alpha_X(\mathsf{app}(t,u),m) = \mathsf{app\text{-}cong}(\alpha_X(t,m),\alpha_X(u,m))$, where $\mathsf{app\text{-}cong} : R\times R \to R$ is the action of the reduction rule of congruence for application defined as $\mathsf{app\text{-}cong}(m_1,m_2) = \mathsf{trans}(\mathsf{app\text{-}cong}_1(m_1), \mathsf{app\text{-}cong}_2(m_2))$, where $\mathsf{trans}$ denotes an action of the transitivity reduction rule with which we can equip $\mathsf{LC}^*_{\beta/\eta}$ (by concatenating sequences of reductions). Accordingly, the morphism is defined as $\mathsf{app\text{-}cong}^R$.

(d) The morphism $R'^{R'} \to R^R$ corresponds to the case of an abstraction. We expect that $\alpha_X(\mathsf{abs}(t),m) = \mathsf{abs\text{-}cong}_X(\alpha_{X+1}(t, R\iota_X(m)))$, where $\iota : \mathsf{Id} \to \mathsf{Id}'$ is the canonical inclusion. Accordingly, we take $\mathsf{abs\text{-}cong}^{R\iota}$ as the the required morphism.

By initiality, we get an algebra morphism from $\mathsf{LC}\cdot j$ to $R^R$, which by uncurrying yields a morphism $\alpha : \mathsf{LC}' \cdot j \times R \to R$.

2. Upgrading $\alpha$ into a module morphism from $\mathsf{LC}' \times R$ to $R$ consists in showing compatibility with substitution in the following sense: for any map $f : X \to \mathsf{LC}(Y)$, for any pair $(t,m) \in \mathsf{LC}'(X) \times R(X)$, the equality $\alpha_X(t,m)\{f\} = \alpha_Y(t\{f\}, m\{f\})$ is satisfied. This is shown by induction on $t \in \mathsf{LC}'(X)$. The case of variables requires a preliminary step: for $t = \eta(x)$, the equation amounts to $\mathsf{refl}(f(x)) = \alpha(\mathsf{LC}i(f(x)),m)$, which is not straightforward. We hence first prove by induction on $t \in \mathsf{LC}(X)$ that $\alpha(\mathsf{LC}i(t),m) = \mathsf{refl}(t)$. We do not detail these straightforward inductions, but rather explain the general methodology to perform induction on $\mathsf{LC}'$ (the case of $\mathsf{LC}$ is similar). Suppose given, for each $t \in \mathsf{LC}'(X)$, a predicate $P_X(t)$. Then, one can form the functor $\mathsf{LC}'_{|P} : \mathsf{Set}_0 \to \mathsf{Set}$ mapping a set $X$ to the subset of $\mathsf{LC}'(X)$ satisfying the predicate $P_X$. It follows that $\mathsf{LC}'_{|P}$ embeds into $\mathsf{LC}$, and if $\mathsf{LC}'_{|P}$ inherits the algebra structure for $\Psi$ through this embedding, then by initiality we get a section of the embedding, which exactly translates the fact that any term $t \in \mathsf{LC}'(X)$ satisfies the property.

3. It remains to show the commutation of Diagram 5.7. Again, an induction on the first argument (thus exploiting initiality of $\mathsf{LC}' \cdot j$) is enough to conclude. □

CHAPTER 6

# OPERATIONAL MONADS AND THEIR SIGNATURES

In this chapter, we consider *operational monads*, extending the notion of *reduction monad* of Chapter 5. Our goal is to deal with programming languages whose terms may not form a monad, or whose reductions are not stable under substitution of terms. Consider the example of the call-by-value lambda calculus: variables can be replaced with values only, rather than any term. Yet, we are interested in reductions between terms rather than values. In this situation, there is a monad of values, and terms form a module over this monad. Although terms form the monad of lambda calculus, reductions are not stable under its monadic substitution: they are stable under the monadic substitution of values.

To be more precise, we limit ourselves to languages whose terms form a free module over a monad $R$, that is, a module of the shape $T \cdot R$ for some endofunctor $T : \mathsf{Set} \to \mathsf{Set}$. This is indeed the case of the call-by-value lambda calculus: each term can be decomposed uniquely as a binary tree whose leaves are values. For this specific example, $R$ is the monad of values of the lambda calculus, and $T$ is the endofunctor of binary trees with leaves in its argument.

We also generalize to the case of *heterogeneous reductions*. This allows us to cover the call-by-value lambda calculus with big-step operational semantics, where terms reduce to values.

We adopt the same terminology and notation as in Chapter 5. We are quicker here, as all the definitions and proofs are straightforward generalizations.

## Plan of the chapter

In Section 6.1, we define operational monads, a generalization of the reduction monads of Chapter 5. In Section 6.2, we adapt the definition of reduction rules to the new setting. This enables us to define signatures for operational monads—*operational signatures*—in Section 6.3, that are particular signatures over the category of operational monads. Then, in Section 6.4, we give some examples of operational signatures.

## 6.1 Operational monads

We define *operational monads* in Section 6.1.1, and then consider some examples of operational monads, in Section 6.1.2.

### 6.1.1 Operational monads

Before defining operational monads, we need the intermediate notion of $T$-reduction monads, when $T$ is a pair of endofunctors on $\mathsf{Set}$.

**Definition 181.** Let $T = (T_1, T_2)$ be a pair of endofunctors on $\mathsf{Set}$. A **$T$-reduction monad** $R$ is given by:

1. a monad on sets (the monad of *terms*), that we still denote by $R$, or by $\underline{R}$ when we want to be explicit;
2. an $R$-module $\mathsf{Red}(R)$ (the module of *reductions*);
3. a morphism of $R$-modules $\mathsf{red}_R : \mathsf{Red}(R) \to (T_1 \cdot R) \times (T_2 \cdot R)$ (*source* and *target* of rules).

We set $\mathsf{source}_R := \pi_1 \circ \mathsf{red}_R : \mathsf{Red}(R) \to T_1 \cdot R$, and $\mathsf{target}_R := \pi_2 \circ \mathsf{red}_R : \mathsf{Red}(R) \to T_2 \cdot R$.

**Notation 182.** *For a $T$-reduction monad $R$, a set $X$, and elements $s \in T_1(R(X))$, $t \in T_2(R(X))$, we think of the fiber $\mathsf{red}_R(X)^{-1}(s, t)$ as the set of "reductions from $s$ to $t$". We sometimes write $m : s \blacktriangleright t : T(R(X))$, or even $m : s \blacktriangleright t$ when there is no ambiguity, instead of $m \in \mathsf{red}_R(X)^{-1}(s, t)$.*

**Remark 183.** We recover Chapter 5 by taking $T = (\mathsf{Id}_{\mathsf{Set}}, \mathsf{Id}_{\mathsf{Set}})$.

**Definition 184.** Let $T = (T_1, T_2)$ be a pair of endofunctors on $\mathsf{Set}$. A **morphism of $T$-reduction monads** from $R$ to $S$ is given by a pair $(f, \alpha)$ of

1. a monad morphism $f : R \to S$, and

2. a natural transformation $\alpha : \mathsf{Red}(R) \to \mathsf{Red}(S)$

satisfying the following two conditions:

3. $\alpha$ is an $R$-module morphism between $\mathsf{Red}(R)$ and $f^*\mathsf{Red}(S)$, and

4. the square

$$\begin{array}{ccc} \mathsf{Red}(R) & \xrightarrow{\alpha} & \mathsf{Red}(S) \\ \downarrow{\scriptstyle \mathsf{red}_R} & & \downarrow{\scriptstyle \mathsf{red}_S} \\ (T_1 \cdot R) \times (T_2 \cdot R) & \xrightarrow[T_1 f \times T_2 f]{} & (T_1 \cdot S) \times (T_2 \cdot S) \end{array}$$

commutes in the category of natural transformations.

**Proposition 185** (Category of $T$-reduction monads)**.** *Let $T = (T_1, T_2)$ be a pair of endofunctors on* $\mathsf{Set}$*. Then, $T$-reduction monads and their morphisms, with the obvious composition and identity, form a category $T$-*$\mathsf{RedMon}$*, equipped with a forgetful functor to the category of monads.*

We get an analogue of Theorem 123:

**Theorem 186.** *Let $T = (T_1, T_2)$ be a pair of endofunctors on* $\mathsf{Set}$*. The category of $T$-reduction monads is isomorphic to the category of monads relative to the functor mapping a set to its discrete $T$-graph, where the category of $T$-graphs is the comma category* $\mathsf{Set}/(T_1 \times T_2)$*.*

**Definition 187.** An **operational monad** is a pair $(T, R)$ where $R$ is a $T$-reduction monad. The pair $T = (T_1, T_2)$ of endofunctors on $\mathsf{Set}$ is called the pair of **state functors** of the operational monad.

Operational monads may be organized into a category:

**Definition 188** (Morphism of operational monads)**.** A **morphism between operational monads** $((T_1, T_2), R)$ **and** $((T'_1, T'_2), R')$ is a triple $(\alpha_1, \alpha_2, f)$ consisting of

- a natural transformation $\alpha_1 : T_1 \to T'_1$;

- a natural transformation $\alpha_2 : T_2 \to T'_2$;

- a $(T'_1, T'_2)$-reduction monad morphism $f$ between $(\alpha_1, \alpha_2)_! R$ and $R'$, where the $(T'_1, T'_2)$-reduction monad $(\alpha_1, \alpha_2)_! R$ is defined as follows:
  - the underlying monad is $R$;
  - $\mathsf{Red}((\alpha_1, \alpha_2)_! R)$ is defined as $\mathsf{Red}(R)$;
  - $\mathsf{red}_{(\alpha_1,\alpha_2)_! R}$ is defined as the composition

$$\mathsf{Red}(R) \xrightarrow{\mathsf{red}_R} (T_1 \cdot R) \times (T_2 \cdot R) \xrightarrow{(\alpha_1 \cdot R)\times(\alpha_2 \cdot R)} (T'_1 \cdot R) \times (T'_2 \cdot R)\,.$$

**Proposition 189** (Category of operational monads)**.** *Operational monads and their morphisms with the obvious composition and identity morphisms form a category* $\mathsf{OpMon}$*.*

### 6.1.2 Examples of operational monads

Any example of reduction monad $R$ of Chapter 5 yields an operational monad whose state functors are the identity endofunctors.

**Example 190** (Call-by-value lambda calculus)**.** In call-by-value, reductions between terms are not stable under substitution of variables with terms in general. However, they are stable under substitution of variables with values. A value is either a variable or a lambda abstraction of an arbitrary term. Note that values are stable under substitution: they induce a monad $\mathsf{LC}_v = \mathsf{Id} + \mathsf{LC}'$ equipped with a monad morphism $i : \mathsf{LC}_v \to \mathsf{LC}$.

Now, reductions are between terms rather than between values, so we need to devise a functor $B : \mathsf{Set} \to \mathsf{Set}$ such that $B \cdot \mathsf{LC}_v$ is isomorphic to LC: then, the call-by-value lambda calculus is a $(B, B)$-reduction monad. Note that a lambda term can always be decomposed as a binary tree whose leaves are values, that is variables or lambda abstractions: each node of this tree is an application. Hence, we choose $B$ to be the functor underlying the monad of binary trees specified by the 1-signature $\Theta \times \Theta$: then, there is an isomorphism $\alpha : \mathsf{LC} \to B \cdot \mathsf{LC}_v$ in the category of $\mathsf{LC}_v$-modules (LC is indeed equipped with a structure of $\mathsf{LC}_v$-module thanks to the inclusion $i$ of $\mathsf{LC}_v$ in LC as a monad morphism). The operational monad of call-by-value lambda calculus is the pair $((B, B), \mathsf{LC}_v)$ where the $(B, B)$-reduction monad $\mathsf{LC}_v$ is defined as follows:

1. the underlying monad is $\mathsf{LC}_v = \mathsf{Id} + \mathsf{LC}'$;
2. $\mathsf{red}_{\mathsf{LC}_v} : \mathsf{Red}(\mathsf{LC}_v) \to (B{\cdot}\mathsf{LC}_v) \times (B{\cdot}\mathsf{LC}_v)$ is generated by the following constructions:

(a) for $t \in \mathsf{LC}'(X)$ and $u \in \mathsf{LC}_v(X)$, we have $\beta(t,u) : \alpha(\mathsf{app}(\mathsf{abs}(t), i(u))) \blacktriangleright \alpha(t)\{* := u\}$;

(b) for $m : \alpha(u) \blacktriangleright \alpha(v) : B(\mathsf{LC}_v(X))$ in $\mathsf{Red}(\mathsf{LC}_v)$ and $t \in \mathsf{LC}(X)$, we have $\mathsf{app\text{-}cong}_1(m,t) : \alpha(\mathsf{app}(u,t)) \blacktriangleright \alpha(\mathsf{app}(v,t))$;

(c) for $m : \alpha(u) \blacktriangleright \alpha(v) : B(\mathsf{LC}_v(X))$ in $\mathsf{Red}(\mathsf{LC}_v)$ and $t \in \mathsf{LC}(X)$, we have $\mathsf{app\text{-}cong}_2(t,m) : \alpha(\mathsf{app}(t,u)) \blacktriangleright \alpha(\mathsf{app}(t,v))$;

(d) for $m : \alpha(u) \blacktriangleright \alpha(v) : B(\mathsf{LC}'_v(X))$ in $\mathsf{Red}(\mathsf{LC}_v)$ we have $\mathsf{abs\text{-}cong}(m) : \alpha(\mathsf{abs}(u)) \blacktriangleright \alpha(\mathsf{abs}(v))$.

We call this operational monad *the operational monad of the call-by-value lambda calculus*. A signature for it is given in Section 6.4.1.

**Example 191** (Big-step call-by-value lambda calculus)**.** In the call-by-value lambda calculus with big-step operational semantics, reductions happen between a term and a value. The underlying monad is $\mathsf{LC}_v$, and the pair of state functors is $(B, \mathsf{Id})$, where $B$ is defined in Example 190. We denote $\mathsf{abs}_v : \mathsf{LC}' \to \mathsf{LC}_v$ the inclusion of lambda abstractions into values. The module of reductions is defined inductively as follows:

- for each value $v \in \mathsf{LC}_v(X)$, the induced term $\alpha(i(v))$ reduces to $v$;
- given a reduction $m_t : \alpha(t) \blacktriangleright \mathsf{abs}_v(t')$, a reduction $m_u : \alpha(u) \blacktriangleright u'$, and a reduction $m : \alpha(t')\{* := u'\} \blacktriangleright v$, we get a reduction $\beta(m_t, m_u, m) : \alpha(\mathsf{app}(t,u)) \blacktriangleright v$.

We specify more formally this operational monad in Section 6.4.2.

**Example 192.** Let us recall the following simple variant of $\pi$-calculus. The syntax for *processes* is given by

$$P, Q ::= 0 \mid (P|Q) \mid !P \mid \nu a.P \mid \overline{a}\langle b\rangle.P \mid a(b).P,$$

where $a$ and $b$ range over a fixed, countable set of *channel names*. In $\nu a.P$, the channel name $a$ is bound, and in $a(b).P$, the channel name $b$ is bound. Processes will be considered equivalent up to *structural congruence*, the smallest equivalence relation $\equiv$ stable under context and such that $0|P \equiv P$, $P|(Q|R) \equiv (P|Q)|R$, $(\nu a.P)|Q \equiv \nu a.(P|Q)$ when $a$ does not occur free in $Q$, and $!P \equiv P|!P$. Reduction is then given by the following

inductive rules, the last one enforcing the quotient by structural congruence.

$$\frac{}{\overline{a}\langle b\rangle.P|a(c).Q \rightsquigarrow P|(Q\{c := b\})} \qquad \frac{P \rightsquigarrow Q}{P|R \rightsquigarrow Q|R} \qquad \frac{P \rightsquigarrow Q}{\nu a.P \rightsquigarrow \nu a.Q}$$

$$\frac{P \equiv P' \qquad P' \rightsquigarrow Q' \qquad Q' \equiv Q}{P \rightsquigarrow Q}$$

In $\pi$-calculus all we substitute is channel names, so the monad is just the identity. There is no need for process variables: no binding of process variable is involved, nor substitution of variables with processes.

The state functor $T$ maps any $X \in \mathsf{Set}$ to the set of processes with free channel names in $X$, considered equivalent up to structural congruence. Finally, let $\mathsf{Red}(R)(X)$ denote the set of reductions between (equivalence classes of) processes in $T(X)$.

We specify more formally this operational monad in Section 6.4.3.

## 6.2 $\mathbb{S}$-reduction rules

This section is a straightforward generalization of the notion of reduction rules of Section 5.3 to the setting of operational monads, when $\mathbb{S}$ is a pair of signatures over the category of endofunctors on $\mathsf{Set}$ (in the sense of Chapter 2): $\mathbb{S}$-reduction rules are defined in Section 6.2.1, and the associated notion of action in a $\mathbb{S}$-reduction monad is explained in Section 6.2.2, after introducing $\mathbb{S}$-reduction $\Sigma$-monads in Section 6.2.2, when $\Sigma$ is a signature for monads.

### 6.2.1 Definition of $\mathbb{S}$-reduction rules

In this subsection, $\Sigma$ is a signature for monads, and $\mathbb{S} = (\mathbb{S}_1, \mathbb{S}_2)$ is a pair of signatures over the category of endofunctors on $\mathsf{Set}$. We present our notion of $\mathbb{S}$*-reduction rule over* $\Sigma$, from which we build *operational signatures* in Section 6.3.

We need first to generalize the notion of $\Sigma$-module from Definition 76 to that of $(\Sigma, \mathbb{S})$-module:

**Definition 193.** We define a $(\Sigma, \mathbb{S})$**-module** to be a functor $T$ from the product of the three categories of models of $\Sigma$, $\mathbb{S}_1$, and $\mathbb{S}_2$ to the category $\int \mathsf{Mod}$ commuting with the forgetful functors to the category $\mathsf{Mon}$ of monads,

$$\begin{array}{ccc} \mathsf{Mon}^{\Sigma} \times [\mathsf{Set},\mathsf{Set}]^{\mathbb{S}_1} \times [\mathsf{Set},\mathsf{Set}]^{\mathbb{S}_2} & \xrightarrow{\quad T \quad} & \int \mathsf{Mod} \\ & \searrow \qquad \swarrow & \\ & \mathsf{Mon} & \end{array}$$

**Example 194.** To each $\Sigma$-module $\Psi$ is associated, by precomposition with the projection to $\mathsf{Mon}^{\Sigma}$, a $(\Sigma, \mathbb{S})$-module still denoted $\Psi$.

**Definition 195.** Let $S$ and $T$ be $(\Sigma, \mathbb{S})$-modules. We define a **morphism of $(\Sigma, \mathbb{S})$-modules** from $S$ to $T$ to be a natural transformation from $S$ to $T$ which becomes the identity when postcomposed with the forgetful functor $\int \mathsf{Mod} \to \mathsf{Mon}$.

Then we define $\mathbb{S}$-*term-pairs*:

**Definition 196.** Given a $\Sigma$-module $\mathcal{V}$, a **$\mathbb{S}$-term-pair from $\mathcal{V}$** is a pair $(n, p)$ of a natural number $n$ and a morphism of $(\Sigma, \mathbb{S})$-modules $p : \mathcal{V} \to (\mathcal{T}_1 \cdot \Theta^{(n)}) \times (\mathcal{T}_2 \cdot \Theta^{(n)})$, where $(\mathcal{T}_i \cdot \Theta^{(n)})$ is the $(\Sigma, \mathbb{S})$-module mapping a triple $(R, T_1, T_2)$ to the $R$-module $T_i \cdot R$.

We now give our definition of $\mathbb{S}$-*reduction rule*.

**Definition 197.** A **$\mathbb{S}$-reduction rule** $\mathcal{A} = (\mathcal{V}, (n_i, h_i)_{i \in I}, (n, c))$ **over** $\Sigma$ is given by:

- Metavariables: a $\Sigma$-module $\mathcal{V}$ of metavariables, that we sometimes denote by $\mathsf{MVar}_{\mathcal{A}}$;
- Hypotheses: a finite family of $\mathbb{S}$-term-pairs $(n_i, h_i)_{i \in I}$ from $\mathcal{V}$;
- Conclusion: a $\mathbb{S}$-term-pair $(n, c)$ from $\mathcal{V}$.

### 6.2.2 $\mathbb{S}$-Reduction $\Sigma$-monads

Similarly to Chapter 5, the purpose of a $\mathbb{S}$-reduction rule is to be validated in a $\mathbb{S}$-reduction monad $R$, that is, an operational monad whose underlying state functors are models of $\mathbb{S}$. Here again, as the hypotheses or the conclusion of the $\mathbb{S}$-reduction rule may refer to some operations specified by a signature $\Sigma$ for monads, this $\mathbb{S}$-reduction monad $R$ must be equipped with an action of $\Sigma$, hence the following definition.

**Definition 198.** Let $\Sigma$ be a signature for monads, and $\mathbb{S} = (\mathbb{S}_1, \mathbb{S}_2)$ be a pair of signatures over the category of endofunctors on $\mathsf{Set}$. The **category** $\mathbb{S}\text{-}\mathsf{RedMon}^{\Sigma}$ **of** $\mathbb{S}$**-reduction** $\Sigma$**-monads** is the category of models of the product of the following signatures:

- the pullback of $\Sigma$ along the forgetful functor from operational monads to monads;
- the pullback of $\mathbb{S}_1$ along the functor mapping an operational monad to its first state functor;
- the pullback of $\mathbb{S}_2$ along the functor mapping an operational monad to its second state functor.

In other words, it is the following pullback.

$$\begin{array}{ccc} \mathbb{S}\text{-}\mathsf{RedMon}^{\Sigma} \lrcorner & \longrightarrow & \mathsf{OpMon} \\ \downarrow & & \downarrow \\ \mathsf{Mon}^{\Sigma} \times [\mathsf{Set}, \mathsf{Set}]^{\mathbb{S}_1} \times [\mathsf{Set}, \mathsf{Set}]^{\mathbb{S}_2} & \longrightarrow & \mathsf{Mon} \times [\mathsf{Set}, \mathsf{Set}]^2 \end{array}$$

More concretely,

- a $\mathbb{S}$**-reduction** $\Sigma$**-monad** is an operational monad $(R, T_1, T_2)$ equipped with
  - an action $\rho$ of $\Sigma$ in $R$, thus inducing a $\Sigma$-monad that we denote also by $R$, or by $\underline{R}$ when we want to be explicit;
  - an action of $\mathbb{S}_i$ in $T_i$ for $i \in \{1, 2\}$, inducing a model of $\mathbb{S}_i$ that we denote also by $T_i$;
- a **morphism of** $\mathbb{S}$**-reduction** $\Sigma$**-monads** $R \to S$ is a morphism $f : R \to S$ of operational monads compatible with the actions of $\Sigma$, $\mathbb{S}_1$, and $\mathbb{S}_2$, i.e, whose underlying monad morphism is a $\Sigma$-monad morphism and whose underlying natural transformations between state functors are model morphisms.

### 6.2.3 Action of a $\mathbb{S}$-reduction rule

Let $\Sigma$ be a signature for monads and $\mathbb{S} = (\mathbb{S}_1, \mathbb{S}_2)$ be a pair of signatures over the category of endofunctors on $\mathsf{Set}$. In this section, we introduce the notion of *action of a* $\mathbb{S}$*-reduction rule over* $\Sigma$ *in a* $\mathbb{S}$*-reduction* $\Sigma$*-monad*. Intuitively, such an action is a “map from the hypotheses to the conclusion" of the $\mathbb{S}$-reduction rule. To make this precise, we need to first take the product of the hypotheses; this product is, more correctly, a *fibred* product.

**Definition 199.** Let $(n, p)$ be a $\mathbb{S}$-term-pair from a $\Sigma$-module $\mathcal{V}$, and $R$ be a $\mathbb{S}$-reduction $\Sigma$-monad. We denote by $p^*(\mathsf{Red}(R)^{(n)})$ the pullback of $\mathsf{red}_R^{(n)} : \mathsf{Red}(R)^{(n)} \to (T_1 \cdot R^{(n)}) \times (T_2 \cdot R^{(n)})$ along $p_R : \mathcal{V}(R) \to (T_1 \cdot R^{(n)}) \times (T_2 \cdot R^{(n)})$:

$$\begin{array}{ccc} p^*(\mathsf{Red}(R)^{(n)}) & \longrightarrow & \mathsf{Red}(R)^{(n)} \\ \downarrow & & \downarrow \mathsf{red}_R^{(n)} \\ \mathcal{V}(R) & \xrightarrow[p_R]{} & (T_1 \cdot R^{(n)}) \times (T_2 \cdot R^{(n)}) \end{array} .$$

We denote by $p^*(\mathsf{red}_R^{(n)}) : p^*(\mathsf{Red}(R)^{(n)}) \to \mathcal{V}(R)$ the projection morphism on the left.

**Definition 200.** Let $\mathcal{A} = (\mathcal{V}, (n_i, h_i)_{i \in I}, (n, c))$ be a $\mathbb{S}$-reduction rule, and $R$ be a $\mathbb{S}$-reduction $\Sigma$-monad.

The **$R$-module $\mathsf{Hyp}_\mathcal{A}(R)$ of hypotheses of $\mathcal{A}$** is $\prod_{i \in I / \mathcal{V}(R)} h_i^* \mathsf{Red}(R)^{(n_i)}$ the fiber product of all the $R$-modules $h_i^* \mathsf{Red} R^{(n_i)}$ along their projection to $\mathcal{V}(R)$. It thus comes with a projection $\mathsf{hyp}_\mathcal{A}(R) : \mathsf{Hyp}_\mathcal{A}(R) \to \mathcal{V}(R)$.

The **$R$-module $\mathsf{Con}_\mathcal{A}(R)$ of conclusion of $\mathcal{A}$** is $c^* \mathsf{Red}(R)^{(n)}$. It comes with a projection $\mathsf{con}_\mathcal{A}(R) : \mathsf{Con}_\mathcal{A}(R) \to \mathcal{V}(R)$.

**Definition 201.** Let $\mathcal{A}$ be a $\mathbb{S}$-reduction rule over $\Sigma$. An **action of $\mathcal{A}$ in a $\mathbb{S}$-reduction $\Sigma$-monad** $R$ is a morphism, in the slice category $\mathsf{Mod}(R)/\mathsf{MVar}_\mathcal{A}(R)$, between $\mathsf{hyp}_\mathcal{A}(R)$ and $\mathsf{con}_\mathcal{A}(R)$, that is, a morphism of $R$-modules

$$\tau : \mathsf{Hyp}_\mathcal{A}(R) \to \mathsf{Con}_\mathcal{A}(R)$$

making the following diagram commute:

$$\begin{array}{ccc} \mathsf{Hyp}_\mathcal{A}(R) & \xrightarrow{\tau} & \mathsf{Con}_\mathcal{A}(R) \\ & \searrow \quad \swarrow & \\ & \mathsf{MVar}_\mathcal{A}(R) & \end{array} . \tag{6.1}$$

**Remark 202.** Any $\mathbb{S}$-reduction rule $\mathcal{A}$ over $\Sigma$ defines an arity $(\mathsf{D}, a, u, v)$ over the category of $\mathbb{S}$-reduction $\Sigma$-monads (in the sense of Chapter 2) as follows:

- $a : \mathsf{D} \to \mathbb{S}\text{-}\mathsf{RedMon}^{\Sigma}$ is the fibration corresponding (through the Grothendieck construction) to the functor mapping a $\mathbb{S}$-reduction $\Sigma$-monad $R$ to $\mathsf{Mod}(R)/\mathsf{MVar}_{\mathcal{A}}(R)$.
- $u$ maps a $\mathbb{S}$-reduction $\Sigma$-monad to $\mathsf{hyp}_{\mathcal{A}}(R) : \mathsf{Hyp}_{\mathcal{A}}(R) \to \mathsf{MVar}_{\mathcal{A}}(R)$;
- $v$ maps a $\mathbb{S}$-reduction $\Sigma$-monad to $\mathsf{con}_{\mathcal{A}}(R) : \mathsf{Con}_{\mathcal{A}}(R) \to \mathsf{MVar}_{\mathcal{A}}(R)$.

This arity yields the same notion of action.

## 6.3 Signatures for operational monads and Initiality

In this section, we define the notion of *operational signature*, as particular signatures over the category of operational monads. They consist of three parts:

- a specification of the pair $T$ of state functors, as a pair of signatures $\mathbb{S} = (\mathbb{S}_1, \mathbb{S}_2)$ over the category of endofunctors on $\mathsf{Set}$,
- a signature for the monad underlying $R$,
- and a family of $\mathbb{S}$-reduction rules over $\Sigma$.

In Section 6.3.4, we review Fiore and Hur's notion of equational system [FH09] that yield effective signatures for specifying endofunctors. In Sections 6.3.1 and 6.3.3, we describe the *category of models* associated to each operational signature. Our main result, Theorem 207, states that an operational signature is effective as soon as its underlying signatures for monads and for endofunctors are effective.

### 6.3.1 Operational signatures

We define here *operational signatures* and their *models*.

**Definition 203.** An **operational signature** is a signature over the category of operational monads consisting of:

- the product signature of
  - the pullback of a signature $\Sigma$ for monads along the forgetful functor from operational monads to monads;

- the pullback of a signature $\mathbb{S}_1$ for endofunctors along the functor mapping an operational monad to its first state functor;
  - the pullback of a signature $\mathbb{S}_2$ for endofunctors along the functor mapping an operational monad to its second state functor.

- a family of arities induced by $(\mathbb{S}_1, \mathbb{S}_2)$-reduction rules over $\Sigma$, as explained in Remark 202.

An operational signature is thus determined by two signatures for endofunctors $\mathbb{S}_1$ and $\mathbb{S}_2$, a signature for monads $\Sigma$, and a family of reduction rules $\mathfrak{R}$ over $\Sigma$. We denote $(\mathbb{S}_1, \mathbb{S}_2, \Sigma, \mathfrak{R})$ the induced operational signature.

Examples are given in Section 6.4.

We unfold the definition of model for operational signatures:

**Remark 204.** Let $\mathcal{O} = (\mathbb{S}_1, \mathbb{S}_2, \Sigma, \mathfrak{R})$ be an operational signature. Let $\mathbb{S} := (\hat{\mathbb{S}}_1, \hat{\mathbb{S}}_2)$. A **model of** $\mathcal{O}$ is a $\mathbb{S}$-reduction $\Sigma$-monad equipped with an action of each reduction rule of $\mathfrak{R}$.

### 6.3.2 Morphisms of models

Here we unfold the definition of morphism between models of a operational signature: it relies on the functoriality of the assignments $R \mapsto \mathsf{Hyp}_{\mathcal{A}}(R)$ and $R \mapsto \mathsf{Con}_{\mathcal{A}}(R)$, for a given reduction rule $\mathcal{A}$ on a signature $\Sigma$ for monads.

**Definition 205.** Let $\mathbb{S}$ be a pair of signatures over the category of endofunctors on $\mathsf{Set}$. Let $\Sigma$ be a signature for monads, and $\mathcal{A}$ be a $\mathbb{S}$-reduction rule over $\Sigma$. Definition 200 assigns to each $\mathbb{S}$-reduction $\Sigma$-monad $R$ the $R$-modules $\mathsf{Hyp}_{\mathcal{A}}(R)$ and $\mathsf{Con}_{\mathcal{A}}(R)$. These assignments extend to functors $\mathsf{Hyp}_{\mathcal{A}}, \mathsf{Con}_{\mathcal{A}} : \mathbb{S}\text{-}\mathsf{RedMon}^{\Sigma} \to \int \mathsf{Mod}$.

**Proposition 206.** *Given the same data, the functors* $\mathsf{Hyp}_{\mathcal{A}}$ *and* $\mathsf{Con}_{\mathcal{A}}$ *commute with the forgetful functors to* $\mathsf{Mon}$*:*

$$\begin{array}{ccc} \mathbb{S}\text{-}\mathsf{RedMon}^{\Sigma} & \overset{\mathsf{Hyp}_{\mathcal{A}}}{\underset{\mathsf{Con}_{\mathcal{A}}}{\rightrightarrows}} & \int \mathsf{Mod} \\ & \searrow \quad \swarrow & \\ & \mathsf{Mon} & \end{array}$$

Now, a morphism between models $R$ and $S$ of a signature $\mathcal{O} = (\mathbb{S}_1, \mathbb{S}_2, \Sigma, \mathfrak{R})$ is a morphism $f$ of $\mathbb{S}$-reduction $\Sigma$-monads commuting with the action of any $\mathbb{S}$-reduction rule, in the sense that for any $\mathbb{S}$-reduction rule $\mathcal{A} \in \mathfrak{R}$, the following diagram of natural transformations commutes:

$$\begin{array}{ccc}
\mathsf{Hyp}_{\mathcal{A}}(R) & \longrightarrow & \mathsf{Con}_{\mathcal{A}}(R) \\
{\scriptstyle \mathsf{Hyp}_{\mathcal{A}}(f)}\downarrow & & \downarrow{\scriptstyle \mathsf{Con}_{\mathcal{A}}(f)} \\
\mathsf{Hyp}_{\mathcal{A}}(S) & \longrightarrow & \mathsf{Con}_{\mathcal{A}}(S)
\end{array}$$

### 6.3.3 The main result

We state our main result, Theorem 207, which gives a sufficient condition for an operational signature to be effective.

**Theorem 207.** *Let $\mathcal{O} = (\mathbb{S}_1, \mathbb{S}_2, \Sigma, \mathfrak{R})$ be an operational signature. If $\mathbb{S}_1$, $\mathbb{S}_2$, and $\Sigma$ are effective, then so is $\mathcal{O}$.*

*Proof.* This is a close adaptation of the proof detailed in Section 5.5. Therefore, we only sketch the main arguments.

Consider the two following pullbacks, the left one defining $\mathsf{OpMon}^{\Sigma,\mathbb{S}}$:

$$\begin{array}{ccccc}
\mathsf{OpMon}^{\Sigma,\mathbb{S}} & \longrightarrow & \mathsf{OpMon} & \longrightarrow & V(\int \mathsf{Mod}) \\
\downarrow & \lrcorner & \downarrow & \lrcorner & \downarrow{\scriptstyle \mathrm{cod}} \\
\mathsf{Mon}^{\Sigma} \times [\mathsf{Set}, \mathsf{Set}]^{\mathbb{S}_1} \times [\mathsf{Set}, \mathsf{Set}]^{\mathbb{S}_2} & \longrightarrow & \mathsf{Mon} \times [\mathsf{Set}, \mathsf{Set}]^2 & \xrightarrow[(R,T_1,T_2)\mapsto(T_1\cdot R, T_2\cdot R)]{} & \int \mathsf{Mod}
\end{array}$$

Then, the left vertical arrow is a fibration by the same argument detailed in Lemma 164. Moreover, the category of models of $\mathcal{O}$ is isomorphic to the category of vertical algebras of an endofunctor $F_{\mathcal{O}}$ on $\mathsf{OpMon}^{\Sigma,\mathbb{S}}$, by the same argument as in Proposition 161. Then, this category of models is fibered over $\mathsf{Mon}^{\Sigma} \times [\mathsf{Set}, \mathsf{Set}]^{\mathbb{S}_1} \times [\mathsf{Set}, \mathsf{Set}]^{\mathbb{S}_2}$ by Lemma 163, and the fiber category over $(R, T_1, T_2)$ is the category of algebras of the restriction of $F_{\mathcal{O}}$ as a finitary endofunctor on $\mathsf{Mod}(R)/(T_1 \cdot R \times T_2 \cdot R)$, following an argument similar to that of Lemmas 166 and 169.

Now, we apply Lemma 165: we only need to show that the fiber category over $(\hat{\Sigma}, \widehat{\mathbb{S}}_1, \widehat{\mathbb{S}}_2)$ has an initial object, that is, we must construct an initial algebra for a finitary endofunctor thanks to the previous remark. Adámek's theorem (Lemma 167) allows to conclude.

□

We already know a class of effective signatures for monads: they are the algebraic 2-signatures (Theorem 107). In the next section, we introduce a similar notion of algebraic signature for endofunctors based on equational systems whose effectivity is ensured by [FH09]. This will allow us to state Corollary 220, as a consequence of Theorem 207, that we will use for our examples.

### 6.3.4 Specifying the state functors

We choose to rely on Fiore and Hur's notion of equational system [FH09]. Thanks to Example 14, they can be considered as signatures in the sense of Chapter 2.

A first definition of signature for endofunctors on $\mathsf{Set}$ that we can think of is an endofunctor on $[\mathsf{Set}, \mathsf{Set}]$ whose algebras are the models of the signature.

**Definition 208.** A **1-signature for endofunctor** (or just 1-signature in this subsection) is a signature over the category of endofunctors on $\mathsf{Set}$ consisting of a single arity induced by an endofunctor $\Sigma$ on $\mathsf{Set}$, as described in Example 3. A 1-signature is thus uniquely determined by an endofunctor $\Sigma$ on $[\mathsf{Set}, \mathsf{Set}]$, and for this reason we identify 1-signatures with endofunctors on the category $[\mathsf{Set}, \mathsf{Set}]$ in the following.

**Proposition 209.** *The category of endofunctors on* $[\mathsf{Set}, \mathsf{Set}]$ *is complete and cocomplete: limits and colimits are computed pointwise.*

**Notation 210.** *The following is reminiscent of 1-signatures for monads:*

- $\Theta$ *is the identity endofunctor* $\mathsf{Id}$ *on the category of endofunctors on* $\mathsf{Set}$*;*
- *if* $\Sigma$ *is a 1-signature for endofunctors, we denote by* $\Sigma^{(n)}$ *its* $n^{th}$ *derivative, where* $\Sigma^{(n)}(F) := \Sigma(F)^{(n)}$ *mapping a set* $X$ *to the set* $\Sigma(F)(X+n)$*, for any endofunctor* $F$*;*
- $\Sigma' := \Sigma^{(1)}$*, for any 1-signature* $\Sigma$ *for endofunctors;*
- *if* $F$ *is an endofunctor on* $\mathsf{Set}$*, the constant 1-signature mapping any functor to* $F$ *is denoted by* $\underline{F}$*;*
- *if* $\Sigma = \coprod_{i \in I} \Theta^{(n_{j_1})} \times \cdots \times \Theta^{(n_{j_i})}$ *is an algebraic parametric module, we denote* $\overline{\Sigma}$ *the 1-signature for endofunctors defined by the same formula.*

The clash of notations with parametric modules is justified by the fact that given any monad $R$ and algebraic parametric module $\Sigma$ for monads, the functor underlying the image of $R$ by $\Sigma$ is the image of the functor underlying $R$ by $\overline{\Sigma}$. More formally:

**Proposition 211.** *Let $\Sigma$ be an algebraic parametric module. Then, the following diagram commutes:*

$$\begin{array}{ccc} \mathsf{Mon} & \xrightarrow{\Sigma} & \int \mathsf{Mod} \\ \downarrow & & \downarrow \\ [\mathsf{Set}, \mathsf{Set}] & \xrightarrow[\overline{\Sigma}]{} & [\mathsf{Set}, \mathsf{Set}] \end{array}$$

We now introduce algebraic 1-signatures for endofunctors, which satisfy an initiality theorem:

**Definition 212.** A 1-signature $\Sigma$ for endofunctors is said **algebraic** if it is a coproduct of endofunctors of the shape $\underline{\mathsf{Id}_{\mathsf{Set}}}^{(i_1)} \times \cdots \times \underline{\mathsf{Id}_{\mathsf{Set}}}^{(i_n)} \times \Theta^{(j_1)} \times \cdots \times \Theta^{(j_m)}$.

**Example 213.** If $\Sigma$ is an algebraic 1-signature for monads, then $\overline{\Sigma}$ is an algebraic 1-signature for endofunctors.

**Proposition 214.** *Any algebraic 1-signature $\Sigma$ for endofunctors is effective.*

*Proof.* This follows from Adámek's theorem (Lemma 167), as the endofunctor underlying an algebraic 1-signature is finitary. □

Actually, the functor underlying the initial model of an algebraic 1-signature for monads is specified by an appropriate algebraic 1-signature for endofunctors:

**Proposition 215.** *Let $\Sigma$ be an algebraic 1-signature for monads. Then the functors underlying $\hat{\Sigma}$ and $\widehat{\overline{\Sigma} + \underline{\mathsf{Id}_{\mathsf{Set}}}}$ are isomorphic.*

Next, we would like to specify equations. To this end, we use equational systems [FH09], that we rephrase using our terminology. We focus on equations of the shape $u_R = v_R : \Psi(R) \to R$, where

- $R$ is an algebra of a an endofunctor $\Sigma$ (that is, $R$ is a monad of the induced 1-signature by $\Sigma$);
- $\Psi(R)$ is an endofunctor on $\mathsf{Set}$;
- $u_R$ and $v_R$ are parallel natural transformations between $\Psi(R)$ and $R$.

The assignment $R \mapsto (\Psi(R), u_R, v_R)$ is required to be functorial.

**Definition 216.** Let $\Sigma$ be an endofunctor on $[\mathsf{Set}, \mathsf{Set}]$. A **$\Sigma$-equation** is an equational $\Sigma$-arity induced by an equational system $[\mathsf{Set}, \mathsf{Set}] : \Sigma \triangleright \Gamma \vdash L = R$ as described in Example 10.

It is uniquely determined by a triple $(\Gamma, u, v)$ consisting of:

- an endofunctor $\Gamma$ on $[\mathsf{Set}, \mathsf{Set}]$;
- a pair of parallel natural transformations $u, v$ from $\Gamma$ to $\Theta$;

Thus we denote by $(\Gamma, u, v)$ the induced $\Sigma$-equation. It is said **algebraic** if $\Gamma$ is.

**Definition 217.** A **2-signature for endofunctors** is a signature over the category of endofunctors consisting of a 1-signature $\Sigma$ and a family of $\Sigma$-equations. It is said **algebraic** if $\Sigma$ is and $E$ consist of algebraic $\Sigma$-equations.

The associated category of models is the full subcategory of models $R$ of $\Sigma$ satisfying $u_R = v_R$ for each $(\Psi, u, v) \in E$.

**Proposition 218.** *Any algebraic 2-signature for endofunctors is effective.*

*Proof.* A first step consists in combining all the equations of the 2-signature into a single one, using coproducts. Then, the result follows from [FH09, Theorem 4.7]. □

**Example 219.** Recalling Example 192, we would like to specify the state functor $T : \mathsf{Set} \to \mathsf{Set}$, such that $T(A)$ denotes the set of $\pi$-calculus processes up to structural congruence with free channels in $A$. For this, let us consider the endofunctor on $[\mathsf{Set}, \mathsf{Set}]$ defined for all $X \in [\mathsf{Set}, \mathsf{Set}]$ and $\gamma \in \mathsf{Set}$ by

$$\begin{array}{rclccccccccccc} \Sigma(X)(\gamma) & = & 1 & + & X(\gamma)^2 & + & X(\gamma) & + & X(\gamma+1) & + & \gamma^2 \times X(\gamma) & + & \gamma \times X(\gamma+1) \\ P, Q & ::= & 0 & \mid & (P|Q) & \mid & !P & \mid & \nu a.P & \mid & \overline{a}\langle b\rangle.P & \mid & a(b).P \\ (\Sigma & = & 1 & + & \Theta^2 & + & \Theta & + & \Theta' & + & \mathsf{Id}^2 \times \Theta & + & \mathsf{Id} \times \Theta') \end{array}$$

(with corresponding pieces of syntax below each term of the sum, and using the notations of Definition 212 for the last line). As an example of $\Sigma$-equation $(\Psi, u, v)$, commutativity of parallel composition would have $\Psi(X)(\gamma) = X(\gamma)^2$. The first natural transformation $u : \Psi \to \Theta$ maps any $\Sigma$-alg, say $\rho : \Sigma X \to X$, to the natural transformation with component $u_{X,\rho,\gamma} : \Psi(X)(\gamma) = X(\gamma)^2 \to X(\gamma)$ at $\gamma$ given by parallel composition, i.e., the composite

$$X(\gamma)^2 \xhookrightarrow{(inj_2)_\gamma} \Sigma(X)(\gamma) \xrightarrow{\rho_\gamma} X(\gamma) ,$$

and $\upsilon_{X,\rho,\gamma}$ given by swapping, and then parallel composition. All equations may be treated similarly, forming a signature $\mathbb{S}$ for endofunctors which is algebraic, whose initial algebra is the desired $T$.

Proposition 218 and Theorem 207 entail the following corollary of Theorem 207.

**Corollary 220.** *Let $\mathcal{O} = (\mathbb{S}_1, \mathbb{S}_2, \Sigma, \mathfrak{R})$ be an operational signature. If $\mathbb{S}_1$ and $\mathbb{S}_2$ are algebraic 2-signatures for endofunctors and $\Sigma$ is an algebraic 2-signature for monads, then $\mathcal{O}$ is effective.*

All the examples of operational signatures considered here satisfy the algebraicity condition of Corollary 220.

## 6.4 Examples of operational signatures

We give examples of operational signatures for the call-by-value lambda calculus in Section 6.4.1, for the big-step operational semantics variant in Section 6.4.2, and for the $\pi$-calculus in Section 6.4.3. The reduction rules are given following the schematic presentation described in Section 5.3.5. These signatures all satisfy the hypotheses of Corollary 220, and thus are effective.

### 6.4.1 Call-by-value lambda calculus

We give an operational signature for the operational monad $((B, B), \mathsf{LC}_v)$ of the call-by-value lambda calculus from Example 190. Recall that $B$ is the monad of binary trees underlying the initial model of the 1-signature $\Theta \times \Theta$. By Proposition 215, the corresponding algebraic signature for endofunctor is $\mathbb{S} = (\Theta \times \Theta + \underline{\mathsf{Id}}_{\mathsf{Set}}, \emptyset)$. A model $T$ of $\mathbb{S}$ comes equipped with a binary operation $\mathsf{app} : T \times T \to T$ and a variable operation $v : \underline{\mathsf{Id}}_{\mathsf{Set}} \to T$. We denote by $j_T : B \to T$ the initial morphism.

Now, we give a signature for the monad $\mathsf{LC}_v$ of values of the lambda calculus. A value is either a variable or an abstracted lambda term, and we argued in Example 190 that a lambda term can be specified as a binary tree whose leaves are values. Thus, we choose the 1-signature $B \cdot \Theta'$ to specify the monad of values. Note that this is algebraic as $B$ is polynomial: we have

$$B \cdot \Theta' = \coprod_{n \in \mathbb{N}} B_n \times (\Theta')^n,$$

where $B_n$ is the set of binary trees with $n$ leaves. A model of this signature is a monad $R$ with a module morphism $\mathsf{abs}_R : B \cdot R' \to R$.

We give now the family $\mathfrak{R}$ of reduction rules:

$$\mathsf{app}(v(\mathsf{abs}(T)), v(U))) \rightsquigarrow j(T)\{* := U\} \qquad \frac{j(T) \rightsquigarrow j(T')}{v(\mathsf{abs}(T)) \rightsquigarrow v(\mathsf{abs}(T'))}$$

$$\frac{T \rightsquigarrow T'}{\mathsf{app}(T, U) \rightsquigarrow \mathsf{app}(T', U)} \qquad \frac{U \rightsquigarrow U'}{\mathsf{app}(T, U') \rightsquigarrow \mathsf{app}(T, U')}.$$

To conclude, the operational signature is $(\mathbb{S}, \mathbb{S}, B \cdot \Theta', \mathfrak{R})$.

### 6.4.2 Call-by-value lambda calculus with big-step operational semantics

We give a signature for the operational monad $((B, \mathsf{Id}), \mathsf{LC}_v)$ of the call-by-value lambda calculus with big-step operational semantics (Example 191). The first state functor $B$ is specified by $\mathbb{S}$ as in Section 6.4.1, whereas the second one $\mathsf{Id}$ is specified by the algebraic 1-signature for functors $\underline{\mathsf{Id}}_{\mathsf{Set}}$: a model $M$ is an endofunctor with a natural transformation $w : \mathsf{Id}_{\mathsf{Set}} \to M$

The underlying signature for monads is the same as the one of the call-by-value lambda calculus (Section 6.4.1). Now, we give the family $\mathfrak{R}$ of reduction rules, using the same notations as in Section 6.4.1:

$$\frac{}{v(T) \rightsquigarrow w(T)} \qquad \frac{T \rightsquigarrow w(\mathsf{abs}(T')) \quad U \rightsquigarrow w(U') \quad j(T')\{* := U'\} \rightsquigarrow V}{\mathsf{app}(T, U)) \rightsquigarrow V}.$$

To conclude, the operational signature is $(\mathbb{S}, \underline{\mathsf{Id}}_{\mathsf{Set}}, B \cdot \Theta', \mathfrak{R})$, where $\mathbb{S}$ is defined in Section 6.4.1.

### 6.4.3 $\pi$-calculus

We give an operational signature for the operational monad $((T, T), \mathsf{Id})$ of the $\pi$-calculus described in Example 192. In Example 219, we gave an algebraic signature $\mathbb{S}$ for specifying the endofunctor $T$. A model $M$ of $\mathbb{S}$ comes equipped with the following natural transformations:

- $\mathsf{par} : M \times M \to M$ mapping $(P, Q)$ to $P|Q$;
- $\mathsf{out} : \mathsf{Id} \times \mathsf{Id} \times M \to M$ mapping $(a, b, P)$ to $\overline{a}\langle b\rangle.P$;
- $\mathsf{in} : \mathsf{Id} \times M' \to M$ mapping $(a, P) \in X \times M(X + \{*\})$ to $a(*).P$;
- $\mathsf{abs} : M' \to M$ mapping $P \in M(X + \{*\})$ to $\nu * .P$.

Note that the monad $\mathsf{Id}_{\mathsf{Set}}$ is the initial model of the empty signature $0$.
The family $\mathfrak{R}$ of reduction rules over $0$ consists of the following:

| Original rule | As a $\mathbb{S}$-reduction rule |
|---|---|
| $\overline{a}\langle b\rangle.P\vert a(c).Q \rightsquigarrow P\vert(Q\{c := b\})$ | $\mathsf{par}(\mathsf{out}(a, b, j_1(P)), \mathsf{in}(a, j_1(Q))) \rightsquigarrow \mathsf{par}(j_2(P), j_2(T[\mathtt{id}, \underline{b}](Q)))$ |
| $\dfrac{P \rightsquigarrow Q}{P\vert R \rightsquigarrow Q\vert R}$ | $\dfrac{P \rightsquigarrow Q}{\mathsf{par}(P, j_1(R)) \rightsquigarrow \mathsf{par}(Q, j_2(R))}$ |
| $\dfrac{P \rightsquigarrow Q}{\nu a.P \rightsquigarrow \nu a.Q}$ | $\dfrac{P \rightsquigarrow Q}{\mathsf{abs}(P) \rightsquigarrow \mathsf{abs}(Q)}$ |

where:

- $j_i : T \to M_i$ is the initial $\mathbb{S}$-model morphism to the underlying state functor $M_i$ of a $\mathbb{S}$-reduction $\Sigma$-monad, for $i \in \{1, 2\}$;
- $T[\mathtt{id}, b]$ is the function $T[\mathtt{id}, \underline{b}] : T(R(X) + 1) \to T(R(X))$, for $\underline{b} : 1 \to R(X)$ the function selecting $b \in R(X)$.

To conclude, the operational signature is $(\mathbb{S}, \underline{\mathsf{Id}}_{\mathsf{Set}}, 0, \mathfrak{R})$.

# Conclusion

In this thesis, we have studied different notions of signatures that are all particular instances of the general definition that we give in Chapter 2.

In Chapters 3 and 4, we have presented notions of signatures for monads and their models. More precisely, in Chapter 3, we have defined the class of *presentable* signatures, which are quotients of traditional algebraic signatures. Presentable signatures are closed under various operations, including colimits. One of the main results of this chapter says that any presentable signature is effective. Despite the fact that the constructions in Section 3.7 make heavy use of quotients, there is no need to appeal to the axiom of choice. While a previous version of the formalisation did use the axiom of choice to show that certain functors preserve epimorphisms, we managed subsequently to prove this without using the axiom of choice. This analysis, and subsequent reworking, of the proof was significantly helped by the formalisation.

One difference to other work on Initial Semantics, e.g., [MU04; GU03; Fio08; FM10], is that we do not rely on the notion of strength. However, a signature endofunctor with strength as used in the aforementioned articles can be translated to a high-level signature as presented in this work (Proposition 44).

In Chapter 4, we extend the notion of signature for monads to take into account more general equations. This yields the definition of a 2-signature, as a pair of a 1-signature $\Sigma$ (that is, a signature in the sense of Chapter 3) and a set of $\Sigma$-*equations* that must be satisfied.

Finally, in Chapter 5, we have introduced the notions of reduction monad and reduction signature: they are meant to model syntax with a notion of reduction. For each such signature, we define a category of models, equipped with a forgetful functor to the category of reduction monads. We say that a reduction signature is effective if its associated category of models has an initial object; in this case, we say that the reduction monad underlying the initial object is generated by the signature. We identify a simple sufficient condition for a reduction signature to be effective. Chapter 6 generalizes the notion of reduction monad to that of operational monad, and adapts the statement of Chapter 5 accordingly. These two chapters is a first step towards a theory for the al-gebraic specification of programming languages and their semantics. In future work, we aim at generalizing our notion of signature to encompass richer languages and to present a notion of signature that allows for the specification of equalities between reductions (cf. Remark 153).

We anticipate that our work extends to simply-typed languages, by changing the base category $\mathsf{Set}$ to a presheaf category $\mathsf{Set}^T$, where $T$ is the set of simple types that we are interested in.

UNIVERSITE BRETAGNE LOIRE / MATHSTIC

UNIVERSITÉ SAVOIE MONT BLANC

IMT Atlantique
Bretagne-Pays de la Loire
École Mines-Télécom

**Titre :** Signatures et modèles pour la syntaxe et la sémantique opérationnelle en présence de liaison de variables

**Résumé :** Cette thèse traite de la spécification et la construction de la syntaxe et sémantique opérationnelle d'un langage de programmation. Nous travaillons avec une notion générale de "signature" pour spécifier des objets d'une catégorie donnée comme des objets initiaux dans une catégorie appropriée de modèles. Cette caractérisation, dans l'esprit de la sémantique initiale, donne une justification du principe de récursion.

Les languages avec liaisons, telles que le lambda calcul pur, sont des monades sur la catégorie des ensembles spécifiées par les signatures algébriques classiques. Les premières extensions de syntaxes avec des équations que nous considérons sont des "quotients" de ces signatures algébriques. Ils permettent, par exemple, de spécifier une opération commutative binaire. Cependant, certaines équations, comme l'associativité, semblent hors d'atteinte. Ceci motive la notion de 2-signature qui complète la définition précédente avec la donnée d'un ensemble d'équations. Nous identifions la classe des "2-signatures algébriques" pour lesquelles l'existence de la syntaxe associée est garantie.

Finalement, nous abordons la spécification de la sémantique opérationnelle d'un langage de programmation tel que le lambda calcul avec $\beta$-réduction. Nous introduisons à cette fin la notion de monade de réduction et leur signatures, puis les généralisons pour aboutir à la notion de monade opérationnelle.

**Title:** Signatures and models for syntax and operational semantics in the presence of variable binding

**Abstract:** This thesis deals with the specification and construction of syntax and operational semantics of a programming language. We work with a general notion of "signature" for specifying objects of a given category as initial objects in a suitable category of models. This characterization, in the spirit of Initial Semantics, gives a justification of the recursion principle. Languages with variable binding, such as the pure lambda calculus, are monads on the category of sets specified through the classical algebraic signatures. The first extensions to syntaxes with equations that we consider are "quotients" of these algebraic signatures. They allow, for example, to specify a binary commutative operation. But some equations, such as associativity, seem to remain out of reach. We thus introduce the notion of 2-signature, consisting in two parts: a specification of operations through a usual signature as before, and a set of equations among them. We identify the class of "algebraic 2-signatures" for which the existence of the associated syntax is guaranteed.

Finally, we takle the specification of the operational semantics of a programming language such as lambda calculus with $\beta$-reduction. To this end, we introduce the notion of reduction monad and their signatures, then we generalize them to get the notion of operational monad.